\pdfoutput=1
\documentclass[PAPER, atlasdraft=false, UKenglish, cernpreprint, texmf, orcidlogo]{atlasdoc}
\usepackage[subcaption, rotating, longtable, multirow, block=space]{atlaspackage}
\usepackage{standalone}
\usepackage{adjustbox}
\usepackage{enumerate}
\usepackage{lscape}
\usepackage{array}
\newcolumntype{H}{>{\setbox0=\hbox\bgroup\let\pm\relax}c<{\egroup}@{}}

\usepackage{atlasbiblatex}

\usepackage[hepparticle, hepprocess, snippets, xref=false, unit=false]{atlasphysics}
\newcolumntype{~}{>{\global\let\currentrowstyle\relax}}
\newcolumntype{^}{>{\currentrowstyle}}
\newcommand{\rowstyle}[1]{\gdef\currentrowstyle{#1}%
#1\ignorespaces
}

\newif\iftHqOutput
\makeatletter
\tHqOutputfalse
\makeatother

\newif\iftHqtrilep
\newif\iftHqdilepSS
\newif\iftHqdilepSStau
\newif\iftHqdilepOStau
\newif\iftHqlepditau
\newif\iftHqlepbb
\tHqtrileptrue
\tHqdilepSStrue
\tHqdilepSStaufalse
\tHqdilepOStaufalse
\tHqlepditaufalse
\tHqlepbbtrue

\graphicspath{{logos/}{figures/}}

\usepackage{ANA-HIGP-2024-03-PAPER-defs}

\AtlasTitle{Search for the production of a Higgs boson in association
with a single top quark in $pp$ collisions at $\sqrt{s} =
\SI{13}{\TeV}$ with the ATLAS detector}

\AtlasAbstract{%
A search for the production of a Higgs boson in association with a
single top quark, \tH, is presented. The analysis uses proton--proton
collision data corresponding to an integrated luminosity of
\SI{140}{\per\fb} at a centre-of-mass energy of $\SI{13}{\TeV}$,
collected by the ATLAS detector at the LHC\@.
The search targets Higgs-boson decays into $b\bar{b}$, $WW^{*}$, $ZZ^{*}$, and
$\tau\tau$, accompanied by an isolated lepton (electron or muon) from the
top-quark decay. Multivariate techniques are employed to enhance the
separation between signal and background processes.
The observed signal strength, $\mu_{\tH}$, defined as the ratio between the measured cross-section and the
predicted Standard Model value, is \mutHresult.
The significance of the observed (expected) signal above the background-only expectation is \significanceObs\ (\significanceExp) standard deviations.
The corresponding observed (expected) upper limit at the 95$\%$ confidence level on
the \tH cross-section is found to be \upperLimitObs~(\upperLimitExp) times the value
predicted by the Standard Model.
An interpretation with an inverted sign of the top-quark Yukawa coupling is performed, and the signal strength and corresponding limit are reported.
}

\AtlasRefCode{HIGP-2024-03}

\PreprintIdNumber{CERN-EP-2025-178}

\AtlasJournal{JHEP 10 (2025) 093}
\AtlasDOI{10.1007/JHEP10(2025)093}

\hypersetup{pdftitle={ATLAS document},pdfauthor={The ATLAS Collaboration}}

\begin{document}

\maketitle

\tableofcontents

%
%
%

%
%

%

%
\section{Introduction}%
\label{sec:introduction}

The discovery of a particle with properties consistent with the Higgs
boson~\cite{PhysRevLett.13.321,PhysRevLett.13.508} of the Standard Model (SM) in 2012 by the ATLAS and CMS
experiments~\cite{HIGG-2012-27,CMS-HIG-12-028} opened a new field of
exploration in the realm of particle physics.
Studying the Higgs-boson couplings is essential to test whether the
observed particle corresponds to the Higgs boson predicted by the
SM. In the SM, the strength of these interactions with other
elementary particles is proportional to their masses (or masses
squared).

Specifically, the Higgs boson interacts with fermions via Yukawa
couplings~\cite{Peskin:1995ev}. Among these, the Yukawa coupling to the top quark, $y_t$, is of
special interest as it is the largest with a value close to unity.
Given its large
value, $y_t$ plays a crucial role in radiative corrections to the
Higgs potential~\cite{Degrassi:2012ry,Buttazzo:2013uya}, influencing the shape of the potential and,
consequently, the stability of the electroweak (EW) vacuum~\cite{Englert_2014}.
In addition, it significantly affects the running of SM parameters, impacting the
scale at which new physics might appear~\cite{Buttazzo:2013uya}.
For all these reasons the Yukawa coupling of the Higgs boson to the top quark is considered
a good probe for new physics. In particular, the study of its charge-conjugation and parity (CP) properties
could provide insight into the observed baryon asymmetry of the
universe~\cite{Gunion:1996xu,Ellis:2013yxa,He:2014xla}.

The magnitude of $y_t$ can be determined indirectly from Higgs-boson
production via a top-quark loop or directly through cross-section
measurements of top-quark pair production in association with a Higgs
boson (\ttH).
The \ttH\ production process was
observed~\cite{HIGG-2018-13,CMS-HIG-17-035} in 2018.
The production cross-section of the Higgs boson in association with a
single top quark is sensitive to both the magnitude of $y_t$ and its
sign~\cite{Tait:2000sh,Barger:2009ky,Biswas:2012bd,Farina:2012xp,Chang:2014rfa}. While the sign of $y_t$ is not defined in an absolute sense, it is
instead defined relative to the coupling strength of the \PW boson to
the Higgs boson.
In the following, this relative-sign convention is adopted throughout,
and the sign of $y_t$ is
referred to accordingly, without explicitly restating its dependence on
the \PW-boson coupling.
Measurements of Higgs-boson processes conducted by the ATLAS
and CMS Collaborations, such as the Higgs-boson decay into photon pairs
(\Hgg)~\cite{HIGG-2019-01,CMS-HIG-19-015},
provide insight into the sign of the $y_t$ coupling. Similar
information comes from the combination of other Higgs-boson
measurements. These include the simplified template cross-section
(STXS) combination for gluon--gluon fusion (ggF), vector-boson fusion
(VBF), and associated production with vector bosons ($V$\PH, where $V$
refers to the \PW and \PZ bosons) or \ttH in different
decay channels~\cite{HIGG-2021-23,CMS-HIG-22-001}.
These measurements probe indirect effects in loop interactions and
currently disfavour a negative value of the
coupling~\cite{HIGG-2015-07}, under the assumption that only SM
particles contribute to the loops.

Higgs-boson production in association with a single top quark occurs
via the EW interaction through three distinct production modes: $t$-channel (\tHq), \tWH\ and $s$-channel~\cite{demartin2015higgs,Demartin:2016axk}.
The main production mode in proton--proton (\textit{pp}) collisions at
the Large Hadron Collider (LHC) ~\cite{Evans:2008zzb} is the \tHq\ process, followed by the associated \tWH process
and the $s$-channel. In this paper, the \tHq and \tWH processes are
collectively referred to as \tH production and are the ones used for the
measurements presented, while $s$-channel production is neglected due to its extremely low
production cross-section (3\% of the total \tH cross-section).

\begin{figure}[!hbtp]
\centering
\subcaptionbox{\label{fig:tHq_feynman_diagrams}}{\includegraphics[width=0.40\linewidth]{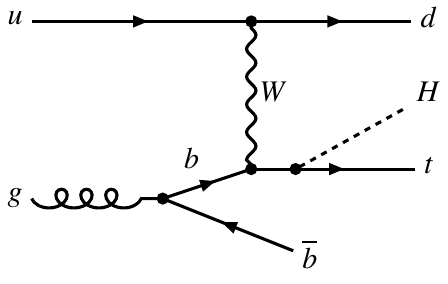}}
\qquad
\subcaptionbox{}{\includegraphics[width=0.30\linewidth]{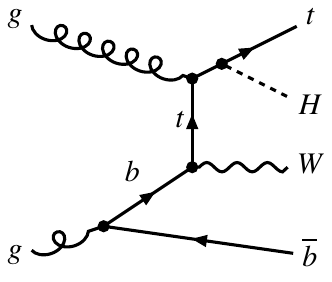}}
\caption{Representative dominant Feynman diagrams for (a) \tHq and (b) \tWH production at
LO, in the 4FS. The \tHq event topology (a) is characterised by two heavy objects, the
top quark and the Higgs boson plus an energetic light-flavour quark
(dominantly a $u$-quark), referred to as the spectator quark, and an extra soft
\bquark from gluon splitting. The \tWH event topology (b) is characterised by three heavy objects, the
top quark, the Higgs boson and a \PW{} boson and an extra soft \bquark
from gluon splitting.}%
\label{fig:tH_feynman_diagrams}
\end{figure}

In the \tHq\ process, an incoming light-flavour quark (i.e. up-type quark or down-type antiquark) from one of
the colliding protons interacts with a \bquark by exchanging a
space-like virtual \PW{} boson.
The origin of the \bquark\ depends on the flavour scheme (FS) used in the calculation:
in the five-flavour scheme (5FS), it is treated as a massless
parton originating from the proton sea, while in the four-flavour scheme (4FS),
where \bquarks are not included in the proton parton distribution functions (PDFs), it arises from gluon splitting into a
\bbbar pair~\cite{PhysRevLett.102.182003,Maltoni:2012pa,Bothmann:2017jfv}.
In the \tWH\ process, a gluon from one of the protons
interacts with a \bquark\ from the other proton or from gluon
splitting, via the exchange of a time-like virtual top quark $t$.
Two examples of dominant leading-order (LO) Feynman diagrams for the \tHq\ and
\tWH\ processes, both in the 4FS, are shown in
\cref{fig:tH_feynman_diagrams}.

Under the SM hypothesis ($\kappat = y_t / y_t^{\text{SM}} = +1$), the interference
between diagrams in which the Higgs boson couples to the top quark and
those where it couples to the \PW{} boson is destructive, leading to a
strongly suppressed cross-section. In contrast, in scenarios beyond
the SM, such as those involving CP-even and CP-odd admixtures in the
top-quark Yukawa coupling structure, the relative sign between the
Higgs-boson couplings to the top quark and the \PW{} boson can be altered,
enhancing the \tH cross-section up to an order of
magnitude~\cite{demartin2015higgs,Demartin:2016axk}. The maximum
enhancement occurs when $\kappat = -1$, corresponding to fully
constructive interference between the contributing diagrams.

The \tHq event topology is characterised by two heavy objects, a
top quark and a Higgs boson, which is radiated from the latter, plus an energetic light-flavour quark
(dominantly a $u$- or $\bar{d}$-quark), known as the spectator quark, and an extra soft
\bquark from gluon splitting.
The top-quark and Higgs-boson systems are produced in the central part
of the detector recoiling against each other.
The production of \tWH\ is characterised by three heavy objects, the
top quark, a \PW{} boson, a Higgs boson that is radiated from one of
these two particules and an extra soft \bquark from gluon
splitting. This event topology closely resembles \ttH production and has an identical final state.

The CMS Collaboration searched for $\tHq+\tWH$ production in $pp$
collisions at a centre-of-mass energy of $\sqrt{s}=\SI{13}{\TeV}$ with an integrated luminosity
of \SI{35.9}{\per\fb}~\cite{CMS-HIG-18-009}. The observed
cross-section limits are $\SI{1.94}{\pb}$ for the SM and
$\SI{0.74}{\pb}$ for the inverted top-quark Yukawa coupling (ITC)
(i.e. $\mathrm{sign}(y_t)=\ITC$) hypothesis.
More recently, using the full Run-2 dataset, the CMS Collaboration
reported updated $\ttH$ and $\tHq+\tWH$ results in the multi-lepton
and \bbbar\ final states in Ref.~\cite{CMS-HIG-19-008} and
Ref.~\cite{CMS-HIG-19-011}, respectively.
In the multi-lepton final state, the measured signal strength is
$\mu_{\tHq+\tWH}= 5.7 \pm 2.7\stat \pm 3.0\syst$.
In the \bbbar\ final state, the observed upper limit relative to the SM prediction is 14.6 at 95$\%$ confidence level.

The analysis presented in this paper uses an integrated luminosity of \lumi of data
collected with the ATLAS detector at $\sqrt{s} = \SI{13}{\TeV}$ between
2015 and 2018 at the LHC\@.
The search targets signatures that are primarily sensitive to \Hbb,
$\PH \to WW^*$, $\PH \to ZZ^*$ and \Htautau decays.
Backgrounds are estimated with a combination of simulation and data-driven techniques,
and a global fit in all final states is used to extract the best
estimate for the \tH production rate under both the SM and ITC hypotheses.


%
%

%

%
\section{Analysis strategy}%
\label{sec:strategy}
The search described in this paper considers three mutually orthogonal final states, hereafter referred to as channels, distinguished by the number of light-flavour leptons ($\ell$), i.e. electron (\Pe) or muon (\Pgm):

\begin{itemize}
\item one lepton (\onelep);
\item two same-charge leptons (\dilepSS);
\item three leptons (\trilep).
\end{itemize}
Events with \Pgt-leptons decaying into an electron or a muon are part of selection, while events containing hadronically decaying \Pgt-leptons (\tauhad) are vetoed in all three channels.

Each channel targets different Higgs-boson decay modes.
The \onelep channel is most sensitive to the \Hbb decay, while the \dilepSS and \trilep channels mostly probe
\HWW.
In the \dilepSS and \trilep channels, subleading contributions arise from \Htautau and from \HZZ decays.

These three channels are characterised by different background compositions, therefore they are
analysed and fitted separately and combined afterwards.%
Each channel defines multiple analysis regions: signal regions (SRs), which are enriched in signal events, and control regions (CRs), which are designed to constrain the dominant background contributions.
In the \onelep channel, two regions are defined: one SR targeting the signal, and a CR dedicated to top-quark pair (\ttbar) production in association with \bjets, denoted CR(\ttb).
In both the \dilepSS and \trilep channels, five regions are defined: one SR, three CRs to constrain backgrounds from non-prompt and fake leptons, and one CR targeting the associated production of a \PW{} boson with a top-quark pair (\ttW).

To enhance signal-to-background separation, all channels employ multivariate analysis techniques, specifically Boosted Decision Trees (BDTs). These methods are used both to discriminate signal from background and to separate different background components to better constrain them, as well as to define the analysis regions. Details are provided in \cref{sec:sel}.

A profile likelihood fit, implemented using the \trexfitter framework~\cite{aly_2025_14849541}, is used to simultaneously extract the signal strength, $\mu_{\tH}$, and the normalisations of the dominant backgrounds in each channel.
The signal strength is first fitted separately in each channel, and then in a combined fit.

The manuscript is structured as follows: \cref{sec:detector} describes the ATLAS detector, while details of the data and MC simulated event samples used in the analysis are provided in \cref{sec:datamc}.
\Cref{sec:object} outlines the object definitions.
The event selection, multivariate analysis techniques, and analysis region definitions for each channel are presented in \cref{sec:sel}, while \cref{sec:back} describes the background estimates and its results.
All uncertainties considered in this measurement are discussed in \cref{sec:systs}.
The resulting constraints on the signal process in both the SM and ITC scenarios are shown in \cref{sec:result}, for each channel and their combination.
Finally, the conclusions are presented in \cref{sec:conclusion}.


%
%

%

%
\newcommand{\AtlasCoordFootnote}{%
ATLAS uses a right-handed coordinate system with its origin at the nominal interaction point (IP)
in the centre of the detector and the \(z\)-axis along the beam pipe.
The \(x\)-axis points from the IP to the centre of the LHC ring,
and the \(y\)-axis points upwards.
Cylindrical coordinates \((r,\phi)\) are used in the transverse plane,
\(\phi\) being the azimuthal angle around the \(z\)-axis.
The pseudorapidity is defined in terms of the polar angle \(\theta\) as \(\eta = -\ln \tan(\theta/2)\).
Angular distance is measured in units of \(\Delta R \equiv \sqrt{(\Delta\eta)^{2} + (\Delta\phi)^{2}}\).}

\section{ATLAS detector}
\label{sec:detector}

The ATLAS detector~\cite{PERF-2007-01} at the LHC covers nearly the entire solid angle around the collision point.\footnote{\AtlasCoordFootnote}
It consists of an inner tracking detector surrounded by a thin superconducting solenoid, electromagnetic and hadronic calorimeters,
and a muon spectrometer incorporating three large superconducting air-core toroidal magnets.

The inner-detector system (ID) is immersed in a \qty{2}{\tesla} axial magnetic field
and provides charged-particle tracking in the range of \(|\eta| < 2.5\).
The high-granularity silicon pixel detector covers the vertex region and typically provides four measurements per track,
the first hit generally being in the insertable B-layer (IBL) installed before the LHC Run~2~\cite{ATLAS-TDR-19,PIX-2018-001}.
It is followed by the SemiConductor Tracker (SCT), which usually provides eight measurements per track.
These silicon detectors are complemented by the Transition Radiation Tracker (TRT),
which enables radially extended track reconstruction up to \(|\eta| = 2.0\).
The TRT also provides electron identification information
based on the fraction of hits (typically 30 in total) above a higher energy-deposit threshold corresponding to transition radiation.

The calorimeter system covers the pseudorapidity range \(|\eta| < 4.9\).
Within the region \(|\eta|< 3.2\), electromagnetic calorimetry is provided by barrel and
endcap high-granularity lead/liquid-argon (LAr) calorimeters,
with an additional thin LAr presampler covering \(|\eta| < 1.8\)
to correct for energy loss in material upstream of the calorimeters.
Hadronic calorimetry is provided by the steel/scintillator-tile calorimeter,
segmented into three barrel structures within \(|\eta| < 1.7\), and two copper/LAr hadronic endcap calorimeters.
The solid angle coverage is completed with forward copper/LAr and tungsten/LAr calorimeter modules
optimised for electromagnetic and hadronic energy measurements, respectively.

The muon spectrometer (MS) comprises separate trigger and
high-precision tracking chambers measuring the deflection of muons in a magnetic field generated by the superconducting air-core toroidal magnets.
The field integral of the toroids ranges between \num{2.0} and \qty{6.0}{\tesla\metre}
across most of the detector.
Three layers of precision chambers, each consisting of layers of monitored drift tubes, cover the region \(|\eta| < 2.7\),
complemented by cathode-strip chambers in the forward region, where the background is highest.
The muon trigger system covers the range \(|\eta| < 2.4\) with resistive-plate chambers in the barrel, and thin-gap chambers in the endcap regions.

The luminosity is measured mainly by the LUCID--2~\cite{LUCID2} detector that records Cherenkov light produced in the quartz windows of photomultipliers located close to the beampipe.

Events were selected by the first-level trigger system implemented in custom hardware,
followed by selections made by algorithms implemented in software in the high-level trigger~\cite{TRIG-2016-01}.
The first-level trigger accepted events from the \qty{40}{\MHz} bunch
crossings at a rate close to \qty{100}{\kHz}, which the high-level
trigger further reduced in order to record complete events to disk at
about \qty{1.25}{\kHz}.

A software suite~\cite{SOFT-2022-02} is used in data simulation, in the reconstruction
and analysis of real and simulated data, in detector operations, and in the trigger and data acquisition
systems of the experiment.


%
%

%

%
\section{Data and simulated event samples}%
\label{sec:datamc}

\subsection{Data event samples}%
\label{subsec:data}

This analysis uses data from $pp$ collisions at $\sqrt{s} = \SI{13}{\TeV}$ collected with the ATLAS detector
at the LHC during 2015--2018. Stringent detector and data-quality requirements were applied~\cite{DAPR-2018-01},
resulting in a data sample corresponding to a total integrated luminosity of \lumi.
The events were selected by single-lepton triggers with variable electron and muon transverse momentum ($\pT$)
thresholds, and various identification and isolation criteria depending on the lepton flavour and the data-taking period~\cite{TRIG-2018-05, TRIG-2018-01}.
The lowest $\pT$ threshold at trigger level used for electrons was \SI{24}{\GeV} (\SI{26}{\GeV}),
while for muons the threshold was \SI{20}{\GeV} (\SI{26}{\GeV}) in 2015 (2016--2018).

\subsection{Simulated event samples}%
\label{subsec:mc}

Samples of simulated events were produced using different Monte Carlo (MC) event generators including parton shower (PS) and hadronisation models for both signal and background processes.
The effect of multiple interactions in the same and neighbouring bunch crossings (pile-up)
was modelled by overlaying the hard-scattering event with simulated minimum-bias events
generated with \PYTHIA[8.186]~\cite{Sjostrand:2007gs, Sjostrand:2014zea} using the \NNPDF[2.3LO] set of PDFs~\cite{Ball:2012cx} and the A3 set of tuned parameters (A3 tune)~\cite{ATL-PHYS-PUB-2016-017}.
Events were reweighted such that the
distribution of the average number of interactions per bunch crossing
matches that observed in data.

\Cref{tab:mcconfig} presents a summary of all configurations used to generate events for signal (\tHq\ and \tWH) and background processes, for both nominal and alternative simulated samples. In all simulated signal and background samples, the top-quark mass was set to $\mtop = \SI{172.5}{\GeV}$ and the Higgs-boson mass to $\mH = \SI{125.0}{\GeV}$. Moreover, the strong coupling constant was set to $\alphas = 0.118$~\cite{ParticleDataGroup:2024cfk} at the reference scale of the \PZ{}-boson mass. These values are also used in the computation of production cross-sections for processes involving these particles. The bottom and charm hadron decays were simulated with \EVTGEN~\cite{Lange:2001uf} for all event samples, except those generated with \SHERPA.

\begin{table}[!htb]
\begin{center}
\caption{
The configurations used to generate events for signal (\tHq\ and \tWH) and background processes.
The simulation samples used to estimate the systematic uncertainties are shown in parentheses and in grey.
}
\label{tab:mcconfig}
\vspace{0.25cm}
\small
\setlength\tabcolsep{1.5pt}
\begin{tabular}{~l^l^l^l^l^l}
\toprule
Process & Generator & ME order & Parton shower & PDF set & Tune \\
\midrule
\tHq                   & MG5\_aMC & NLO & \PYTHIA[8]\ & \NNPDF[3.0nlo] & A14 \\
\rowstyle{\color{gray}}
& (MG5\_aMC) & (NLO) & (\Herwig[7]) &  (\NNPDF[3.0nlo]) &  (\HERWIG[7.1]) \\
\midrule
\tWH                   & MG5\_aMC & NLO & \PYTHIA[8]\ & \NNPDF[3.0nlo] & A14 \\
\rowstyle{\color{gray}}
& (MG5\_aMC) & (NLO) & (\Herwig[7]) &  (\NNPDF[3.0nlo]) &  (\HERWIG[7.1]) \\
\midrule
$\ttbar$               & \POWHEGBOX & NLO & \PYTHIA[8] & \NNPDF[3.0nlo] & A14  \\
\rowstyle{\color{gray}}
& (\POWHEGBOX) & (NLO) & (\Herwig[7]) & (\NNPDF[3.0nlo]) &  (\HERWIG[7.1]) \\
\rowstyle{\color{gray}}
& (\POWHEGBOX) & (NLO) & (\PYTHIA[8]) & (\NNPDF[3.0nlo]) & (A14 $\pT^\text{hard}$)  \\
\rowstyle{\color{gray}}
& (MG5\_aMC) & (NLO) & (\PYTHIA[8]) & (\NNPDF[3.0nlo]) & (A14)  \\
\rowstyle{\color{gray}}
\ttbb             & (\POWHEGBOXRES) & (NLO) & (\PYTHIA[8]) & (\NNPDF[3.0nlo]) & (A14)  \\
\midrule
\ttH                   & \POWHEGBOX  & NLO & \PYTHIA[8]\ & \NNPDF[3.0nlo] & A14 \\
\rowstyle{\color{gray}}
& (\POWHEGBOX) & (NLO) & (\Herwig[7]) &  (\NNPDF[3.0nlo]) & (\HERWIG[7.1]) \\
\rowstyle{\color{gray}}
& (MG5\_aMC)  & (NLO) & (\PYTHIA[8]) & (\NNPDF[3.0nlo]) & (A14)   \\
\midrule
$\ttbar W$             & \Sherpa[2.2.10] & NLO QCD+EW & \Sherpa &  \NNPDF[3.0nnlo] & default \\
\rowstyle{\color{gray}}
& (MG5\_aMC)  & (NLO QCD+EW) & (\PYTHIA[8]) & \NNPDF[3.0nnlo] & (A14)   \\
\midrule
$\ttbar W$ (EW)        & \Sherpa[2.2.10] & LO & \Sherpa &  \NNPDF[3.0nnlo] & default \\
\rowstyle{\color{gray}}
& (MG5\_aMC)  & (LO) & (\PYTHIA[8]) & (\NNPDF[3.0nlo]) & (A14)   \\
\midrule
\ttZ                   & MG5\_aMC & NLO & \PYTHIA[8] & \NNPDF[3.0nlo] & A14  \\
\rowstyle{\color{gray}}
& (\Sherpa[2.2.10]) & (LO) & (\Sherpa) &  (\NNPDF[3.0nnlo]) & (default) \\
\rowstyle{\color{gray}}
& (MG5\_aMC) & (NLO) & (\Herwig[7]) & (\NNPDF[3.0nlo]) & (\HERWIG[7.1])  \\
\midrule
Single top quark            & \POWHEGBOX & NLO & \PYTHIA[8] & \NNPDF[3.0nlo] & A14   \\
($t$-ch., $\Wt$, $s$-ch.) &  &  &  &  \\
\rowstyle{\color{gray}}
& (\POWHEGBOX) & (NLO) & (\Herwig[7]) & (\NNPDF[3.0nlo]) & (\HERWIG[7.1])  \\
\rowstyle{\color{gray}}
& (MG5\_aMC) & (NLO) & (\Herwig[7]) & (\NNPDF[3.0nlo]) & (\HERWIG[7.1])  \\
\midrule
$V+\text{jets}$                & \Sherpa[2.2.1] & \MEPSatNLO & \Sherpa & \NNPDF[3.0nlo] & default \\
\midrule
$VV$, $qqVV$            & \Sherpa[2.2.2]  & \MEPSatNLO & \Sherpa & \NNPDF[3.0nnlo] & default  \\
\rowstyle{\color{gray}}
& (\POWHEGBOX)  & (NLO) & (\PYTHIA[8]) & (\NNPDF[3.0nlo]) & (A14)   \\
\midrule
$VVV$                  & \Sherpa[2.2.2]  & \MEPSatNLO & \Sherpa & \NNPDF[3.0nnlo] & default  \\
\midrule
$tZ$                  & MG5\_aMC & NLO & \PYTHIA[8] & \NNPDF[3.1nlo] & A14 \\
\midrule
$tWZ$                  & MG5\_aMC & NLO DR1 & \PYTHIA[8] & \NNPDF[2.3lo] & A14 \\
\rowstyle{\color{gray}}
&  (MG5\_aMC) & (NLO DR2) & (\PYTHIA[8]) & \NNPDF[2.3lo] & (A14) \\
\midrule
$t\bar t t, t\bar t t\bar t$      & MG5\_aMC & QCD NLO & \PYTHIA[8] & \NNPDF[3.1nlo] & A14 \\
\midrule
$\ggH, \qqH$ & \POWHEGBOX & QCD NLO & \PYTHIA[8] & \CTEQ[6L1] & \AZNLO \\
\midrule
$\WH, \ZH$ & \PYTHIA[8] & QCD NLO & \PYTHIA[8] & \NNPDF[2.3lo] & A14 \\
\bottomrule
\end{tabular}
\end{center}
\end{table}


Systematic uncertainties from different modelling choices are evaluated by comparing alternative simulated samples to the nominal ones.
These alternative samples are generated either through reweighting or with dedicated event generators. Each variation is applied independently, unless explicitly stated otherwise, to ensure a consistent evaluation of its individual impact on the analysis.

\subsubsection{Simulated signal event samples}%

The production of a Higgs boson with a single top quark in the $t$-channel, \tHq, was modelled using \MGNLO[2.6.2]~\cite{Alwall:2014hca}
with matrix elements (MEs) at next-to-leading-order (NLO) accuracy in QCD in the 4FS using the \NNPDF[3.0nlo]~nf4~\cite{Ball:2014uwa} PDF set.
The renormalisation (\muR) and factorisation (\muF) scales were set to $0.5\times \sum_i \sqrt{m^2_i+p^2_{\text{T},i}}$, summing over all ME-generated particles with mass $m$ and transverse momentum \pT.
Events were interfaced with \PYTHIA[8.230] using the A14 set of tuned parameters (A14 tune)~\cite{ATL-PHYS-PUB-2014-021} and the \NNPDF[2.3LO] PDF set.
The simulated \tHq\ signal samples were normalised to the NLO QCD theory prediction without NLO EW corrections at $\rts = \SI{13}{\TeV}$ and in the 5FS of $74.3^{+6.5\%}_{-14.9\%}~(\text{QCD})~\pm~3.7\%~(\text{PDF+\alphas})\,\si{\fb}$~\cite{deFlorian:2016spz,YR4twiki}.
The PDF and \alphas uncertainties are calculated using the
\textsc{PDF4LHC15\_nlo\_30\_pdfas} PDF sets~\cite{Butterworth:2015oua} and added in quadrature.
The QCD uncertainty is computed from a combination of QCD scale variations and the difference of predictions in the 4FS and 5FS schemes, using the \textsc{PDF4LHC15\_nlo\_nf4\_100} PDF (central) set.
The associated production of a Higgs boson with a single top quark, \tWH, was modelled using \MGNLO[2.8.1] with MEs at NLO accuracy in QCD in the 5FS using the \NNPDF[3.0NLO] PDF set.
The functional forms of the \muR and \muF scales were set the same as for the \tHq simulation sample.
The diagram-removal (DR) scheme (DR1)~\cite{Frixione:2008yi} was employed to
remove the overlap between \tWH\ and \ttH simulation samples~\cite{ATL-PHYS-PUB-2016-020}.
The events were interfaced with \PYTHIA[8.245] using the A14 tune and the \NNPDF[2.3lo] PDF set.
The simulated \tWH\ signal samples were normalised to the NLO QCD theory prediction in
the 5FS \cite{deFlorian:2016spz,YR4twiki} and using the
DR2 scheme~\cite{Demartin:2016axk} to handle the overlap between
the \tWH\ and \ttH\ processes.
At $\rts = \SI{13}{\TeV}$ the predicted cross-section is $\sigma(\tWH)_{\text{NLO}} = 15.2^{+4.9\%}_{-6.7\%}~(\text{QCD scale})~\pm~6.3\%~(\text{PDF+\alphas})\,\si{\fb}$~\cite{deFlorian:2016spz,YR4twiki}.
To assess the modelling uncertainty due to the choice of the PS and hadronisation model, alternative event samples were produced for signal processes.
Both \tHq and \tWH processes were produced using \MGNLO\ with the nominal set-up but interfaced with \HERWIG[7.1.6]~\cite{Bahr:2008pv,Bellm:2015jjp} using the \MMHT[nnlo]~\cite{Harland-Lang:2014zoa} PDF set, with the \HERWIG[7.1] default set of tuned parameters (\HERWIG[7.1] tune)~\cite{Bellm:2015jjp}.
Moreover, simulated signal samples for both \tHq\ and \tWH\ with the ITC hypothesis were produced using similar set-ups as the ones used to produce SM simulated samples. In this case, the samples were normalised to the cross-section, for $pp$ collisions at $\sqrt{s}=\SI{13}{\TeV}$, of \(\sigma(\tHq)_{\text{NLO}}^{\text{\ITC}} = 873\,\si{\fb}\) and \(\sigma(\tWH)_{\text{NLO}}^{\text{\ITC}} = 114\,\si{\fb}\), respectively.
For both nominal and ITC simulated signal samples, alternative variations of \muR\ and \muF\ scales and initial-/final-state radiation (ISR/FSR) were produced within the nominal set-up (i.e. \MGNLO+\PYTHIA[8]) as reweighted events.

\subsubsection{Simulated background event samples}%

The production of \ttbar\ events were modelled using \POWHEGBOX[v2]~\cite{Frixione:2007nw,Nason:2004rx,Frixione:2007vw,Alioli:2010xd} at NLO QCD in 5FS with the \NNPDF[3.0NLO] PDF set.
In this case, the \hdamp parameter\footnote{The \hdamp parameter
controls the \pT\ of the first additional emission beyond the LO Feynman diagram
in the PS and therefore regulates the  high-\pT\ emission against which the \ttbar\ system recoils.} was set to $1.5 \times \mtop$~\cite{ATL-PHYS-PUB-2016-020}.
Events were interfaced with \PYTHIA[8.230] using the A14 tune and the \NNPDF[2.3LO] PDF set.
The simulated samples are normalised to the theory prediction at next-to-next-to-leading-order (NNLO) in QCD including the resummation of
next-to-next-to-leading logarithmic (NNLL) soft-gluon terms calculated using \textsc{Top++2.0}~\cite{Beneke:2011mq,Cacciari:2011hy,Baernreuther:2012ws,Czakon:2012zr,Czakon:2012pz,Czakon:2013goa,Czakon:2011xx},
equals to $834^{+29}_{-37}$~pb,
where the uncertainty is the sum in quadrature of scale, PDF and $\alphas$ uncertainties.

A series of alternative \ttbar\ simulated event samples were produced to evaluate the impact of modelling uncertainties.
The effect of the choice of the PS and hadronisation model is quantified by comparing the nominal simulation with one produced using the same ME generator but interfaced with \HERWIG[7.1.6].
To evaluate the impact of the choice of the NLO matching scheme two samples were produced: one where the parameter $\pT^\text{hard}$, which regulates the definition of the vetoed region of the \PYTHIA8 showering~\cite{Hoche:2021mkv}, was halved, and other event sample generated by \MGNLO and matched to \HERWIG[7].
Moreover, to assess the choice of the value of the \hdamp parameter, alternative simulated samples generated using the nominal set-up but with $\hdamp = 3.0 \times \mtop$ were considered.
A 4FS \ttbb\ simulated sample was produced using \POWHEGBOXRES with the \OPENLOOPS~\cite{Buccioni:2019sur,Cascioli:2011va,Denner:2016kdg} library and the \NNPDF[3.1nnlo]~nf4 PDF set, interfaced with \PYTHIA[8.244].
The simulation used a \bquark mass of \qty{4.95}{\GeV} and dynamic scales $\muF$ set to $0.5\times\sum_{i=t,\bar{t},b,\bar{b},j}m_{\mathrm{T},i}$ and $\muR$ set to $0.5\times\sqrt[4]{m_{\text{T}, t} \cdot m_{\text{T},\bar{t}} \cdot m_{\text{T},b} \cdot m_{\text{T},\bar{b}}}$, where $m_{\mathrm{T},i} = \sqrt{m_i^2 + p_{\mathrm{T},i}^2}$ denotes the transverse mass of particle $i$.
The \hdamp parameter was defined as $0.5\times\sum_{i=t,\bar{t},b,\bar{b}}m_{\mathrm{T},i}$, and the parameter $h_{\text{bzd}}$ was set to 5.
Here, $h_{\text{bzd}}$ controls the suppression of the Born-zero configurations in \POWHEG.
This sample is considered to evaluate the impact of using the 4FS instead of the 5FS and
also the effect of generating additional $b$-quarks explicitly at NLO in the ME in the 4FS,
as opposed to their approximate treatment via the PS in the 5FS set-up.

The \ttH events were simulated using \POWHEGBOX[v2] at NLO accuracy in QCD in 5FS with the \NNPDF[3.0NLO] PDF set, and with \muR and \muF scales set to $\sqrt[3]{m_{\text{T},t} \cdot m_{\text{T},\bar{t}} \cdot m_{\text{T}, H}}$~\cite{Hartanto:2015uka}.
Events were interfaced with \PYTHIA[8.230] using the A14 tune and the \NNPDF[2.3LO] PDF set.
The NLO QCD+EW cross-section at $\rts = \SI{13}{\TeV}$ is $507^{+35}_{-50}\,\si{\fb}$~\cite{deFlorian:2016spz}, where the uncertainties were
estimated from variations of \alphas\ and the \muR\ and \muF\ scales.
Additional variations use event weights for QCD scales and radiation, with an alternative PS simulated sample using \HERWIG[7.04].

The \ttW events were produced using \SHERPA[2.2.10]~\cite{Bothmann:2019yzt} at NLO (LO) accuracy in QCD for up to one (two) additional partons.
The \NNPDF[3.0nnlo] PDF set was used along with the dedicated set of tuned PS parameters developed by the \SHERPA authors.
In addition, EW corrections were also produced with \SHERPA[2.2.10] at LO accuracy for up to one additional parton.
The \ttW\ cross-section is calculated at NLO QCD with up to two additional partons using an improved \MGNLO FxFx merging procedure~\cite{Frederix:2012ps,Frederix:2021agh} and combines it with leading \( \mathcal{O}(\alpha^2 \alpha_s^2) \) and subleading \( \mathcal{O}(\alpha^3 \alpha_s) \) NLO EW corrections. This calculation also includes on-shell LO
decays of the \ttW\ including the tree-level spin correlations within the narrow-width approximation.
At $\rts = \SI{13}{\TeV}$, this yields 722$^{+70}_{-78}$ (scale)$^{+7}_{-7}$~(PDF)\,fb~\cite{Frederix:2021agh}.
The choice of the NLO matching scheme was evaluated using an alternative simulated sample produced with \MGNLO[2.3.3] with the FxFx~\cite{Frederix:2012ps} merging scheme simulated samples including EW corrections.

The associated production of the \PZ{} boson with a top-quark pair (\ttZ) was simulated using \MGNLO[2.3.3] at NLO QCD with the \NNPDF[3.0nlo] PDF set, interfaced with \PYTHIA[8.210] using the A14 tune and the \NNPDF[2.3lo] PDF set.
The \ttZ cross-section, including off-shell effects and $Z/\gamma^*$ interference
, was calculated at NLO QCD+EW as
$0.88^{+0.09}_{-0.10}\,\text{pb}$~\cite{ATL-PHYS-PUB-2016-005,deFlorian:2016spz}, where the uncertainties were estimated from variations of \alphas\ and the \muR\ and \muF\ scales.
The choice of the NLO matching scheme was evaluated using an alternative simulated sample produced with \SHERPA[2.2.0] at LO accuracy along with the \NNPDF[3.0nnlo] PDF set, using the \MEPSatLO prescription~\cite{Hoeche:2012yf},
with up to one additional parton for the \ttZll sample and two additional partons for the \ttZqq and \ttZnunu samples.

All single top-quark events were generated with \POWHEGBOX[v2] at NLO in QCD with either the \NNPDF[3.0nlo]~nf4 or \NNPDF[3.0nlo] PDF sets, with \muF and \muR scales set to $\sqrt{m_b^2 + \pTX[2][b]}$~\cite{Alioli:2009je,Re:2010bp}.
The $t$-channel was produced considering the 4FS, while the $s$-channel and associated $\tW$ processes used the 5FS\@. The latter process also employed the DR1 scheme to remove the overlap between \tW\ and \ttbar~\cite{ATL-PHYS-PUB-2016-020}.
Events were interfaced with \PYTHIA[8.230] using the A14 tune and the \NNPDF[2.3lo] PDF set.
Systematic variations include QCD \muR\ and \muF\ scales and radiation uncertainties, alternative PS using \POWHEGBOX[v2] with \HERWIG[7.16],
and matching scheme uncertainties using \MGNLO[2.6.2] with \HERWIG[7.16] and the \MMHT[nnlo] PDF set.
For associated $\tW$, an additional simulated sample using the diagram subtraction (DS) scheme was produced to evaluate the \ttbar\ interference uncertainties.

The production of single-boson ($V+\text{jets}$), diboson ($VV$), and triboson ($VVV$) processes were simulated using either \SHERPA[2.2.1] or \SHERPA[2.2.2], with varying levels of QCD accuracy depending on the process.
The \NNPDF[3.0nnlo] PDF set was used along with a dedicated set of \SHERPA-tuned PS parameters.
For the $V+\text{jets}$ processes, MEs were generated at NLO (LO) accuracy in QCD for up to two (four) partons, using the Comix library~\cite{Gleisberg:2008fv}. Simulated samples were normalised to NNLO predictions~\cite{Anastasiou:2003ds}.
For the $VV$ processes, MEs accounted for off-shell effects and included Higgs-boson contributions where relevant.
Fully leptonic and semileptonic final states, where one boson decays leptonically and the other hadronically, were simulated at NLO (LO) accuracy in QCD for up to one (three) additional partons.
Loop-induced processes \(gg \to VV\) were generated at LO accuracy in QCD with up to one additional parton, for both fully leptonic and semileptonic final states.
The EW production of a $VV$ pair in association with two jets (\(VVjj\)) was simulated at LO accuracy in QCD\@.
For the $VVV$ processes, MEs were generated with factorised gauge-boson decays at NLO accuracy in QCD for the inclusive process and at LO accuracy in QCD for up to two additional parton emissions.
In all cases, NLO MEs were matched to PS using a colour-exact variant of the \MCatNLO algorithm~\cite{Hoeche:2011fd}. Different jet multiplicities were merged via the CKKW matching procedure~\cite{Catani:2001cc,Hoeche:2009rj}, extended to NLO QCD accuracy through the \MEPSatNLO~\cite{Hoeche:2012yf} prescription, with a merging threshold set to \SI{20}{\GeV}. Leading-order MEs were merged using the \MEPSatLO prescription.
These calculations were performed in the $G_{\mu}$ scheme, which uses the Fermi constant as input, ensuring an optimal description of pure EW interactions at the EW scale.

To assess modelling uncertainties, scale uncertainties were evaluated using seven \(\muR\) and \(\muF\) variations (by factors of 0.5 and 2) as event weights. Furthermore, in the particular case of the $VV$ processes, alternative event simulated samples were produced to evaluate the impact of different ME, PS, and hadronisation models by comparing the nominal \diboson \SHERPA event samples with \POWHEGBOX[v2] simulations of $WW$, $WZ$, and $ZZ$~\cite{Nason:2013ydw} processes at NLO accuracy in QCD\@.
The effects of singly resonant amplitudes and interference due to \(Z/\gamma^*\) exchange, as well as same-flavour lepton combinations in the final state, were included where relevant. However, $WW/ZZ$ interference in same-flavour lepton and neutrino final states was neglected.
Events were interfaced with \PYTHIA[8.186] using the \AZNLO\ set of tuned parameters (\AZNLO tune)~\cite{STDM-2012-23}.
The \CT[10] PDF set~\cite{Lai:2010vv} was used for hard-scattering calculations, while the \CTEQ[6L1]~\cite{Pumplin:2002vw} PDF set was applied for PS\@.
The \(\muF\) and \(\muR\) scales were set to the boson-pair invariant mass, with a requirement of \(m_{\ell\ell} > \SI{4}{\GeV}\) at the ME level for same-flavour lepton pairs.

The \tZq event sample was simulated using \MGNLO[2.3.3] at NLO accuracy in QCD in the 4FS with the \NNPDF[3.0nlo] PDF set.
The events were interfaced with \PYTHIA[8.230] using the A14 tune and the \NNPDF[2.3lo] PDF set.
Off-resonance events away from the \PZ{}-mass peak were included.
The functional form of the \muR and \muF scales
was set to $4\sqrt{m_b^2+\pTX[2][b]}$, where the $b$-quark was the one produced by a gluon splitting in the event.
The uncertainty due to initial-state radiation and to higher/lower parton radiation were estimated following exactly the same prescription as with the $t$-channel process.

The \tWZ event production was modelled using \MGNLO[2.3.3] at NLO QCD with the \NNPDF[3.0nlo] PDF set in the 5FS\@.
Events were interfaced with \PYTHIA[8.212], using the A14 tune and the \NNPDF[2.3lo] PDF set.
The DR1 scheme was used to remove the overlap between the \tWZ and \ttZ processes.
Parton radiation uncertainties follow the same prescription as the $t$-channel process.
An alternative DR scheme~\cite{Demartin:2016axk} was simulated to assess systematic uncertainty.

The production of three top-quark process (\ttt) events was modelled using \MGNLO[2.2.2] with NLO accuracy in QCD interfaced with \PYTHIA[8.186],
using the \NNPDF[2.3lo] PDF set and the A14 tune.
The production of four top-quark process (\tttt) events was modelled using \MGNLO[2.3.3] with NLO accuracy in QCD with the \NNPDF[3.1nlo] PDF set.
The events were interfaced with \PYTHIA[8.230] for the PS and hadronisation, using the A14 tune and the \NNPDF[2.3lo] PDF set.

Four SM Higgs-boson processes are considered as minor backgrounds:
gluon--gluon fusion (\ggH), VBF (\qqH), and associated production with \PW{} and \PZ{} bosons (\WH and \ZH).
The \ggH and VBF events were modelled with either \POWHEGBOX[v2] or v1 (r2856)~\cite{Alioli:2008tz,Nason:2009ai}, providing NLO accuracy in QCD with the \CT[10] PDF set.
The \muR and \muF scales were set to $\sqrt{\mH^{2} + p^2_{\text{T},H}}$. These generated events were interfaced with either \PYTHIA[8.210] or 8.186 for PS and hadronisation, using the \AZNLO tune and the \CTEQ[6L1] PDF set. The \WH and \ZH production was modelled with \PYTHIA[8.186], using the A14 tune and the \NNPDF[2.3lo] PDF set.

Most of the MC simulated samples used for the nominal estimates were processed through a simulation~\cite{SOFT-2010-01} of the ATLAS detector based on \GEANT~\cite{Agostinelli:2002hh},
while those used for evaluating systematic uncertainties were processed with a fast simulation~\cite{SOFT-2010-01} that relies on a parameterisation of the calorimeter response~\cite{ATL-PHYS-PUB-2010-013}.
Furthermore, fast-simulated event samples were used for certain nominal samples, particularly for the \tHq\ and \tWH\ signal processes and the \tttt\ background process with no loss of accuracy.


%
%

%

%
\section{Object definitions}
\label{sec:object}
Interaction vertices from the $pp$ collisions are reconstructed from at least two tracks in the ID system with \pt larger than \SI{500}{\MeV}
that are consistent with originating from the beam collision region in the $x$--$y$ plane.
If more than one primary vertex candidate is found in the event,
the candidate for which the associated tracks form the largest sum of squared $\pt$ is selected as the hard-scatter primary vertex~\cite{ATL-PHYS-PUB-2015-026}.

Electron candidates are reconstructed from energy clusters in the electromagnetic calorimeter matched to a track in the ID~\cite{EGAM-2018-01}.
They are required to satisfy $\pt>\SI{10}{\GeV}$ and $|\eta|<2.47$,
excluding the transition region between the endcap and barrel calorimeters ($1.37<|\eta|<1.52$).
\texttt{Loose} and \texttt{Tight} electron identification working points are used, based on a likelihood discriminant employing calorimeter,
tracking and combined variables that provide separation between electrons and jets.

Muon candidates are reconstructed from tracks measured in the ID matched to tracks measured in the MS~\cite{MUON-2018-03,MUON-2022-01}.
The resulting muon candidates are re-fitted using the complete track information from both detector systems.
They are required to satisfy $\pt > \SI{10}{\GeV}$ and $|\eta| < 2.5$.
\texttt{Loose} and \texttt{Medium} muon identification working points are used.

Electron (muon) candidates are matched to the primary vertex by requiring that the significance of their transverse impact parameter,
$d_0$\footnote{The transverse impact parameter, $d_0$, is defined in the $x$--$y$~plane as the distance of closest approach of the track to the beamline.},
satisfies $|d_0/\sigma(d_0)|<5.0\,(3.0)$, where $\sigma(d_0)$ is the measured uncertainty in $d_0$, and by requiring that their longitudinal impact parameter,
$z_0 \sin\theta$\footnote{The longitudinal impact parameter, $z_0$, is defined as the distance in $z$ between the primary vertex and the point on the track used to evaluate $d_0$.},
satisfies $|z_0 \sin\theta|<\SI{0.5}{mm}$~\cite{EGAM-2018-01,MUON-2018-03}.

To further suppress leptons from heavy-flavour hadron decays, misidentified jets, or photon conversions (collectively referred to as \enquote{non-prompt leptons}),
a BDT discriminant, referred to as the non-prompt-lepton BDT~\cite{HIGG-2017-02},
based on isolation and lifetime information associated with a track jet that matches the selected electron or muon,
is employed.
The \texttt{Tight} working point based on the non-prompt-lepton BDT is used,
which allows prompt muons (barrel/endcap electrons) to be selected with an efficiency that is about 60\% ($60\%/70\%$) for $\pt \sim \SI{20}{\GeV}$
and reaches a plateau of 95\% ($95\%/90\%$) for $\pt \sim 40\ (40/65)\SI{}{\GeV}$,
while the rejection factor against leptons from the decay of $b$-hadrons is about 20.

For the \dilepSS channel, to further suppress electrons with incorrect charge assignment, a BDT discriminant based on calorimeter and tracking quantities~\cite{PERF-2017-01} is used, reducing the signal by 1.9\% and the total background by approximately 40\%, both at preselection level (see \cref{sec:sel:2lSS}).
The electron candidates are separated into three classes: \enquote{material conversion}, \enquote{internal conversion}, and \enquote{non-conversion} electron candidates.
Most electrons originating from photon conversions, whether from material conversions in the detector or internal conversions, are rejected by the standard electron identification criteria.
To further suppress remaining conversion electrons in the \dilepSS and \trilep channels, additional vetoes are applied, except in control regions specifically targeting conversion electrons ($e^*$).
Material conversion candidates have a reconstructed displaced vertex with radius $r > 20$~mm that includes the track associated with the electron\footnote{
The beampipe and IBL inner radii are 23.5~mm and 33~mm, respectively.}.
The invariant mass of the associated track and the closest (in $\Delta\eta$) opposite-charge track reconstructed in the silicon detector, calculated at the conversion vertex, is required to be $<\SI{100}{\MeV}$.
Internal conversion candidates must fail to satisfy the requirements for material conversions, and the di-track invariant mass, calculated here at the primary vertex, is required to be $<\SI{100}{\MeV}$.

The various lepton working points used are summarised in Table~\ref{tbl:tightleps}.
Lepton candidates are initially selected using relatively \enquote{\textit{Loose}} criteria (labelled as $L$), and a subset may later be required to satisfy stricter \enquote{\textit{Tight}} criteria (labelled as $T$) to further optimise the event selection.
The choice of working point depends on the main background processes in each category.
The specific selections adopted for the signal and control regions are described in \cref{sec:sel}.
The lepton working points used here (\textit{Loose} and \textit{Tight}) are not to be confused with the identification working points such as \texttt{Loose}, \texttt{Medium}, and \texttt{Tight}.

\begin{table}[!htb]
\begin{center}
\caption{\label{tbl:tightleps} Description of the \textit{Loose} ($L$) and \textit{Tight} ($T$) lepton definitions.
The electron $e^{*}$ is required to fulfil, in addition to the corresponding lepton definition requirements, those corresponding to an internal or material conversion candidate.
}
\resizebox{0.78\textwidth}{!}{
\begin{tabular}{l|c|c|c|c}
\toprule
& \multicolumn{2}{|c|}{Electron} & \multicolumn{2}{c}{Muon} \\
\midrule
Lepton definition                  & $L$ & $T$ &  $L$ & $T$ \\
\midrule
Identification                     & \texttt{Loose} & \texttt{Tight} & \texttt{Loose} & \texttt{Medium} \\
\midrule
Transverse impact parameter        &  \multicolumn{2}{c|}{$<5.0$} & \multicolumn{2}{c}{$<3.0$ } \\
significance $|d_0|/\sigma_{d_0}$  & \multicolumn{2}{c|}{} &  \multicolumn{2}{c}{}  \\
\midrule
Longitudinal impact parameter $z_0$     &  \multicolumn{4}{c}{$|z_0 \sin \theta|< 0.5$~mm}  \\
\midrule
Isolation                          &  No & Yes &  No & Yes  \\
( Non-prompt lepton BDT)            &  & & &  \\
\midrule
Electron charge-misassignment veto &  No & Yes & \multicolumn{2}{c}{--} \\
Electron conversion candidate veto &  No & Yes (except $e^{*}$) & \multicolumn{2}{c}{--} \\
\bottomrule
\end{tabular}
}
\end{center}
\end{table}

The constituents for jet reconstruction are identified by combining measurements from both the ID and the calorimeter using a particle flow (PFlow) algorithm~\cite{PERF-2015-09}.
Jet candidates are reconstructed from these PFlow objects using the \antikt algorithm~\cite{Cacciari:2008gp,Fastjet} with a radius parameter of $\Delta R=0.4$.
They are calibrated using simulation with corrections obtained by using in situ techniques in data~\cite{JETM-2018-05}.
Only jet candidates with $\pT> \SI{20}{\GeV}$ and within $|\eta|<4.5$ are selected.
To reduce the effect of \pileup, each jet with $\pT<\SI{60}{\GeV}$ and $|\eta|<2.4$ must satisfy the \texttt{Tight} working point of the jet vertex tagger (JVT)~\cite{PERF-2014-03} criteria
used to identify jets that originate from the primary vertex.
Moreover, jets within $2.5 < \abseta < 4.5$ (hereafter called ``forward jets'') and $\pT < \SI{120}{\GeV}$ must satisfy the \texttt{Tight} working point of the forward jet vertex tagger (fJVT)~\cite{PERF-2016-06} criteria.
A set of quality criteria is also applied to reject events containing at least one jet arising from non-collision sources or detector noise~\cite{ATLAS-CONF-2015-029}.

Jets containing $b$-hadrons are identified ($b$-tagged) via the \texttt{DL1r}~\cite{FTAG-2019-07} algorithm that uses a deep feed-forward neural network
based on the distinctive features of $b$-hadron decays,
primarily the impact parameters of tracks and the displaced vertices reconstructed in the ID\@.
Additional input to this network is provided by discriminating variables constructed by a recurrent neutral network (RNN)~\cite{ATL-PHYS-PUB-2017-003},
which exploits the spatial and kinematic correlations between tracks originating from the same $b$-hadron.
A multivariate $b$-tagging discriminant value is calculated for each jet.
In this measurement, a jet is considered $b$-tagged if it passes the working point corresponding to 85\%, 77\%, 70\%, or 60\% average expected efficiency to tag a $b$-quark jet,
with a light-jet\footnote{`Light jet' or `Light-flavour jet' refers to a jet originating from the hadronisation of a light-flavour quark ($u$, $d$, $s$) or a gluon.} rejection factor of about 40 to 2500,
and a charm-jet ($c$-jet) rejection factor of about 3 to 40, as determined for jets with $\pt >\SI{20}{\GeV}$ and $|\eta|<2.5$ in simulated $\ttbar$ events.
This discrete version of the $b$-tagging discriminant, named pseudo-continuous $b$-tagging score (PCBT), is also exploited as jet variable.
The reconstruction of the \tauhad is seeded by jets reconstructed via the anti-$k_t$ algorithm, using calibrated energy clusters as inputs, with a distance parameter of $\Delta R = 0.4$~\cite{ATL-PHYS-PUB-2015-045, ATLAS-CONF-2017-029, ATL-PHYS-PUB-2019-033, ATL-PHYS-PUB-2022-044}.
Jets seeding \tauhad candidates are additionally required to have $\pt > \SI{10}{\GeV}$ and $|\eta|< 2.5$.
To separate the \tauhad candidates initiated by hadronic $\tau$-lepton decays from jets initiated by quarks or gluons, a RNN~\cite{ATL-PHYS-PUB-2019-033}
identification method was trained on information from reconstructed charged-particle tracks and clusters of energy in the calorimeter associated with \tauhad candidates as well as high-level discriminating variables.
A separate multivariate discriminant based on a BDT~\cite{ATLAS-CONF-2017-029, ATL-PHYS-PUB-2022-044} is also used to reject backgrounds arising from electrons mimicking a \tauhad.
This discriminant (\textsc{eBDT}) is built using information from the calorimeter and the tracking detector.
Transition radiation information from the TRT system plays a key role in the performance of this discriminant.
Baseline \tauhad are required to have one or three associated tracks, electric charge of $\pm1$, $\pt > \SI{20}{\GeV}$ and $|\eta|< 2.5$, excluding the barrel--endcap transition region.
Each \tauhad must satisfy the \texttt{Medium} RNN identification selection requirement, with efficiencies of 75\% for 1-prong and 60\% for 3-prong taus,
and the \texttt{Loose} \textsc{eBDT} requirement, with an efficiency of 95\%~\cite{ATLAS-CONF-2017-029, ATL-PHYS-PUB-2022-044}.

Ambiguities between independently reconstructed electrons, muons, jets and \tauhad can arise.
A sequential \enquote{overlap removal} procedure is performed to resolve these ambiguities and thus avoid double counting of candidates.
This procedure is applied to electrons and muons satisfying the $L$ criteria.
If two electrons are separated by $\Delta R<0.1$, only the one with the higher \pt is kept.
If an electron and a muon overlap within $\Delta R<0.1$, the muon is removed if it is reconstructed only from an ID track and calorimeter energy deposits consistent with a minimum-ionising particle (i.e.\ if it is \enquote{calorimeter-tagged}),
otherwise the electron is removed.
If an electron and a selected jet are found within $\Delta R < 0.2$, the jet is removed.
Any electron later found within \DeltaR of \num{0.4} of a jet is removed.
Muons associated with a jet must satisfy a jet--muon separation of $\Delta R < 0.4$.
If the overlapping jet contains less than three tracks with $\pt > \SI{500}{\MeV}$, the jet is removed, otherwise the muon is rejected.
Any \tauhad found within a \DeltaR of \num{0.2} of an electron is removed.
Any \tauhad found within a \DeltaR of \num{0.2} of any type of muon with \pT greater than \SI{2}{\GeV} is removed,
although if the \tauhad \pT is greater than \SI{50}{\GeV}, it will only be removed if it is found to overlap with a reconstructed muon.
Finally, if a jet is found within a \DeltaR of \num{0.2} of a \tauhad, the jet is removed.

The missing transverse momentum $\vec{p_\text{T}}^\text{miss}$ (with magnitude \MET) is defined as the negative vector sum of the transverse momenta of all identified and calibrated objects and remaining unclustered energy,
the latter of which is estimated from low-$p_\text{T}$ tracks associated with the primary vertex but not assigned to any lepton or jet candidate~\cite{PERF-2016-07}.
The \MET is taken as a measurement of the undetectable particles, and is affected by energy losses due to detector inefficiencies and acceptance, and by energy resolution.


%
%

%

%
\section{Event selection and classification}%
\label{sec:sel}
Events must satisfy the single-electron or single-muon trigger requirements described in \cref{subsec:data}.
Selected events are required to contain one lepton matched to the
corresponding lepton reconstructed by the trigger within $\Delta R <
0.15$ and with $\pt > \SI{27}{\GeV}$. These criteria ensure operation
on the trigger efficiency plateau, thereby minimising the corrections
needed in simulation to reproduce the per-lepton trigger efficiencies
measured in data~\cite{TRIG-2018-05,TRIG-2018-01}.

Events selected by the trigger are required to have at least one
primary vertex with at least two associated ID tracks with $\pt >
\SI{0.5}{\GeV}$. In each event, the primary vertex is defined as the
reconstructed vertex having the highest scalar sum of squared \pT of
associated tracks~\cite{ATL-PHYS-PUB-2015-026} among the vertices
consistent with the average beam-spot position.
These events are also required to contain either one lepton and at least
three $b$-tagged jets (70\% working point), two leptons with the same electric charge
or three leptons, based on the \textit{Loose} lepton definition
introduced in \cref{sec:object}.
Events containing \tauhad leptons are explicitly vetoed in all three
channels.

The sensitivity of the analysis relies on the discrimination provided by the multivariate response that combines different event-based variables.
In each channel, BDTs with gradient boosting (BDTGs) are trained on simulated events in a preselection region using the \texttt{XGBoost} Python package~\cite{Chen:2016btl}.
The model performance is evaluated using the hold-out method, in which
the considered data sample is randomly divided into an independent
training-data sample and a test-data sample that is used as a proxy to evaluate the performance of the model
on unseen data.
In all channels, only the \tHq process is considered as a signal in the BDTG training.
This decision is driven by the similarities observed between the \tWH and \ttH topologies,
so that including the \tWH process in the training weakens the separation between \tHq and \ttH and other background processes.
While the \tWH sample is not part of the BDTG training, its contribution is fully accounted for in the signal yield and in the limit setting later on.
A strong separation of the signal from the \ttH process leads to a reduced interplay between \ttH and the \tH signal in the results.

Details on the definition of the preselection regions and the background composition are given in the following while the yields are reported in \cref{tab:presel_yields}.

%

\begin{table}[htbp]
\centering
{
\tablesetup
\newcolumntype{C}{>{\centering\arraybackslash}S[table-format=6.4]@{\(\,\pm\,\)}S[table-format=3.4]}
\caption{Preselection region yields as predicted by the MC simulation. The uncertainty is statistical only and the PDG rounding is applied. The \ttbar background is split into \ttb, \ttc and \ttlight categories for the \onelep channel only; therefore the \ttbar contribution for this channel is absent. Moreover, for the \onelep channel only the data-driven estimate of the non-prompt and fake leptons background is given due to lack of a MC-based estimate.}
\label{tab:presel_yields}
\begin{adjustbox}{max width=0.999\textwidth}
\begin{tabular}{l|
*{2}{r@{\(\,\pm\,\)}l |}
*{1}{r@{\(\,\pm\,\)}l }}

\toprule
Process & \multicolumn{2}{c|}{\onelep} & \multicolumn{2}{c|}{\dilepSS}   & \multicolumn{2}{c}{\trilep} \\
\midrule
\tHq    & 91.1 & 1.7 & 12.41 & 0.17 & 3.15 & 0.05\\
\tWH    & 37.21 & 0.26 & 3.97 & 0.09 & 2.36 & 0.07 \\
\ttbar  & \multicolumn{2}{c|}{--}       & 1430 & 14     & 324 & 6   \\
\ttb    & 60170 & 90 & \multicolumn{2}{c|}{--}     & \multicolumn{2}{c}{--}   \\
\ttc    & 33585 & 70 & \multicolumn{2}{c|}{--}     & \multicolumn{2}{c}{--}   \\
\ttlight& 106290  & 120 & \multicolumn{2}{c|}{--}     & \multicolumn{2}{c}{--}   \\
$e/\mu_{\text{Fakes}}$   & 8028 & 80 & \multicolumn{2}{c|}{--}        & \multicolumn{2}{c}{--}       \\
\ttH    & 1676.9 & 1.5& 126.6 & 1.2   & 74.62 & 0.25  \\
\ttW    & 362.0 &1.6& 798.0 & 2.1       & 205.7 & 1.1   \\
\ttZ    & 891.0 & 6& 165.3 & 1.0   & 567.8 & 1.7   \\
\tZq    & 170.5 & 1.3 & 88.9 & 0.5    & 273 & 0.9   \\
\tWZ    & 2.22 & 0.22 & 18.73 & 0.08      & 80.47  & 0.16    \\
\PZ{}/\PW{}+jets & 4681 & 34 & 518 & 30 & 131 & 25 \\
VV, VVV & 292 & 6 & 298.6 & 3.3 & 579 & 4 \\
Single-top & 9331 & 31 & 164 & 4 & 20.1 & 1.5 \\
\Pqt{}\Paqt{}\Pqt{}, \Pqt{}\Paqt{}\Pqt{}\Paqt{} & 8.9 & 0.1   & 4.81 & 0.06 & 4.88 & 0.06  \\
\(V\PH, \text{VBF}\PH, gg\PH\)   & 25.0 & 0.1      & 19 & 4 & 7.4 & 2.8  \\
\midrule
Total & 225640 & 190        & 3650 & 33                    & 2274 & 26 \\
\bottomrule
\end{tabular}
\end{adjustbox}
}
\end{table}


%
%

%

%
\subsection{\onelepsec channel}%
\label{sec:sel:lepbb}
The \onelep preselection region is defined by requiring the presence of exactly one \textit{Tight} lepton (electron or muon) with \pT > \SI{27}{\GeV}.
The \SI{70}{\%} working point of the \texttt{DL1r} tagger is used to select at least three \btagged jets.
All events with at least five jets and at least four \btagged jets are vetoed.
This requirement has a negligible impact on the signal acceptance and will help any future combination with other \ttH analyses.
Given the presence of one neutrino coming from the \PW-boson decay, the event is expected to have a non-zero \MET.
For this reason an \MET>\SI{25}{\GeV} requirement is applied.

As shown in \Cref{tab:presel_yields}, the main contribution to the background in the preselection region comes from \ttjet events, constituting 90\% of the expected total yield.
The \ttjet events are further classified depending on the flavour of the additional jets produced.
Simulated events where at least one of the additional jets is matched to a \Pqb-hadron are classified as \ttb (30\% of the expected \ttjet yield);
if no matched \bjet is found but at least one jet is matched to a \Pqc-hadron the event is classified as \ttc (17\% of the expected \ttjet yield)
while all the other events are classified as \ttlight (53\% of the expected \ttjet yield).
The latter contribution also includes \ttbar events with no additional jets.

In the \onelep channel, the preselected events are used to train a five-dimensional BDTG including the categories \tHq, \ttb, \ttc, \ttlight and others.
The ``others'' category includes all the remaining backgrounds.
A Deep Neural Network (DNN) is also tested to validate the results obtained with the BDTG. Since the BDTG shows slightly better performance, it is chosen for the final results.

In total, 25 variables were used in the training, such as the number of jets, the lepton charge and flavour information, invariant mass and angular separation of various reconstructed jets, the minimum $\chi^2$ values from reconstructing the \PW-bosons and top-quark masses from their decay products under the \ttlight or \ttb hypotheses, as well as the pseudo-continuous $b$-tagging discriminant of selected jets (see \cref{tab:variables}). These variables are motivated by the expected kinematic properties of the signal. The \tHq topology features a forward spectator jet from the scattered incoming light-flavour quark, a hadronically decaying top quark, and a different multiplicity of \bjets compared to the dominant \ttbar background. The most important variables entering the BDTG reflect these features: the number of jets associated with the reconstructed hadronic top quark from different \ttbar reconstruction techniques provides strong separation from the \ttbar background; the lepton charge helps distinguish \tHq events, as the incoming light-flavour quark is dominantly a $u$-quark from one of the colliding protons, leading mostly to positively charged top quarks (and hence positive leptons), from charge-symmetric backgrounds such as \ttbar, \ttW, \ttZ or \ttH; the \pT of the highest-\pT forward jet is characteristic of the spectator jet in \tHq production; and the event sphericity encodes the more isotropic topology of \ttbar events compared to the typically more forward--backward structure of \tHq events~\cite{STDM-2011-33}.
The complete lists of input variables can be found in \cref{appendix}.
Variable importance is evaluated using the ``gain'' metric provided by the \texttt{XGBoost} algorithm, which reflects the relative improvement in model accuracy due to each variable.

The preselection phase-space is then split into two regions, defined by rectangular requirements on the number of forward jets ($N_{\text{forward}}^{\text{jet}}$)
and the number of central jets failing to meet the \texttt{DL1r} tagger 70$\%$ working-point requirement ($N_{\text{not-}b}^{70}$).
In the \onelep channel, a jet is classified as forward if its pseudorapodity is $2.5 < |\eta| < 4.5$ or if its pseudorapodity is $2 <|\eta| < 2.5$ and it fails to meet the 70$\%$ $b$-tagging requirement.
The two regions are enhanced in the \tH contribution and \ttb and they are referred to as SR and CR(\ttb), respectively.
The SR selection requires at least one forward jet and at most one central jet failing to meet the 70$\%$ $b$-tagging requirement, while for CR(\ttb) no forward jet should be present and at least two central jets must fail to meet the 70$\%$ $b$-tagging requirement.
\cref{tab:CRSR} summarises the selections applied to define the SR and CR.
%
%


\begin{table}[!htbp]
\tablesetup
\caption{Regions selection requirements for \onelep, \dilepSS and \trilep channels.
The regions are defined using a mixture of requirements on the
BDTGs outputs, the flavour of the leptons, the number of jets and
$b$-tagged jets, the sum of the transverse momentum of the leptons \HT(\Pl),
the leptons invariant masses ($m(\Pl\Pl)$) and the conversion criteria.}
\label{tab:CRSR}
\begin{adjustbox}{max width=0.999\textwidth}
\begin{tabular}{l | l | lccccc}
\toprule
Channel & Region & \multicolumn{1}{c}{BDT score} & Conversion cut & Jets & Leptons flavour (\pT ordered) & Other\\
\midrule
\multirow{6}{*}{\onelep} & \multirow{3}{*}{SR} & & & $N^{70}_{\text{not-}b}< 2$ & &\\
& & - & - & $N_{\text{forward}}^{\text{jet}}\geq 1$ & - & - \\
& & & & & &\\
\cmidrule{2-7}
& \multirow{3}{*}{CR(\ttb)} & & & $N^{70}_{\text{not-}b} \ge 2$ & &\\
& & - & - & $N_{\text{forward}}^{\text{jet}} = 0$ & - &\\
& & & & & &\\
\midrule
\multirow{18}{*}{\dilepSS} & \multirow{3}{*}{SR} & \(\BDT(\tHq)>0.65\) & & & &\\
& & \(\BDT(\ttbar)<0.5\) & yes & - & - & - \\
& & \(\BDT(\ttW)<0.6\) and \(\BDT(VV)<0.8\) & & & &\\
\cmidrule{2-7}
& \multirow{3}{*}{CR(\muHF)} & \(\BDT(\tHq)<0.65\) & & &\\
& & \(\BDT(\ttbar)>0.3\) & yes & - & \Pgm\Pgm & - \\
& & \(\BDT(\ttW)<0.6\) and \(\BDT(VV)<0.9\) & & & &\\
\cmidrule{2-7}
& \multirow{3}{*}{CR(\elHF)} & \(\BDT(\tHq)<0.65\) & & &  &\\
& & \(\BDT(\ttbar)>0.3\) & yes & - & \Pgm\Pe, \Pe\Pe& $\HT(\Pl)<\SI{225}{\GeV}$\\
& & \(\BDT(\ttW)<0.6\) and \(\BDT(VV)<0.9\) & & & &\\
\cmidrule{2-7}
& \multirow{3}{*}{CR(\elCo)} & & & & &\\
& & \(\BDT(\ttbar)>0.3\) & inverted & - & \Pgm\Pe, \Pe\Pe &m$(\Pl\Pl)<\SI{150}{\GeV}$\\
& & & & &\\
\cmidrule{2-7}
& \multirow{3}{*}{CR(\ttW)} & \(\BDT(\ttW)>0.6\) & & & &\\
& & \(\BDT(\ttbar)<0.3\) & yes & - & - & - \\
& & & & & &\\
\midrule
\multirow{24}{*}{\trilep} & \multirow{3}{*}{SR} & \(\BDT(\tHq)>0.7\) & & & &\\
& & \(\BDT(\ttbar)<0.9\) & yes & - & - & - \\
& & \(\BDT(\ttW)<0.8\) & & & &\\
\cmidrule{2-7}
& \multirow{3}{*}{CR(\muHF)} & \(\BDT(\tHq)<0.7\) & & & &\\
& & \(\BDT(\ttbar)>0.5\) & yes & - & \Pgm\Pgm\Pgm, \Pe\Pgm\Pgm, & - \\
& & \(\BDT(\ttW)<0.8\) & & & \Pgm\Pe\Pgm, \Pe\Pe\Pgm &\\
\cmidrule{2-7}
& \multirow{3}{*}{CR(\elHF)} & \(\BDT(\tHq)<0.7\) & & & &\\
& & \(BDT(\ttbar)>0.5\) & yes & - & \Pgm\Pgm\Pe, \Pe\Pgm\Pe & - \\
& & \(BDT(\ttW)<0.8\) & & & \Pgm\Pe\Pe, \Pe\Pe\Pe &\\
\cmidrule{2-7}
& \multirow{3}{*}{CR(\elCo)} & & & & &\\
& & - & inverted & - & \Pgm\Pgm\Pe, \Pe\Pgm\Pe, & - \\
& & & & & \Pgm\Pe\Pe, \Pe\Pe\Pe& \\
\cmidrule{2-7}
& \multirow{3}{*}{CR(\ttW)} & & & & & \\
& & \(\BDT(\ttW)>0.8\) & yes & - & - & - \\
& & & & & &\\
\bottomrule
\end{tabular}
\end{adjustbox}
\end{table}


%
%

%

\newcommand{\HTFootnode}{
The \HT variable is defined performing the scalar sum of the transverse momentum of all the reconstructed objects in the final state, in this case jets and leptons.
The $\HT(\Pl)$ variable instead is computed including only leptons.
}
\subsection{\dilepSSsec channel}%
\label{sec:sel:2lSS}
The \dilepSS preselection region is defined by requiring exactly two \textit{Tight} light leptons (electrons or muons) with the same electric charge.
For the selected events, the leading lepton must have $\pT>\SI{27}{\GeV}$ and the sub-leading lepton must have $\pT>\SI{20}{\GeV}$.

The number of jets with $|\eta|<4.5$ is required to be between one and six.
The \SI{70}{\%} working point of the \texttt{DL1r} tagger is used to select \bjets.
The number of \bjets is required to be between one and three.

As shown in \Cref{tab:presel_yields}, the dominant source of background in the preselection region stems from \ttbar followed by \ttW and \Zjets production.
Both \ttbar\ and \Zjets enter the preselection region if a non-prompt or fake lepton is in the event or if the charge of one of the electrons is wrongly measured.
The rest of the background processes consists of \ttW, \ttZ or \diboson production.

Four binary-classifier BDTGs are trained in the \dilepSS channel, each targeting the \tHq signal or one of the main background processes (\ttbar, \ttW, and \diboson) against the sum of the remaining three classes.
These classifiers are used to define  mutually exclusive enriched regions.
In total, 52 variables were used in the training: 28, 19, 29 and 27 for the four BDTGs, respectively. These variables include kinematic and topological quantities that characterise the signal and its main backgrounds. The most important variables include the charge of the two leptons, \HT\footnote{\HTFootnode}, the number of $b$-tagged jets, the invariant masses and angular separations of pairs of reconstructed jets and leptons (see \cref{tab:variables}). The charge of the two leptons is highly discriminating, since the \tHq process favours positively charged top quarks as explained before, while dominant backgrounds such as \ttbar, \ttH, and diboson production are charge-symmetric.
Invariant masses of leptons and jets also provide discriminating power, as the decay kinematics and spatial correlations in the \tHq topology produce characteristic patterns between leptons from top-quark decays and the forward spectator jet or \bjet. In contrast, in backgrounds such as \ttbar, \ttH, and dibosons, these correlations are absent or substantially altered due to additional heavy resonances or differing event topologies.
The other variables further exploit differences in global event activity, jet and \bjet multiplicities, and the more isotropic topology of \ttbar, \ttW and \ttZ compared to the forward–-backward structure of \tHq events.
The complete lists of input variables can be found in \cref{appendix}.

Five regions are defined in the \dilepSS channel:
the SR is designed to maximise the \tHq contribution, while the CRs target the main background processes.
The regions are defined using a mixture of requirements on the BDTGs outputs,
selecting a given flavour for the two leptons in the final state or cutting on event kinematic variables such as \HT or leptons invariant masses.
The material- or internal-conversion cuts defined in \cref{sec:object} are applied to all the electrons in all regions,
except for the conversion control region, CR(\elCo), where this requirement is inverted.
The regions described here are orthogonal to one another.
\cref{tab:CRSR} summarises the selections applied to define the SR and CRs.
Three CRs are defined to derive mis-identified leptons background: CR(\muHF), CR(\elHF) and CR(\elCo).
The CR(\elCo) is enriched in events with one electron stemming from \gam-conversion (\elCo), while CR(\muHF) and CR(\elHF) are enriched in contributions from non-prompt electrons or muons stemming from heavy-flavoured (containing a \Pqb- or \Pqc-quark) hadron decay, \elHF and \muHF, respectively.
One CR is defined to target the irreducible \ttW background.
%
%

%
%
%


%
%

%

\newcommand{\DeltaLepFootnode}{
Among the three leptons in the event in the \trilep channel, the lepton with the highest \pT is called $\ell_A$, the one with the second highest \pT is called $\ell_B$ while the lepton with the softest \pT is called $\ell_C$.
The variable $\Delta R(\ell_C,\ell_A+\ell_B)$ is defined computing the $\Delta R$ value of the $\ell_C$ lepton with the system composed by summing the four vectors of leptons $\ell_A$ and $\ell_B$.
}
\subsection{\trilepsec channel}%
\label{sec:sel:3l}
The \trilep preselection region is defined by requiring the presence of exactly three \textit{Tight} light leptons (electrons or muons).
The sum of the leptons' charges is required to be $\pm 1$.
The light leptons are ordered based on their \pT, with the leading lepton having $\pT>\SI{27}{\GeV}$,
the sub-leading lepton required to have $\pT>\SI{20}{\GeV}$ and the softest lepton to have $\pT>\SI{10}{\GeV}$.
The number of jets with $|\eta|<4.5$ is required to be between one and six.
The \SI{70}{\%} working point of the \texttt{DL1r} tagger is used to select \bjets.
The number of \bjets is required to be between one and three.
As shown in \Cref{tab:presel_yields}, the dominant background contributions in the preselection region arise from \diboson, \ttZ, and \ttbar processes.

Three binary-classifier BDTGs are trained in the \trilep channel, each targeting the \tHq signal or one of the main background processes (\ttbar and \ttW) against the sum of the remaining two classes.
These classifiers are used to define regions enriched in \tHq and in each of the background processes included in the training.
A total of 50 input variables are used in the training: 22, 19, and 30 for each of the three classifiers, respectively.
These variables include kinematic and topological quantities that characterise the signal and its main backgrounds.
The most important ones include the invariant masses of leptons, the \pT of leptons, the sum of leptons' charges, and the multiplicity of central and $b$-tagged jets. The dilepton invariant masses are powerful discriminants, since in backgrounds such as \ttZ and diboson production leptons often originate from on-shell $Z$ bosons, producing peaks around the \PZ{}-boson mass, while in \tHq events they mainly arise from \PW{}-boson decays and do not show such resonant structures (see \cref{tab:variables}). The sum of leptons' charges is also discriminant similarly to the other two channels. The lepton transverse momenta provide additional separation, as backgrounds with extra vector bosons typically yield harder leptons than those expected in \tHq events.
Finally, the number of $b$-tagged jets provides discrimination, as \tHq process typically features fewer central and \bjets than backgrounds such as \ttbar, \ttH, \ttW, and \ttZ.
The complete lists of input variables can be found in \cref{appendix}.

Five orthogonal regions are defined in the \trilep channel:
one SR to maximise the \tHq contribution, three CRs to estimate in data the mis-identified leptons background and one CR to estimate the \ttW background.
The regions are defined using a mixture of requirements on the BDTGs outputs, the number of jets and invariant masses.
The material- and internal-conversion requirements defined in \cref{sec:object} are applied to all the electrons in all regions,
except for one region where this requirement is inverted to enhance the contribution of the process targeted by that region.
\Cref{tab:CRSR} summarises the selection criteria applied to define the SR and CRs.

%
%
%

%
%
%
%


%
%
%


%
%
%

%
%

%

%
\section{Background estimation}
\label{sec:back}
The background processes satisfying the SR and CR selections are categorised into irreducible and reducible backgrounds.
Irreducible backgrounds arise from processes that have an identical number of prompt leptons from weak boson or leptonic $\tau$-lepton decays, and a similar number of $b$-jets as the targeted signal.
Reducible backgrounds arise instead from particle mis-identification in the detector, mis-assigning of the lepton charge or mis-tagging a charm or light-flavoured jet as a \bjet.
Fake and non-prompt leptons can originate from different sources like decays of bottom or charm mesons, photon conversions
or light jets creating accidentally a detector signature similar to a light lepton.

\begin{figure}[!htp]
\centering
\subfloat[\label{fig:data_over_prediction_regions_summary1L}]{\includegraphics[height=0.3\textheight]{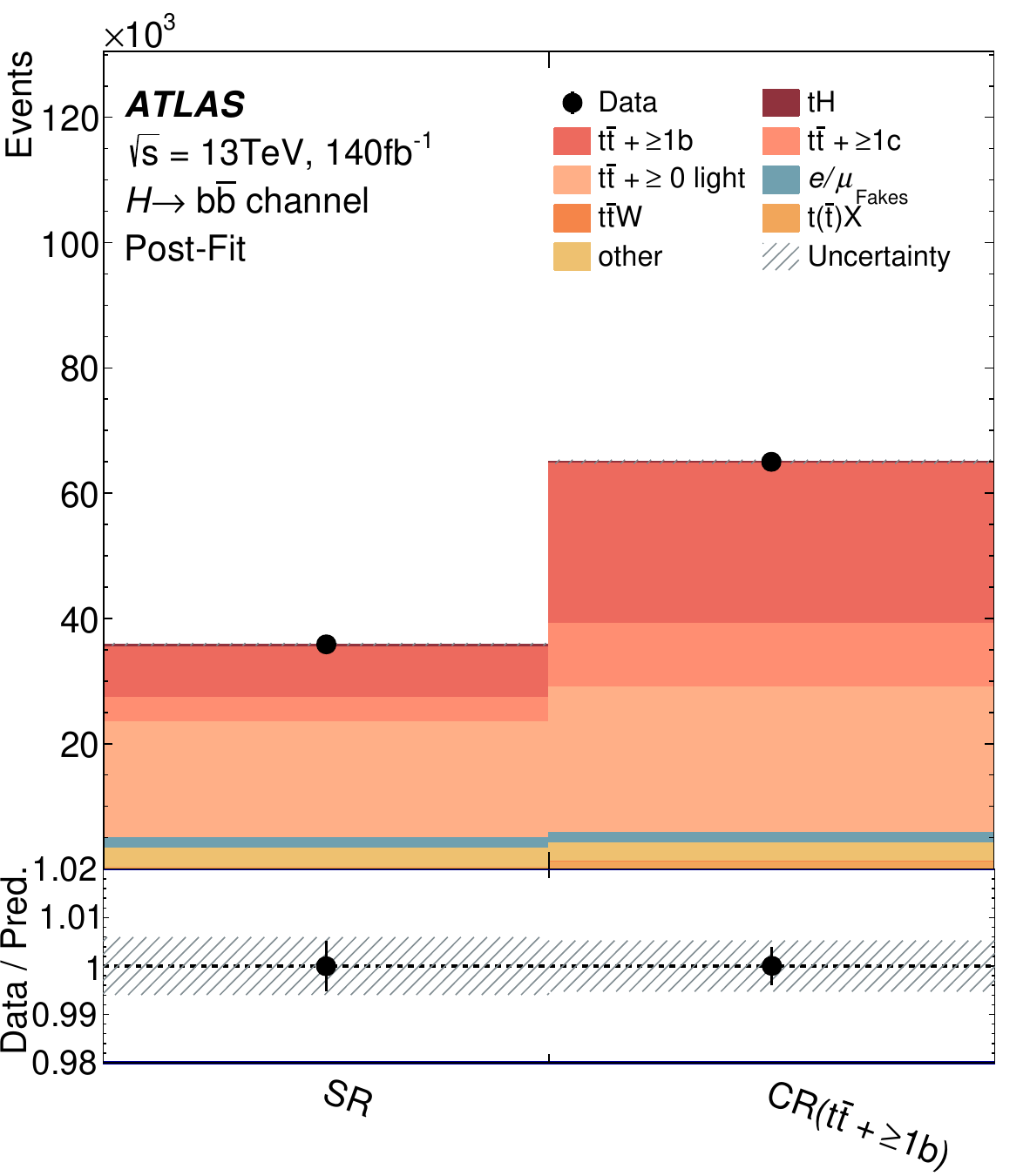}}
\subfloat[\label{fig:data_over_prediction_regions_summaryML}]{\includegraphics[height=0.3\textheight]{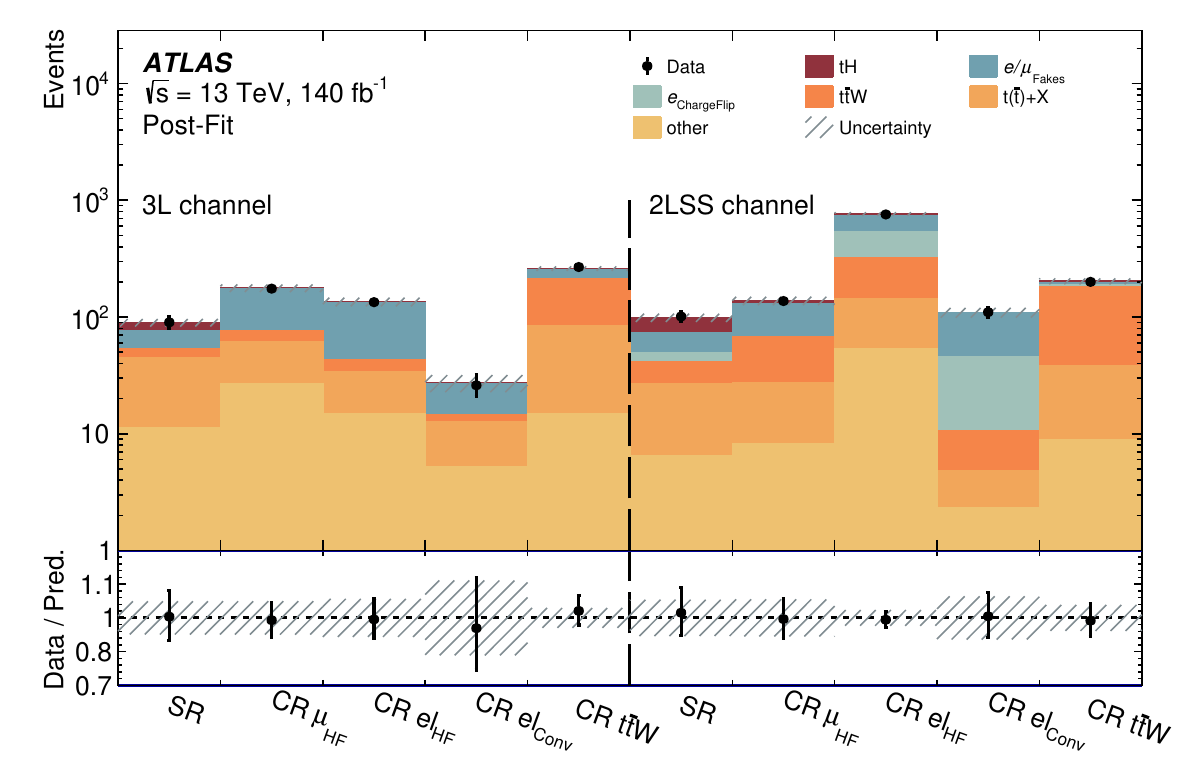}}
\caption{
The data and estimated signal-plus-background yields in the SR and CRs of the (a) \onelep channel and (b) \trilep and \dilepSS channels.
The term $e/\mu_\text{Fakes}$ refers to backgrounds from non-prompt or misidentified leptons, namely contributions from heavy-flavour decays and photon conversions.
Both signal and background events are estimated with a likelihood fit to data (\enquote{Post-Fit}), as described in \cref{sec:back,sec:result}.
The uncertainty band includes statistical and systematic contributions.}
\label{fig:data_over_prediction_regions_summary}
\end{figure}

The contribution of irreducible backgrounds is generally derived through the respective simulated samples and constrained in CRs where needed.
A data-driven estimate is provided for most of the reducible backgrounds and for background processes where the MC prediction is not in good agreement with the data.
\Cref{fig:data_over_prediction_regions_summary} compares the total data yields in all SRs and CRs as defined in \cref{sec:sel} for the three analysis channels to the signal plus background predictions obtained by the likelihood fit to data described in \cref{sec:back} and \cref{sec:result}.

%

%
%

%

%
\subsection{Background estimation in the \onelepsec channel}%
\label{sec:back:lepbb}
As already discussed in \cref{sec:sel:lepbb}, the main source of background in the \onelep channel consists of \ttbar+jets events.
The production of the \ttbar in association with heavy flavour jets is known to be not well modelled by the simulation.
For this reason, simulated \ttbar+jets events are split into \ttb, \ttc and \ttlight categories and the CR(\ttb) is included in the fit to constrain its contribution from data.
The normalisation factor of the \ttb background found in the combined fit to data is $k(\ttb) = 1.24^{+0.17}_{-0.15}$.
The \ttc predicted yield has no dedicated correction factor due to the difficulty in building a pure CR.
Instead, an overall uncertainty on the \ttc prediction is included in the fit as described in \cref{sec:sys:lepbb}.
The remaining irreducible backgrounds are all estimated by using the MC predictions.
The small contribution of reducible background from non-prompt and fake leptons is estimated using the matrix-method technique described in Ref.~\cite{EGAM-2019-01}.
%


%
%

%

%
\subsection{Background estimation in the \dilepSSsec and \trilepsec channels}%
\label{sec:back:2lSS}
In the \dilepSS channel, the main irreducible backgrounds originate from \ttW production, having final states and kinematic properties inherently similar to the \tHq signal.
Smaller contributions originate from the following rare processes: \tZq, \ttH and \tWZ production.
Backgrounds with prompt leptons are directly estimated from simulation using the samples described in \cref{sec:datamc}, except for \ttW, for which the normalisation is fitted.
In the \trilep channel instead, the main irreducible backgrounds originate from \diboson production, \ttZ and \tZq.
Other minor irreducible backgrounds are \triboson, other Higgs-boson production modes such as \ggH, VBF (\qqH) and $V\PH$, and \ttt and \tttt.
A CR dedicated to the \ttW background, CR(\ttW), is defined in both the \dilep and \trilep channel to better constraint its contribution from data.
The normalisation factor associated with the \ttW background as estimated from the combined fit is found to be $k(\ttW) = 1.09\pm0.11$.

The most important sources of reducible background in both the \dilepSS and \trilep channels are the \ttbar, $W/Z$+jets and single-top-quark processes.
These processes can satisfy the preselection requirement due to the presence of a fake or non-prompt lepton in the final state.
These backgrounds are estimated from simulation, with the normalisation determined by the likelihood fit through the use of the dedicated CRs introduced earlier.
The non-prompt leptons in the simulated event samples are labelled according to whether they originate from heavy-flavour or light-flavour hadron decays, or from a material/internal conversion.
The light-flavour category is negligible.
In the \dilepSS channel, normalisation factors are extracted to be $k(\elCo)_{\dilepSS} = 1.20\pm0.47$, $k(\elHF)_{\dilepSS} = 0.55\pm0.65$ and $k(\muHF)_{\dilepSS} = 0.80\pm0.32$.
In the \trilep channel, the normalisation factors extracted are $k(\elCo)_{\trilep} = 0.82\pm0.59$, $k(\elHF)_{\trilep} = 1.47\pm0.42$ and $k(\muHF)_{\trilep} = 0.91\pm0.24$.

Electrons with mis-identified electric charge (\enquote{charge-flip}),  with one electron either undergoing hard bremsstrahlung
followed by an asymmetric conversion ($e^\pm \to e^\pm \gamma^* \to e^\pm e^+ e^-$) or having mismeasured track curvature,
represent another source of reducible background. This contribution is negligible in the \trilep channel and sizeable in the \dilepSS channel.
The muon charge misassignment rate is negligible in the $\pt$ range relevant to this analysis.
The electron charge misassignment rate is measured in data by using samples of $Z \to e^+e^-$ events reconstructed as either same-charge pairs or opposite-charge pairs,
with the non-\PZ{} background subtracted via a sideband method utilising events outside the \PZ{}-boson mass window.
The charge misassignment rate is extracted from the ratio of the numbers of same-charge and opposite-charge events close to the \PZ{}-boson mass
through a likelihood approach taking into account the possibility that both electron charges are misassigned.
The rates are parameterised as a function of electron $\pt$ and $|\eta|$, and vary from about $10^{-5}$ for low-$\pt$ electrons ($17 \leq \pT \leq \SI{50}{\GeV}$) with $|\eta| \leq 1.37$,
to about $4 \times 10^{-3}$ for high-$\pt$ electrons ($\pT \geq \SI{100}{\GeV}$) with $2 \leq |\eta| \leq 2.47$.
To estimate the charge-flip background in each \dilepSS region, the measured charge misassignment rate is applied to data events satisfying the requirements of the region,
except that the two leptons must have opposite charges.



%
%

%

%
\section{Sources of systematic uncertainty}
\label{sec:systs}
Different sources of systematic uncertainty are considered for this measurement.
They are grouped into three categories:
instrumental uncertainties that affect physics object reconstruction;
modelling uncertainties relevant for MC simulation-based estimations;
and uncertainties affecting the estimate of fake and non-prompt lepton backgrounds.
All uncertainties include both shape and normalisation effect, unless otherwise stated in their description.

The description of the sources of systematic uncertainties is organised as follows: first, those common to all three channels are presented, followed by sections addressing uncertainties specific to the \onelep\ channel and the \dilepSSsec\ and \trilepsec channels.

\subsection{Common systematic uncertainties}

Instrumental systematic uncertainties considered are related to: trigger efficiency; lepton reconstruction, identification and isolation efficiencies; lepton calibration; jet calibration; JVT and fJVT efficiencies; $b$-tagging efficiency; \MET calibration.
The experimental systematic uncertainties are applied either as an event-reweighting correction factor or as a rescaling or smearing of the object energy and momentum for scale and resolution uncertainties.

Uncertainties associated with the lepton selection arise from the trigger, reconstruction, identification and isolation efficiencies,
and the lepton momentum scale and resolution as described in Ref.~\cite{EGAM-2018-01,MUON-2018-03,MUON-2022-01}.
Uncertainties on the efficiency of the BDT designed to reject non-prompt leptons are estimated through a $Z\to\ell\ell$ tag-and-probe method and cover contributions related to the $Z(\to\ell\ell)$+jets MC modelling,
the template cut/shape, the $m_{\ell\ell}$ window, the tag-and-probe lepton selections, the multijet background, the non-prompt lepton background, the luminosity,
the cross-sections of the considered processes, and the limited number of events in simulation and data.

Uncertainties associated with jet reconstruction and calibration arise from the jet energy scale (JES), jet energy resolution (JER) and the JVT/fJVT requirements.
The JES uncertainty accounts for contributions from composition of jet flavour, $\eta$-intercalibration, punch-through, single-particle response, calorimeter response and \pileup making up for a total of 31 independent variations~\cite{JETM-2018-05}.
The JER is measured separately for data and MC using an in-situ technique and the associated uncertainties are split into 13 components~\cite{JETM-2018-05}.
The JVT efficiency is measured in data and correction factors are applied to simulated events to correct the simulated performance.
The uncertainty associated with the JVT performance is obtained by varying the correction factors~\cite{PERF-2014-03}.

The uncertainty in \MET results from the propagation of the uncertainties in the energy scales and resolutions
of photons, leptons and jets, and from the modelling of its soft term \cite{JETM-2020-03}.

The efficiency of the $b$-tagging algorithm is measured for each jet flavour using control samples in data and in simulation.
From these measurements, correction factors are derived to correct the tagging rates in the simulation.
For \bjet{}s, the correction factors and their uncertainties are estimated from data using dileptonic \ttbar events~\cite{FTAG-2018-01,FTAG-2019-07}.
For $c$-jets, they are derived from jets arising from \Wboson boson decays in \ttbar events~\cite{FTAG-2020-08}.
For light-flavour jets, the correction factors are derived from \Zjets events~\cite{FTAG-2019-02}.
Uncertainties from sources affecting the \Pqb-tagging efficiencies of $b$-jets, $c$-jets and light-jets are evaluated as a function of jet \pt, including bin-to-bin correlations.

The uncertainty in the reweighting of the MC pile-up distribution to match the distribution in data is evaluated by varying the reweighting factors
and has a very small impact on either the combined or individual results.
The uncertainty in the combined 2015--2018 integrated luminosity is 0.83\% \cite{DAPR-2021-01},
obtained using the LUCID-2 detector for the primary luminosity measurements,
complemented by measurements using the inner detector and calorimeters.

Systematic uncertainties in the \tHq and \tWH MC predictions due to missing higher-order QCD corrections in the modelling are estimated by doubling and halving the \muR\ and \muF\ scales chosen for the nominal sample.
A systematic uncertainty due to the choice of the PS model is estimated as the relative difference between the \POWHEG+\PYTHIA[8] and \POWHEG+\HERWIG[7] predictions and applied to the nominal predictions.
An uncertainty is considered on the cross-section summing in quadrature the contributions from QCD scale, flavour scheme, PDF and $\alpha_s$ from a NLO QCD prediction: the values obtained are equal to 15.4$\%$ and 9.2$\%$ for \tHq and \tWH, respectively~\cite{deFlorian:2016spz,YR4twiki}.
The uncertainty in the \ttbar modelling due to the choice of the $\hdamp$ parameter, which controls the resummation of higher-order terms, is estimated by comparing the nominal prediction with a \POWHEG+\PYTHIA[8] sample where the parameter is set to $3\times \mtop$.
Reweighted events are used to assess the impact of ISR simulation within the nominal \POWHEG+\PYTHIA[8] simulation: \muF\ and \muR scales are varied by 0.5 (high radiation) or 2.0 (low radiation) together with the PS tune \texttt{Var3c}~\cite{ATL-PHYS-PUB-2014-021}.
The \muR\ and \muF\ scales are varied in both the hard-scatter process and PS and they are modified independently.
The hadronisation model uncertainty and other non-perturbative effects in the PS are evaluated by comparing the nominal sample with \POWHER[7.2.1].
The PS scale uncertainty is evaluated using the \texttt{Var3c} eigentune variation of the A14 tune \cite{ATL-PHYS-PUB-2017-007}.
The impact of FSR is evaluated by varying the renormalisation scale for PS emissions by factors of 0.5 and 2.0, using reweighted events.

The uncertainty in \ttH modelling due to the ISR is estimated by using weights in the ME calculation and the PS.
The \muR\ and \muF\ scales are varied, and ISR \alphas variations are evaluated similarly to the \ttbar process.
The impact of FSR is evaluated using PS weights that vary the renormalisation scale for QCD emission in the FSR by a factor of 0.5 and 2.0.
Additionally, uncertainties in the ME (PS, hadronisation and underlying-event modelling) are estimated by comparing the nominal \POWHEG+\PYTHIA 8 MC simulation with \AMCatNLO+\PYTHIA 8 (\POWHEG+\HERWIG 7).
The predicted \ttH cross-section uncertainty is $^{+5.8\%}_{-9.2\%}(\text{scale})+3.6\%(\text{PDF+\alphas})$~\cite{deFlorian:2016spz} and the two components are considered uncorrelated.

Systematic uncertainties in the \ttW MC predictions due to missing higher-order QCD corrections in the modelling are considered.
To estimate the uncertainty due to ambiguities in the ME and PS algorithm and parameter choices, the nominal \Sherpa prediction is compared with the prediction of the \MGNLO+\PYTHIA 8 FxFx sample described in Section~\ref{sec:datamc}.
In addition, a dedicated PS model uncertainty is estimated as the relative difference between the \POWHEG+\PYTHIA 8 and \POWHEG+\HERWIG 7 predictions and applied to the nominal \ttW prediction.
Modelling uncertainties due to \muR\ and \muF\ scales~\cite{Bothmann:2016nao}, and the amounts of ISR/FSR are evaluated similarly to the \ttbar\ and \ttH\ processes.
Uncertainties in the PDF modelling are included by varying the value of \alphas and using different PDF sets.
The predicted \ttW cross-section uncertainty is $^{+9.7\%}_{-10.8\%}(\text{scale})+7.2\%(\text{PDF+\alphas})$~\cite{Frederix:2021agh} and the two components are considered
uncorrelated.
The uncertainty on the theoretical prediction of the \ttW cross-section is dropped in the \trilep and \dilepSS channels as the normalisation of this background is estimated from data.
Uncertainties due to missing higher orders in QCD are estimated for \ttZ production by varying the nominal \muR\ and \muF\ scales are evaluated similarly to the \ttbar, \ttH and \ttW processes.
Uncertainties in additional-jet modelling are estimated with ISR \alphas variations taken from the A14 tune.
To estimate the uncertainty due to ambiguities in the ME and PS algorithm and parameter choices, the nominal prediction is compared with \Sherpa.
PS, hadronisation and underlying-event modelling uncertainties are estimated by comparing the nominal \AMCatNLO+\PYTHIA 8 MC simulation with \MGHER[7].
The predicted \ttZ cross-section uncertainty is $^{+9.6\%}_{-11.3\%}(\text{scale})+4\%(\text{PDF+\alphas})$~\cite{deFlorian:2016spz}.

Systematic uncertainties due to missing higher-order QCD corrections are also considered for the three single-top-quark processes: $t$-channel, \tW and $s$-channel.
The uncertainty due to the choice of the hadronisation model and the other non-perturbative aspects of the PS is evaluated comparing the nominal prediction with that provided by the \POWHEG{}+\HERWIG[7.1.6] simulation.
An uncertainty is also attributed to the choice of the \POWHEG approach to perform the matching between the hard-scatter and the PS. This uncertainty is estimated
comparing the nominal \POWHEG{}+\PYTHIA[8] prediction with the \AMCatNLO{}+\PYTHIA[8] simulation for the \tW process,
and the \POWHEG{}+\HERWIG[7.1.6] with the \AMCatNLO{}+\HERWIG[7.1.6] prediction for the \(t\)-channel and \(s\)-channel.
An additional uncertainty is quoted for the \tW process to estimate the difference between the DS and DR schemes.
Finally, a conservative 5\% uncertainty on the theoretical cross-section of single-top-quark processes (\(t\)-channel, \tW and \(s\)-channel) is considered in the \onelep channel. In the \trilep channel, a 5.4\% uncertainty on the \tW production cross-section~\cite{Kidonakis:2010ux} is considered. For the \dilepSS channel, where single-top-quark contributions are negligible, no uncertainty from these processes is included.

Uncertainties in \diboson modelling due to missing higher orders in QCD are estimated by varying the nominal \muR\ and \muF\ scales by factors of $0.5$ and $2.0$.
In addition, a normalisation uncertainty of 24.5\% is considered, as a result of adding in quadrature the theory uncertainty (5\%) and 24\% per additional jet, accordingly to the Berends–Giele scaling~\cite{Berends:1991}.

For minor backgrounds such as \Zjets, \Wjets, \tZq, \tWZ, rare top-quark (\ttt and \tttt) and other Higgs-boson production modes (\ggH, VBF and $V\PH$) only the theoretical uncertainty on the predicted cross-section is considered.
A normalisation uncertainty of 35\% and 40\% is considered for the \Zjets and \Wjets processes in the \onelep channel only.
This is the result of adding in quadrature the theory uncertainty (4\% for \Wjets and 5\% for \Zjets) and 24\% per additional jet.
In the \dilepSS and \trilep channels, the normalisation of these processes is derived from data as part of the fake and non-prompt leptons estimate.
The predicted \tZq cross-section uncertainty is $^{+7.7\%}_{-7.9\%}(\text{scale})+0.9\%(\text{PDF+\alphas})$~\cite{Alwall:2014hca} and the two components are considered uncorrelated.
The contribution of the \tWZ, rare top-quark and other Higgs-boson production processes is small in all the channels, for this reason a conservative 50$\%$ uncertainty on their cross-section is applied.

%
%

%

%
\subsection{Channel-specific uncertainties: \onelepsec channel}%
\label{sec:sys:lepbb}
In the \onelep channel, the effect of every modelling uncertainty associated to the \ttbar simulation are considered independently for the three categories, \ttb, \ttc and \ttlight.
In the \ttb case, since its yield is estimated from data, the normalisation of alternative samples is rescaled to match the nominal prediction in the preselection region.
In addition to the \ttbar modelling uncertainties common to all the channels, one from the NLO matching scheme in the \ttbar simulated sample is also included comparing the nominal sample to the  $\pT^\text{hard}$ varied one.
A modelling uncertainty due to the differences between the nominal \ttb simulation in the 5FS and an alternative \ttbb simulation in the 4FS is considered~\cite{Ferencz:2023fso}.
The uncertainty is applied independently to three sub-components of the \ttb background, defined according to the number of hadrons matched to the additional jet produced in association with the top-quark pair.
The three sub-components are:
\begin{itemize}
\item \ttbar+1$b$, if one additional jet in the event is matched to a single $b$-hadron,
\item \ttbar+1$B$, if one additional jet in the event is matched to a gluon splitting $g\rightarrow b\bar{b}$,
\item \ttbar+2$b$, if two additional jets in the event are matched to a single $b$-hadron each.
\end{itemize}
A 50\% uncertainty on the normalisation of the \ttc is also considered in the fit model \cite{TOPQ-2021-26}.
Finally, a theoretical uncertainty of 6\% is applied on the \ttlight normalisation as estimated in a dedicated validation region.

A systematic uncertainty is included to account for the difference between lepton fake rates originating from heavy-flavour decays versus those from misidentified jets.
This uncertainty is evaluated separately for muons and electrons and its normalisation component contributes approximately 10$\%$ for muons and 30$\%$ for electrons.


%
%

%

%
\subsection{Channel-specific uncertainties: \dilepSSsec and \trilepsec channels}%
\label{sec:sys:2lSS3l}
For both the \dilepSS and \trilep channels, all the common \ttbar modelling uncertainties are considered, as the \ttbar simulation is used to estimate non-prompt lepton backgrounds originating from heavy-flavour decays and conversions.
The modelling uncertainties associated to \ttbar simulation are considered correlated among the various non-prompt leptons categories and between the \dilepSS and \trilep channels.
Their effect is shape only as the normalisation of these backgrounds is estimated from data.
In addition to the \ttbar modelling uncertainties common to all the channels, one from the NLO matching scheme in the \ttbar simulated sample is also included comparing predictions from \POWHER[7.1.3] and \MGNLOHER[7.1.3] simulations.
A conservative uncertainty of 50\% is assigned to a small fraction of events, roughly 3$\%$ in the preselection region, where the origin of the non-prompt lepton could not be identified.

In the \dilepSS channel a systematic uncertainty of 10\%--60\% is assigned to the estimation of background from electrons with misidentified charge as described in \cref{sec:back:2lSS}.



%
%

%

%
\section{Results}
\label{sec:result}
A maximum-likelihood fit to all bins in the three SRs and the nine CRs (see \cref{tab:CRSR}) is performed to simultaneously determine the background and the \tH signal yields in the data.
The discriminant variable used in the SRs is defined as follows: for the \onelep channel, it is the BDTG response for \tHq\ signal-like events, P(\tHq); for the \dilepSS and \trilep channels, it is given by the output of the BDTG trained to target \tHq events, BDT(\tHq). In the CR(\ttb), the discriminant is the BDTG response for \ttb-like events, P(\ttb).
The \HT variable is used in all the CRs of the \dilepSS channel and in the \trilep channel control region targeting the \ttW process.
Finally, the $\Delta R(\ell_C,\ell_A+\ell_B)$ variable~\footnote{\DeltaLepFootnode} is used in the CR(\muHF) and CR(\elHF) of the \trilep channel while the BDTG response of the binary classifier targetting \ttbar is used in the CR(\elCo).

The likelihood function ${\cal L}(\mu,\vec{k}, \vec{\theta})$ is constructed as a product of Poisson probability terms over all bins considered in the measurement,
and depends on the following: the signal-strength parameter, $\mu$, defined as a multiplicative factor applied to the predicted yield for the \tHq and \tWH signal processes;
$\vec{k}$, the normalisation factors for several backgrounds;
and $\vec{\theta}$, a set of nuisance parameters (NPs) encoding systematic uncertainties in the signal and background expectations~\cite{Cranmer:1456844}.
Systematic uncertainties can impact the estimated signal and background rates, the migration of events between categories, and the shape of the fitted distributions;
they are summarised in \cref{sec:systs}.
Both $\mu$ and $\vec{k}$ are treated as free parameters in the likelihood fit.
The NPs $\vec{\theta}$ allow variations of the expectations for signal and background according to the systematic uncertainties, subject to Gaussian constraints in the likelihood fit.
Their fitted values represent the deviations from the nominal expectations that are needed to provide the best fit to the data.
Statistical uncertainties in each bin due to the limited size of the simulated event samples are taken into account with dedicated parameters, using the Beeston--Barlow \enquote{lite} technique~\cite{Barlow:1993dm}.
All systematic uncertainties are fully correlated across channels, except for those that are specific to a given channel.

The best-fit value of the \tHq and \tWH signal strength, $\mu_{\tH}$, is:
\begin{equation*}
\mutHresult\,.
\label{eq:mu_tH}
\end{equation*}

\begin{figure}[htbp!]
\centering
\subfloat[\label{fig:SM_poi}]{\includegraphics[TRExPOITrim, height=0.29\textheight]{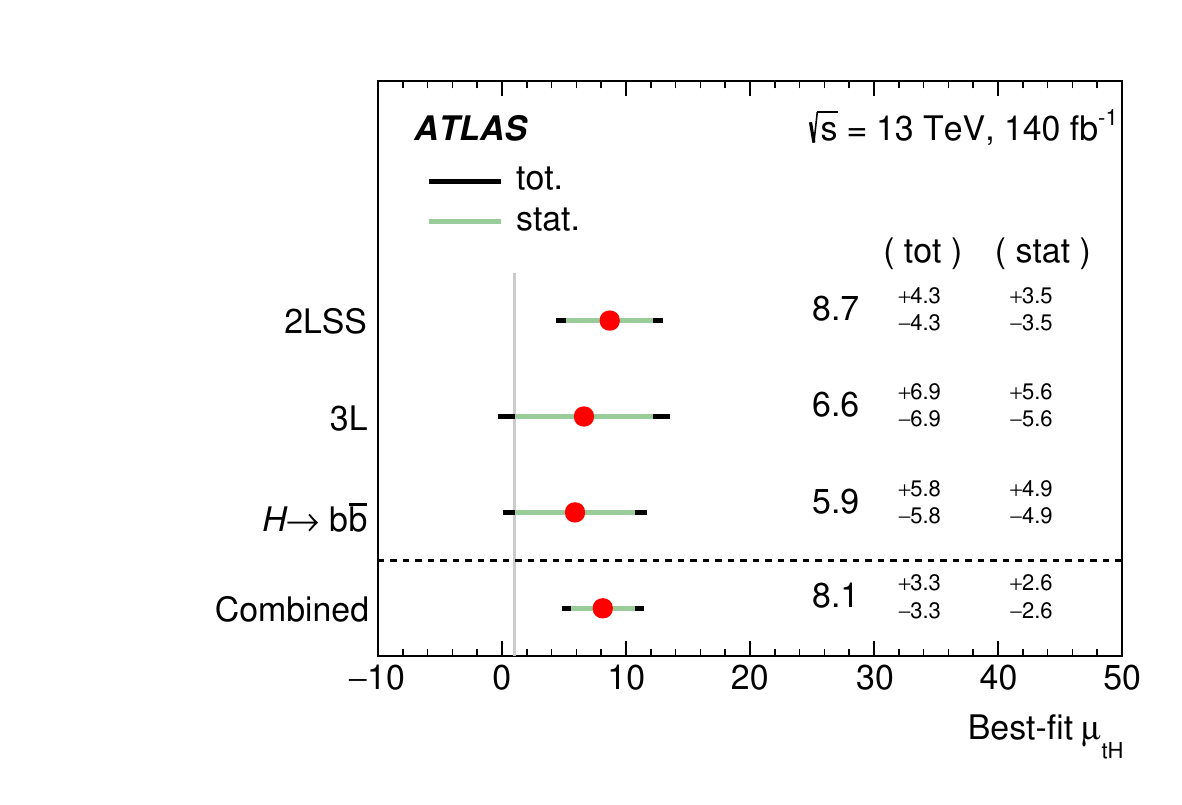}}
\subfloat[\label{fig:SM_limit}]{\includegraphics[TRExNPTrim, height=0.29\textheight]{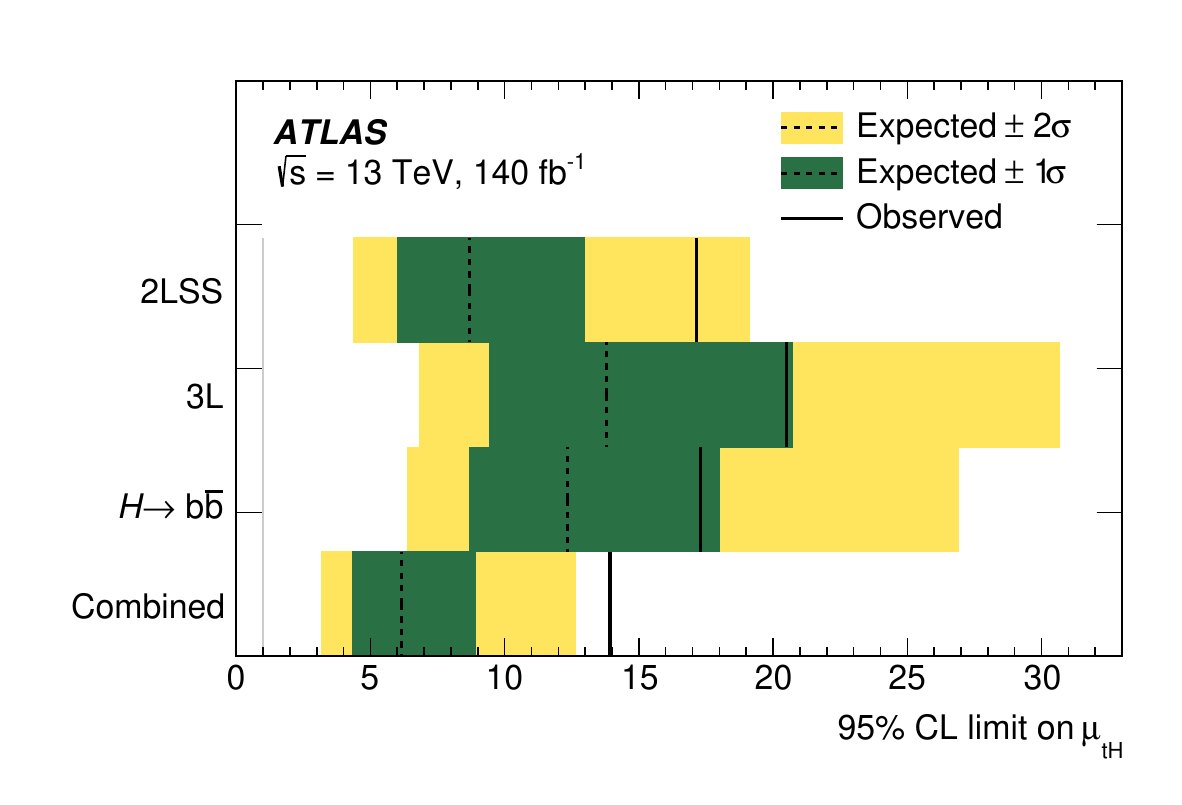}}
\caption{The (a) fitted values of the $\mu_{\tH}$ and (b) 95\% CL upper limits in the individual channels and in the combined measurement.
The values reported for the individual channels are obtained by fitting $\mu_{\tH}$ separately for each channel, whereas the combined measurement is derived by simultaneously fitting all channels together.
The observed limits are shown (solid black lines), together with the expected limits under the SM hypothesis (dotted black lines).
A grey line is added in correspondence of the $\mu_{\tH}$ equal to one.
In the case of the expected limits, the one- and two-standard-deviation uncertainty bands are also shown.}
\label{fig:SM_poi_limit}
\end{figure}

This value corresponds to a measured value of the cross-section of $\sigma(\tH) = \measxsectH\,\mathrm{fb}$.
\Cref{fig:SM_poi} shows the signal strength measured in the different channels together with their combined value.

The compatibility of the signal strength evaluated from the different channels and the combined value is assessed by performing a likelihood ratio test and a p-value of 91\% is obtained.
The significance of the observed (expected) signal above the background-only expectation is \significanceObs\ (\significanceExp) standard deviations.
The significance of the observed signal compared with the signal-plus-background hypothesis is instead 2.4.
In the absence of a significant deviation from the background-only hypothesis, a 95\% CL upper limit on the signal strength is set at \upperLimitObs~(\upperLimitExp) times the observed (expected) SM prediction, as shown in~\cref{fig:SM_limit}.
To quantify how well the fit model describes the data, a goodness-of-fit test is performed.
The test consists of comparing the nominal model to a so-called saturated model \cite{Cousins2013GeneralizationOC}, one constructed with enough degrees of freedom to describe the data perfectly.
The probability obtained for the goodness-of-fit test is 98\%, confirming excellent fit quality.
%

%


\begin{table}[t]
\centering
\caption{The absolute contribution of the main sources of uncertainty on the signal strength $\mu_{\tH}$.
Some NPs with small impacts are grouped into categories, with the impact of the category being the quadrature sum of individual parameter impacts.
Indentation is used to denote subcategories.
}
\begin{tabular}{l|c|c|c|c}
\hline
Source  & \multicolumn{4}{c}{Impact on $\mu_{\tH}$ }\\ \hline
&   Combination  & \onelep  &\dilepSS & \trilep \\  \hline\hline
Theory uncertainties (modelling)                    & 1.69   		 & 2.97  	& 1.64 	  & 2.63\\
\quad Signal										& 1.26   		 & 0.88     & 1.33 	  & 1.22 \\
\quad \ttb 											& 0.70   		 & 2.57     & - 	  & - \\
\quad \ttlight and \ttc                             & 0.54  		 & 1.04     & - 	  & - \\
\quad \ttW                                          & 0.23    		 & <0.01    & 0.52    & 0.18 \\
\quad \ttbar                                        & 0.43     		 & -        & 0.52    & 1.95 \\
\quad Rest of backgrounds                           & 0.50     		 & 0.60     & 0.63    & 1.23 \\
\hline
Experimental uncertainties      					& 1.07           & 0.95   	& 1.61 	  & 2.61  \\
\quad JES and JER				& 1.00           & 0.51 	& 1.57	  & 2.51  \\
\quad Non-prompt lepton bkg.~modelling             & 0.25           & 0.59	    & 0.22	  & 0.07 \\
\quad Flavour tagging                               & 0.17           & 0.52	    & 0.13 	  & 0.17 \\
\quad Muons                                         & 0.12           & < 0.01 	& 0.12	  & 0.40 \\
\quad \MET                                           & 0.10           & 0.04	    & 0.20 	  & 0.03 \\
\quad Pile-up                                       & 0.12           & 0.04	    & 0.05 	  & 0.44\\
\quad Electrons                                     & 0.10           & 0.07	    & 0.11	  & 0.32 \\
\quad Charge mis-identification                     & 0.07           & - 		& 0.06	  & - \\
\quad JVT and fJVT                                  & 0.04           & 0.10 	& 0.11	  & 0.06 \\
\quad Luminosity                                    & 0.09           & 0.09 	& 0.10 	  & 0.14 \\

\hline
Total systematic uncertainty      					&  2.01  		& 3.11  	& 2.30 	  & 4.06  \\
Statistical uncertainty                 			&  2.56 		& 4.87 	    & 3.5 	  & 5.60 \\
\quad Normalisation factors							&  0.50 		& 0.04 	    & 0.84    & 0.73 \\
MC statistics                    					&  0.64  		&  < 0.01 	& 1.05 	  & 1.67 \\
\hline
\textbf{Total}                                		&  3.3          & 5.8       & 4.3     & 6.9  \\ \hline
\end{tabular}
\label{tab:sys_impact}
\end{table}


The contributions from the different sources of uncertainty affecting the measured $\mu_{\tH}$ are evaluated using the covariance-matrix decomposition method~\cite{Pinto:2023yob} and are shown, grouped by category, in \cref{tab:sys_impact}.
The leading source of uncertainty is the statistical uncertainty.
The largest source of systematic uncertainty is the theoretical uncertainty with the biggest contribution coming from the uncertainty on the signal prediction and \ttbar.
Among the experimental uncertainties, the leading contribution stems from the JES and JER.

\begin{figure}[htbp!]
\centering
\includegraphics[width=0.65\textwidth]{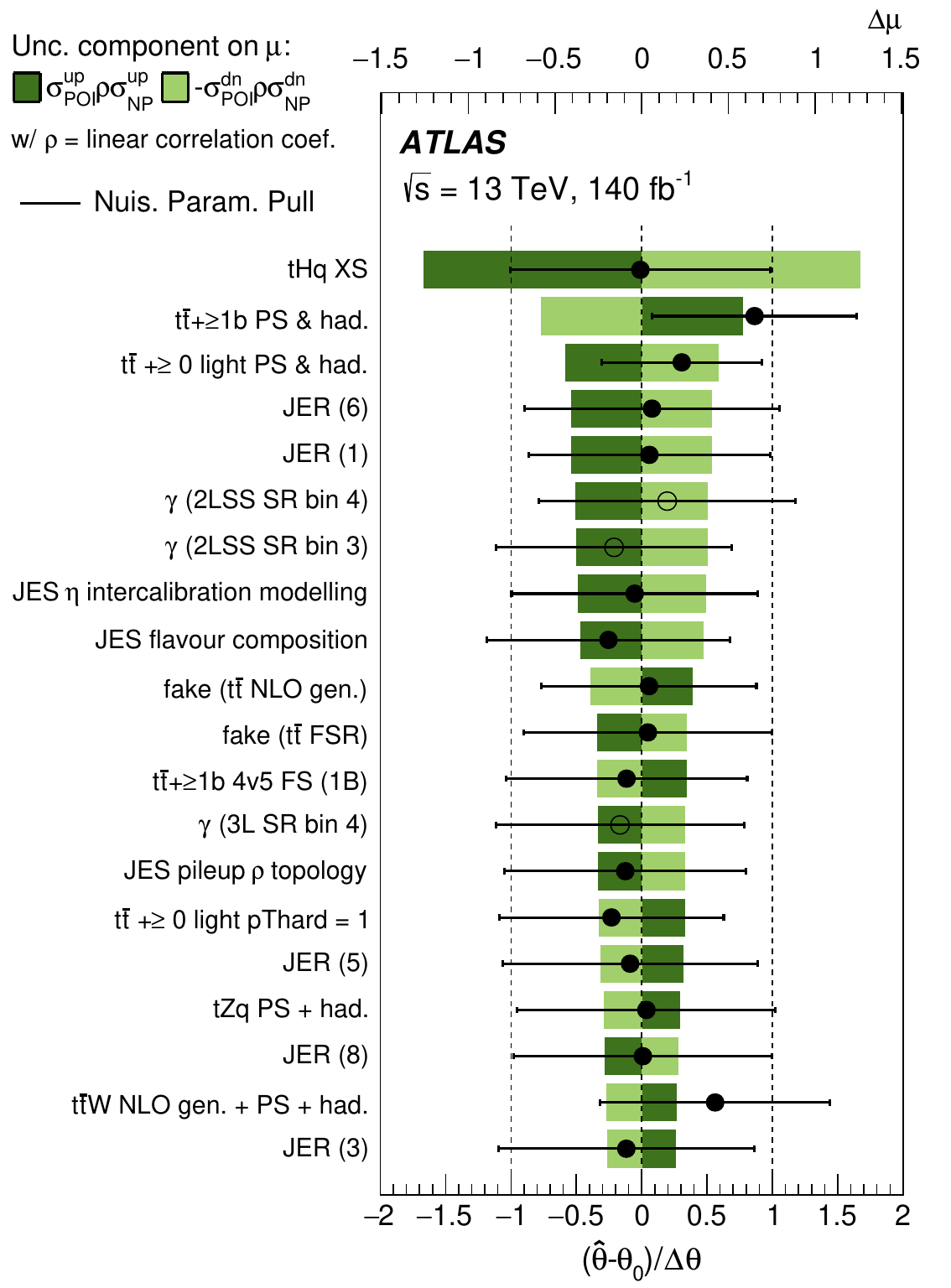}
\caption{The impact of the top 20 most important systematic uncertainties on the fitted $\mu_{\tH}$ in the SM scenario. The uncertainties are listed in decreasing order of their impact on $\mu_{\tH}$. The filled boxes show the relative impact on $\mu_{\tH}$, referring to the top $x$-axis. The points, referring to the bottom $x$-axis, show the deviations on the fitted NPs $\boldsymbol{\theta}_{0}$ from their nominal values, expressed in terms of standard deviations with respect to their nominal uncertainties. The associated black uncertainty bars show the fitted uncertainties of the NPs, relative to their nominal uncertainties. Filled markers refer to systematic uncertainties, while empty black markers refer to statistical uncertainties in the MC simulation in a given bin of an analysis region and they are labelled as $\upgamma$. The results shown are extracted from the fit including all channels.}
\label{fig:SM_ranking}
\end{figure}

The impact of the top 20 NPs in the fit on $\mu_{\tH}$ is shown in \cref{fig:SM_ranking} for the SM scenario and it is assessed as follows: if the uncertainty on the parameter is $\sigma_{k}$, the uncertainty on $\mu_{\tH}$ is $\sigma_\mu$, and the correlation between the parameter and $\mu_{\tH}$ is $\rho_{k,\mu}$, the impact is computed as $\sigma_k \rho_{k,\mu} \sigma_\mu$.
The plot shows the measured NPs and their agreement with the expected values.
The most constrained NP corresponds to the PS uncertainty in the \ttlight category.
All pulls are below one standard deviation with the strongest pull stemming from the PS uncertainty on the \ttb category, suggesting a mild mis-modelling of this contribution in the \onelep channel.
Overall, the lack of significant pulls or constraints in the NPs indicates a good agreement between data and the fit model, supporting the robustness of the fit and the reliability of the extracted signal strength.

\begin{figure}[ht!]
\centering
\subfloat[]{\includegraphics[width=0.33\textwidth]{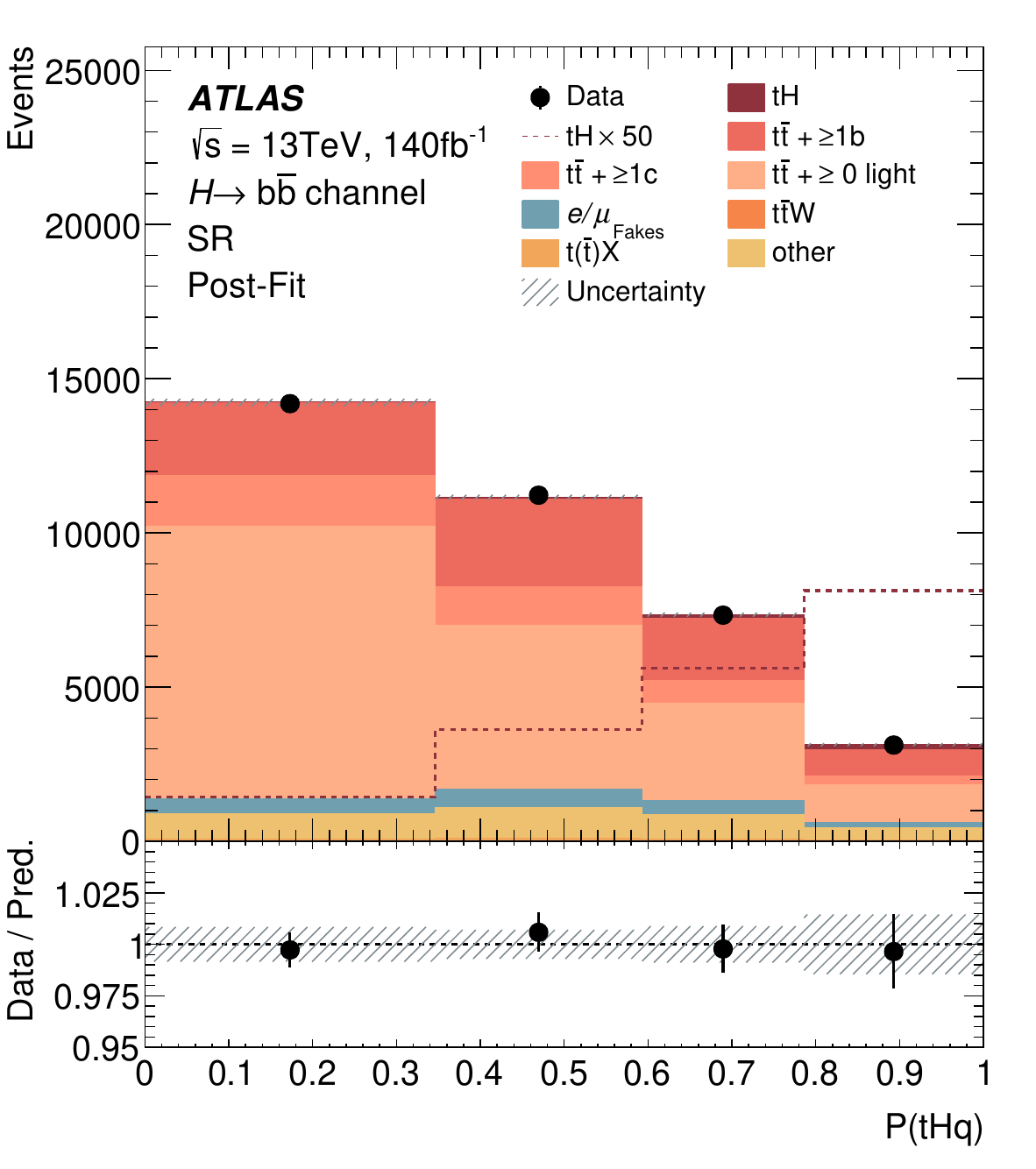}}
\subfloat[]{\includegraphics[width=0.33\textwidth]{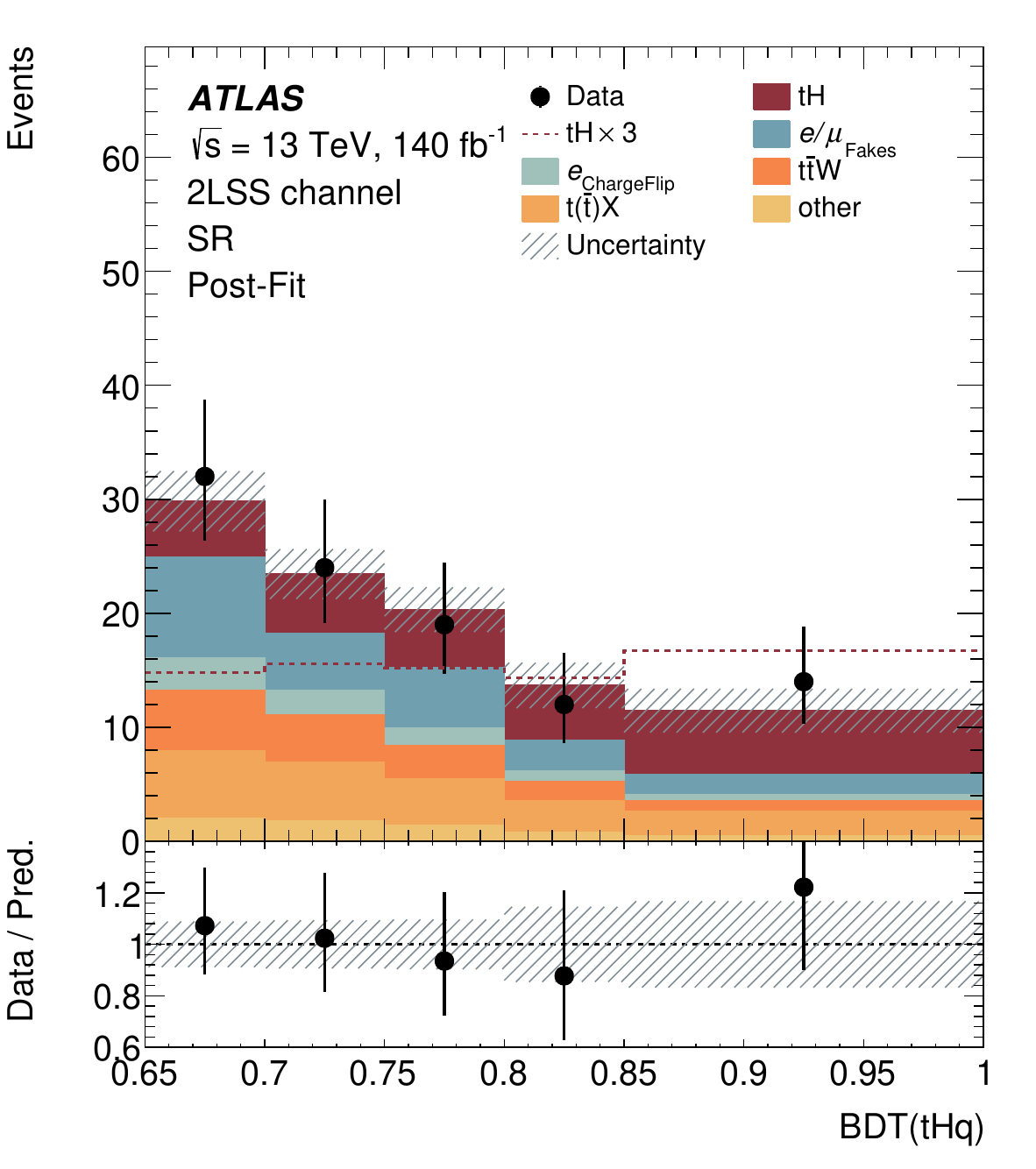}}
\subfloat[]{\includegraphics[width=0.33\textwidth]{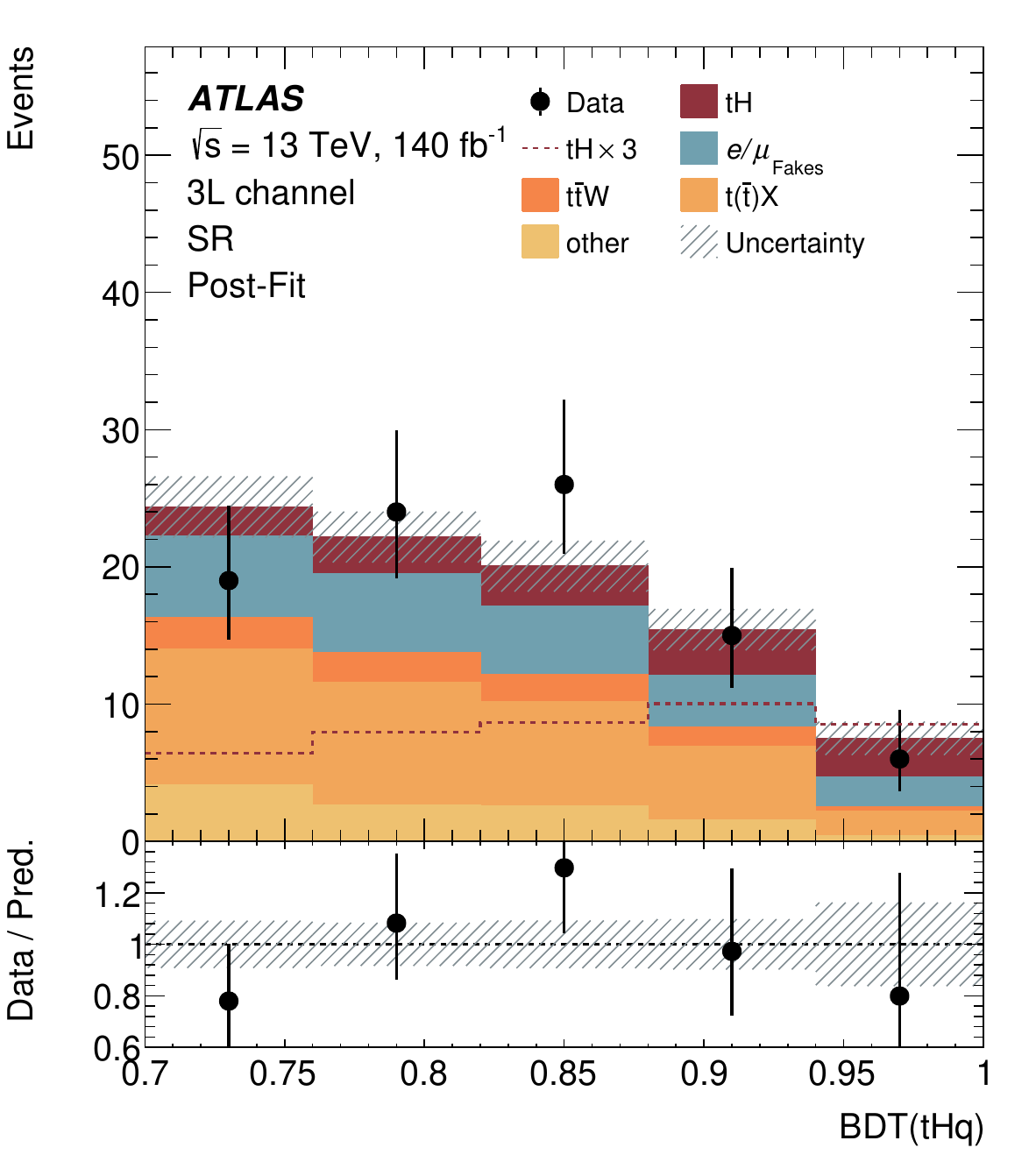}}
\caption{Comparison between data and the signal-plus-background prediction for the event yields in the (a) \onelep, (b) \dilepSS and (c) \trilep SRs.
The term $e/\mu_\text{Fakes}$ refers to backgrounds from non-prompt or misidentified leptons, namely contributions from heavy-flavour decays and photon conversions.
The SM signal and the background contributions after the likelihood fit to data (\enquote{Post-Fit}) under the signal-plus-background hypothesis are shown as filled histograms.
A dashed red line is added to the plot to show the signal contribution normalised to the measured value multiplied by 50 for the \onelep channel and by 3 for the \dilepSS and \trilep channels.
The ratio of the data to the total post-fit prediction (\enquote{Pred.}) is shown in the lower panel.
The combined statistical and systematic uncertainty in the prediction is indicated by the grey hatched band.
}
\label{fig:summary_SR}
\end{figure}

\begin{figure}[ht!]
\centering
\subfloat[]{\includegraphics[width=0.33\textwidth]{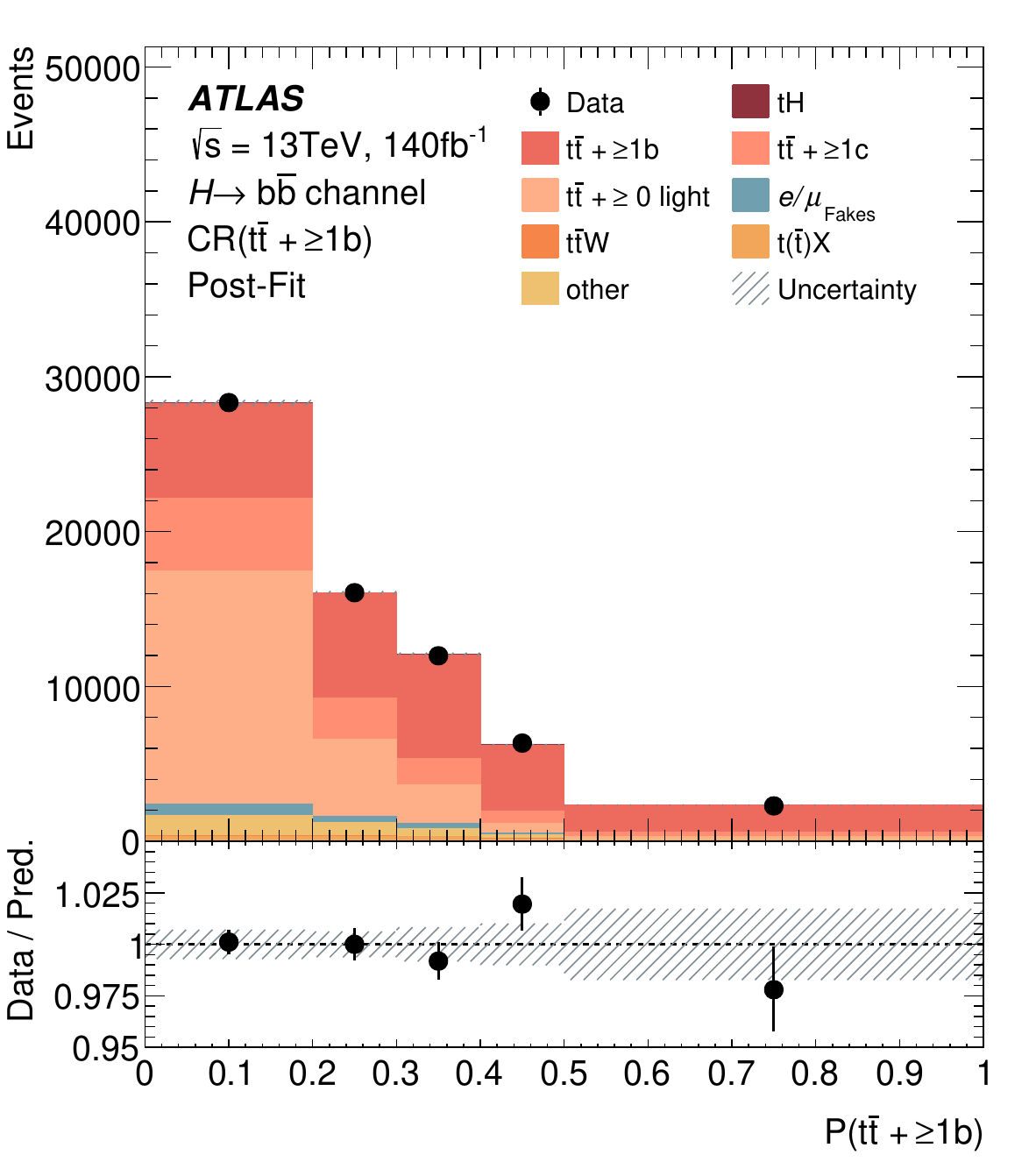}}
\subfloat[]{\includegraphics[width=0.33\textwidth]{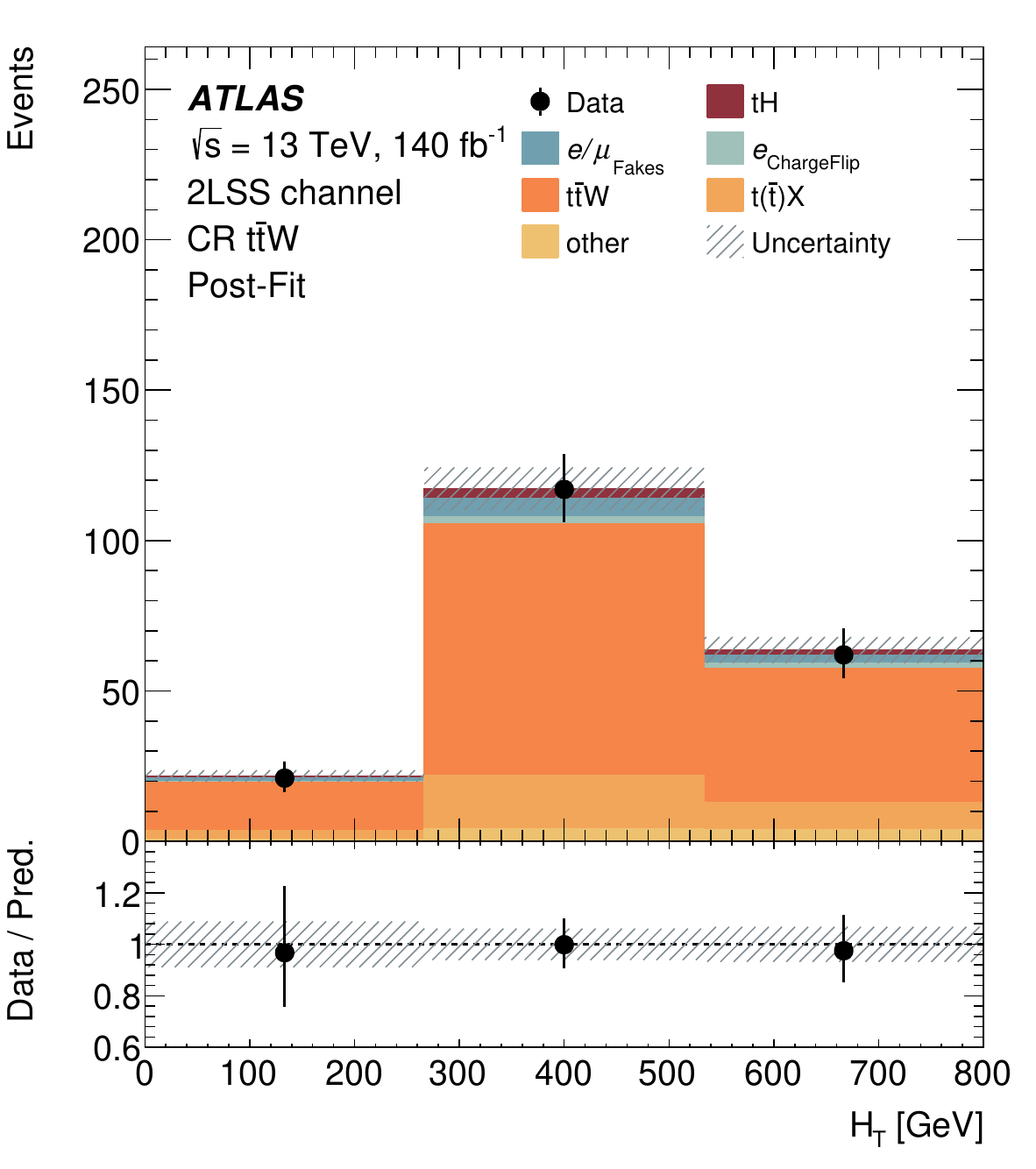}}
\subfloat[]{\includegraphics[width=0.33\textwidth]{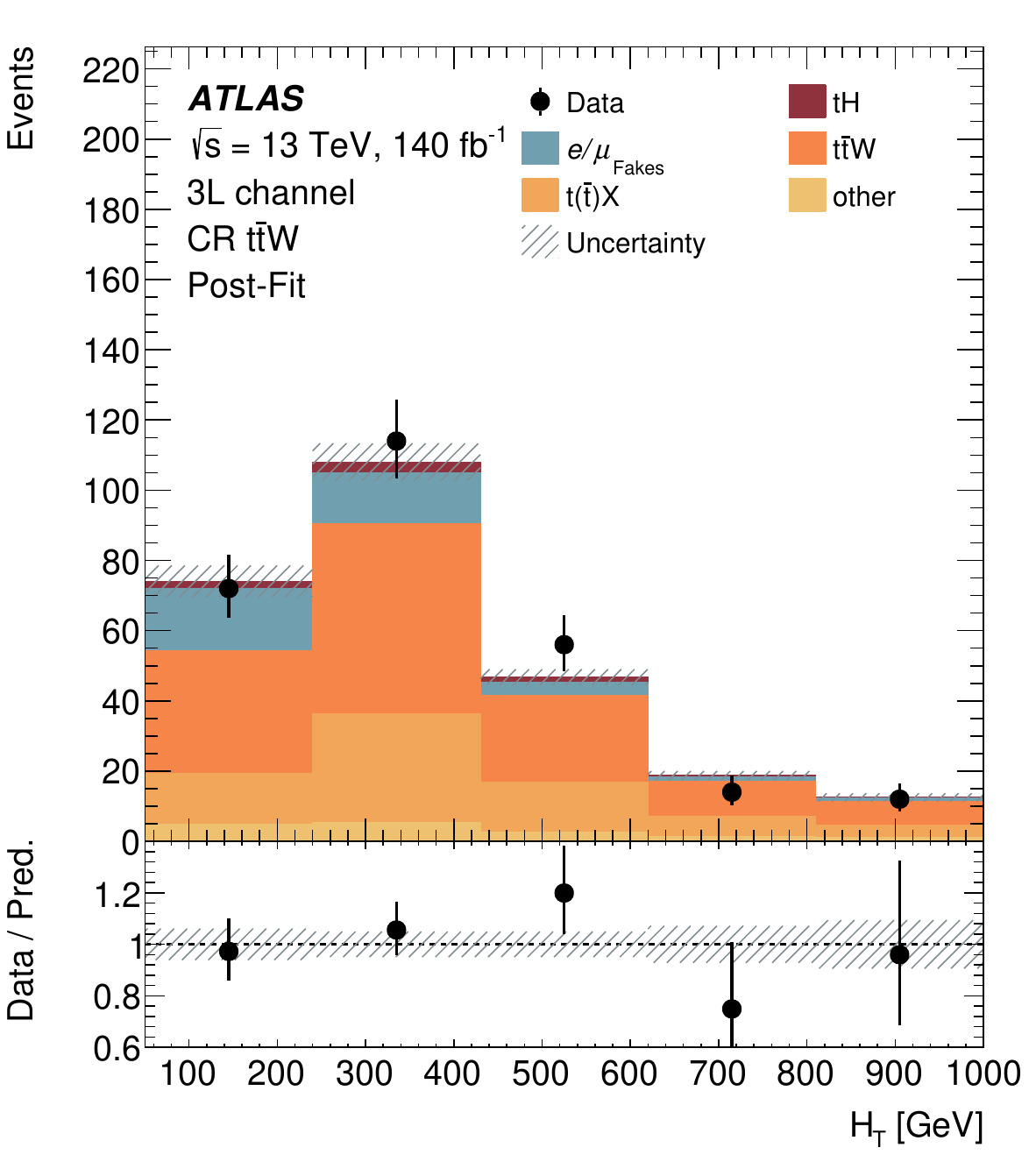}} \\
\subfloat[]{\includegraphics[width=0.33\textwidth]{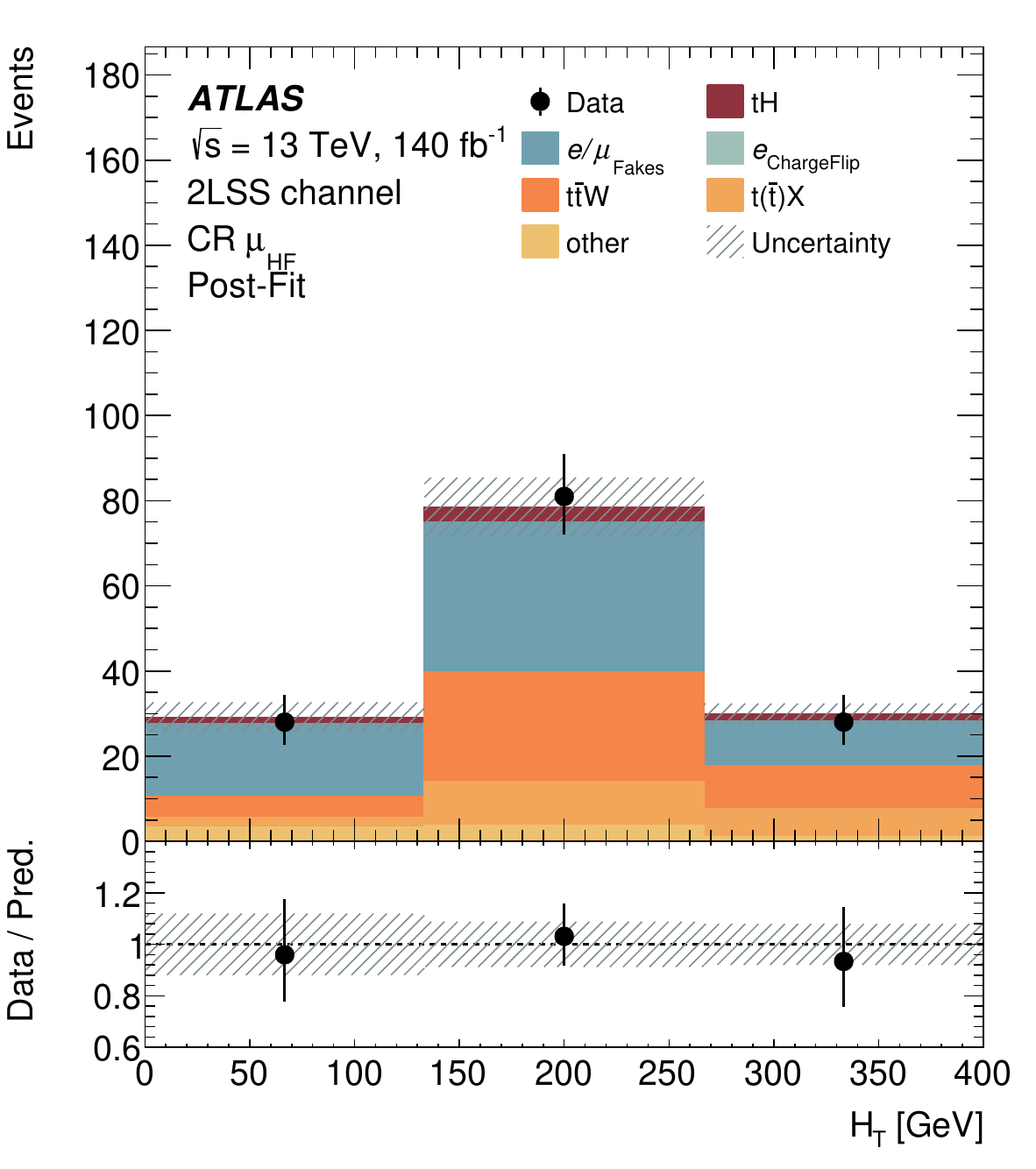}}
\subfloat[]{\includegraphics[width=0.33\textwidth]{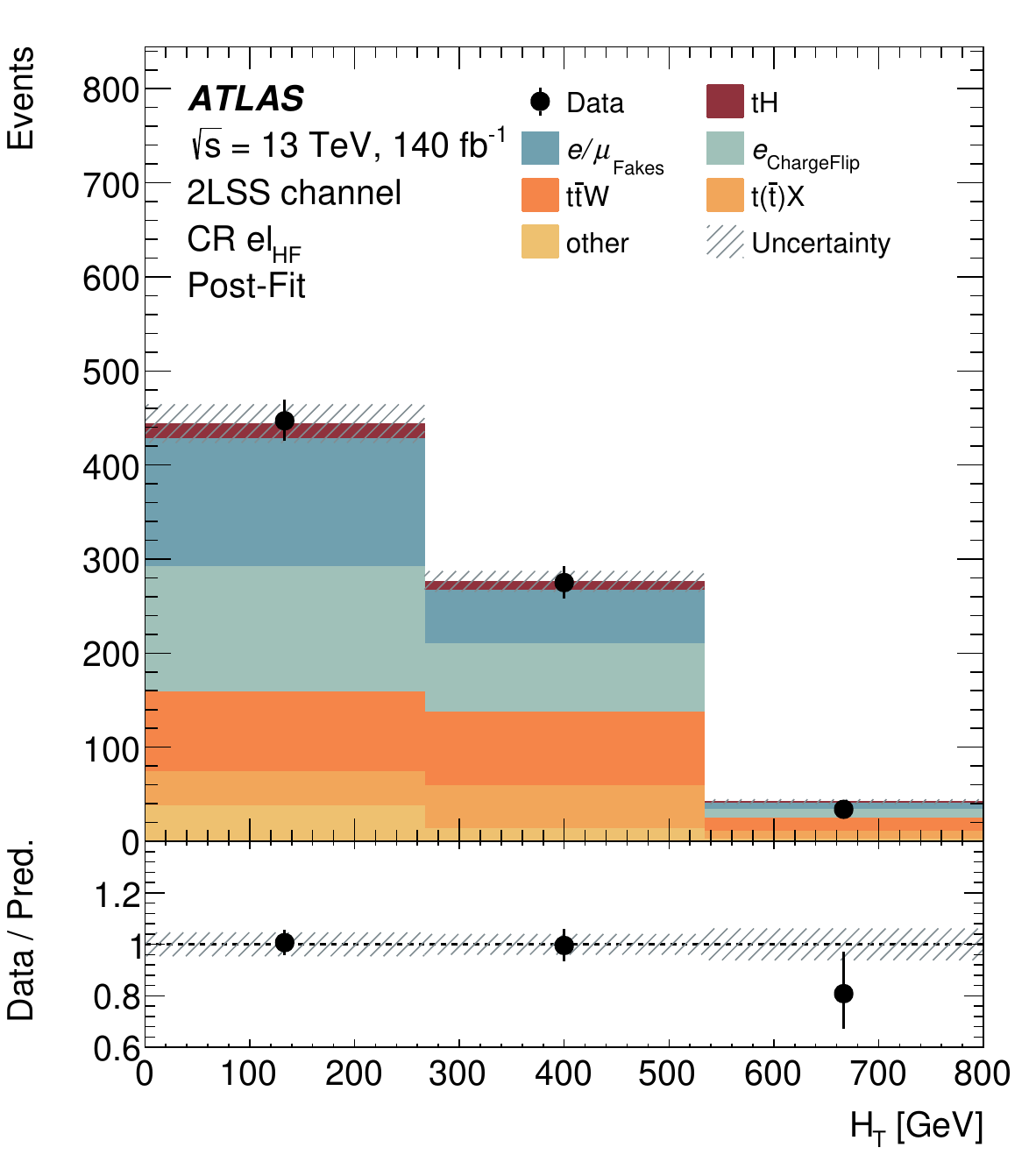}}
\subfloat[]{\includegraphics[width=0.33\textwidth]{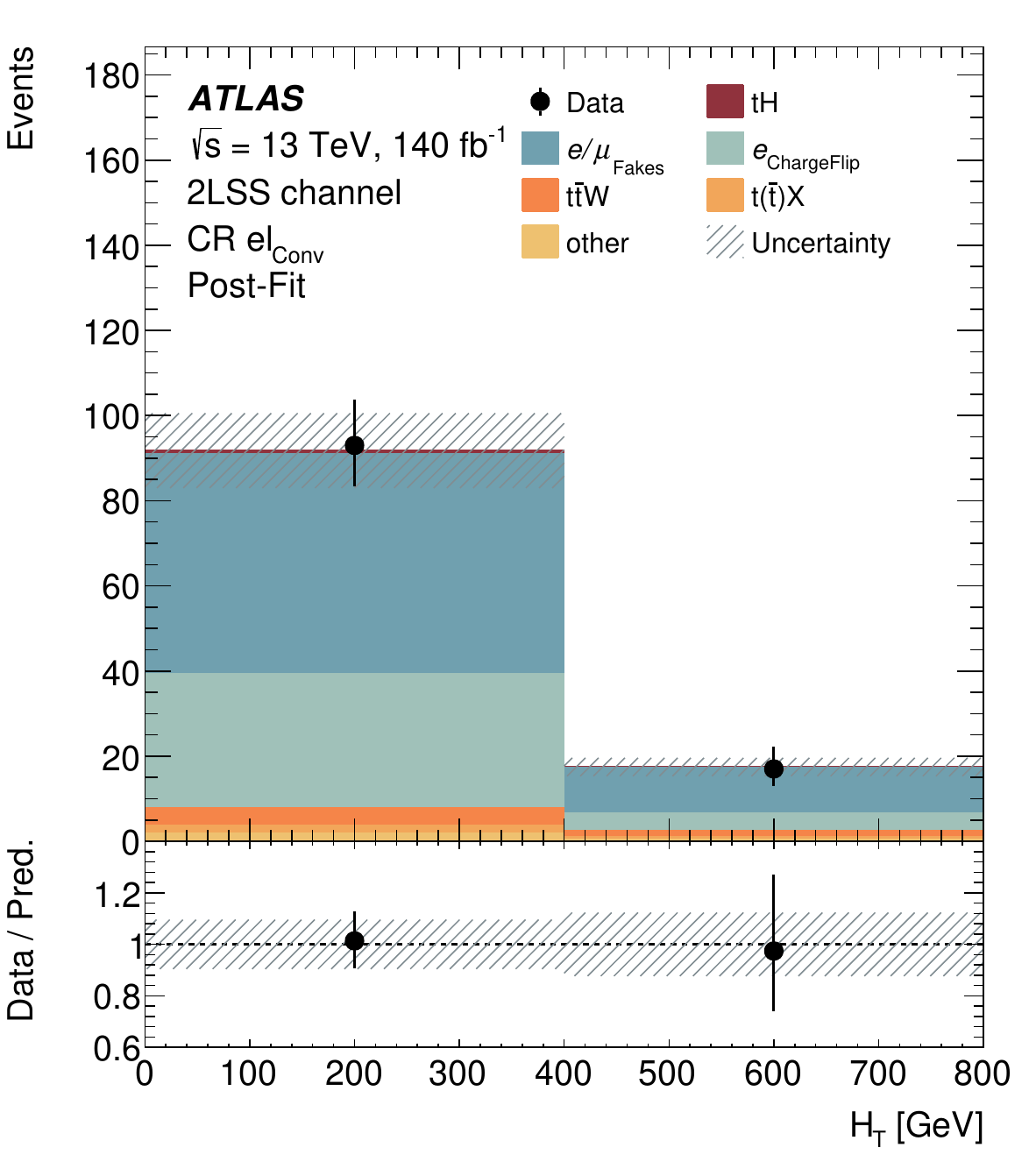}}\\\
\subfloat[]{\includegraphics[width=0.33\textwidth]{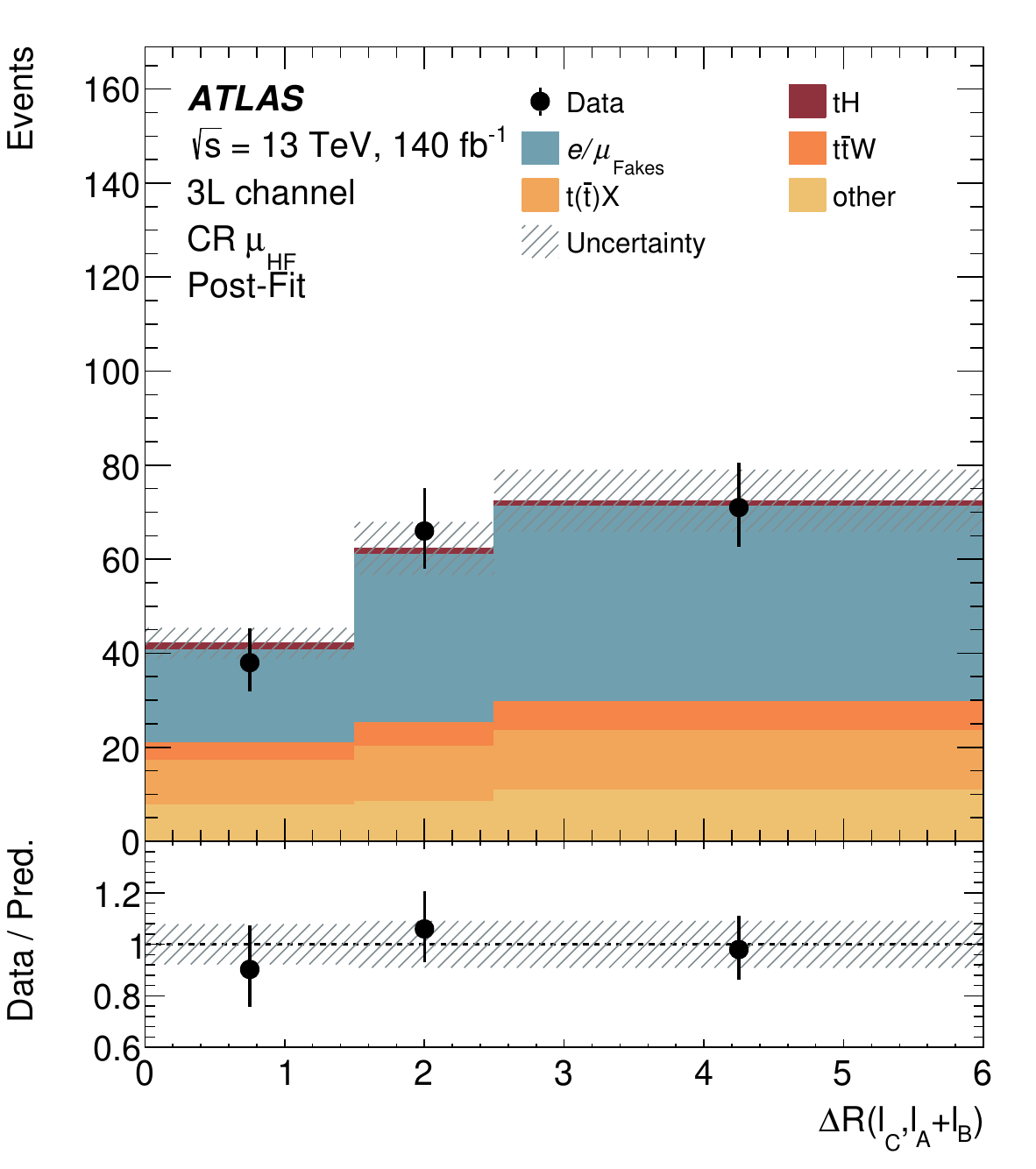}}
\subfloat[]{\includegraphics[width=0.33\textwidth]{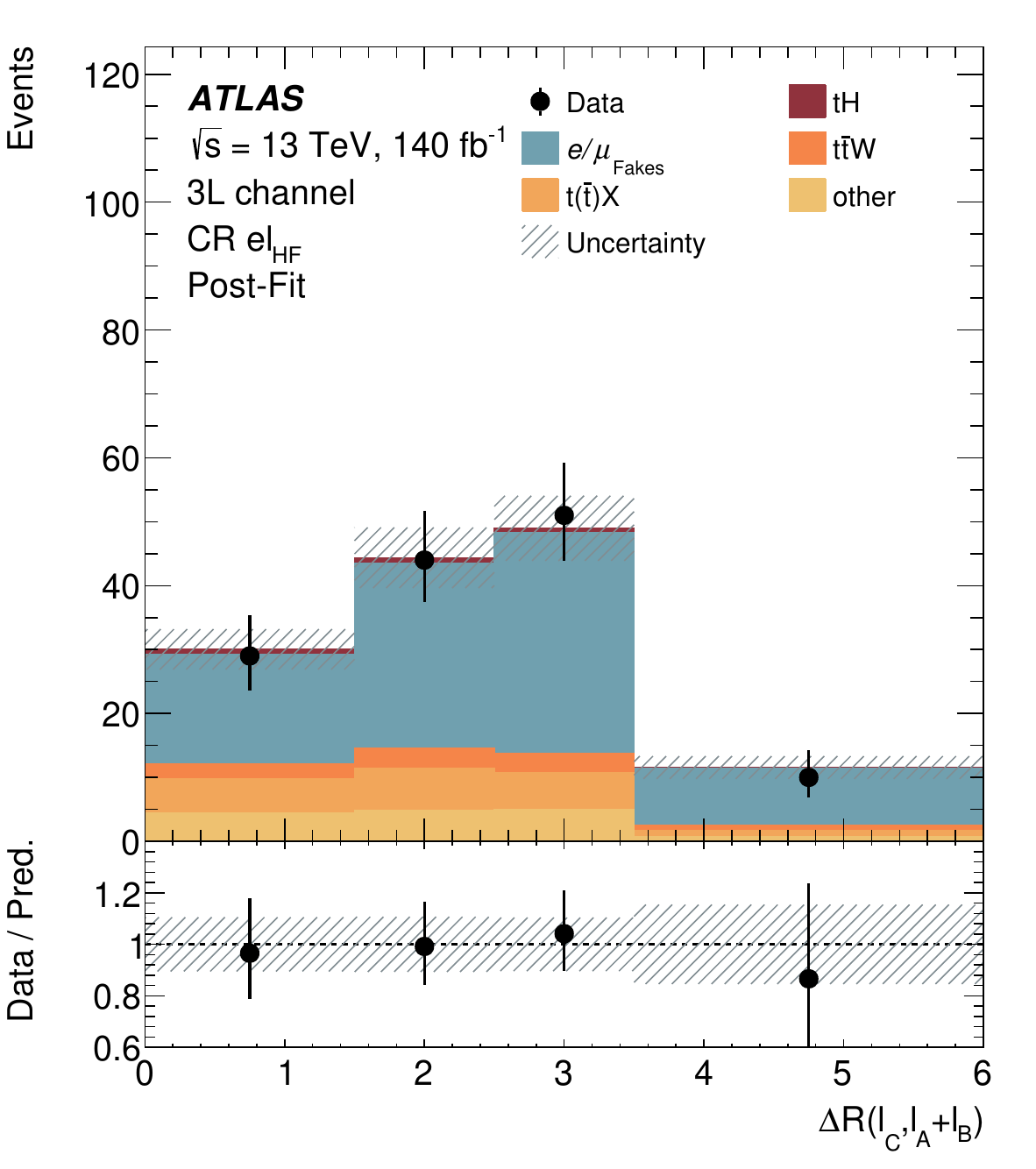}}
\subfloat[]{\includegraphics[width=0.33\textwidth]{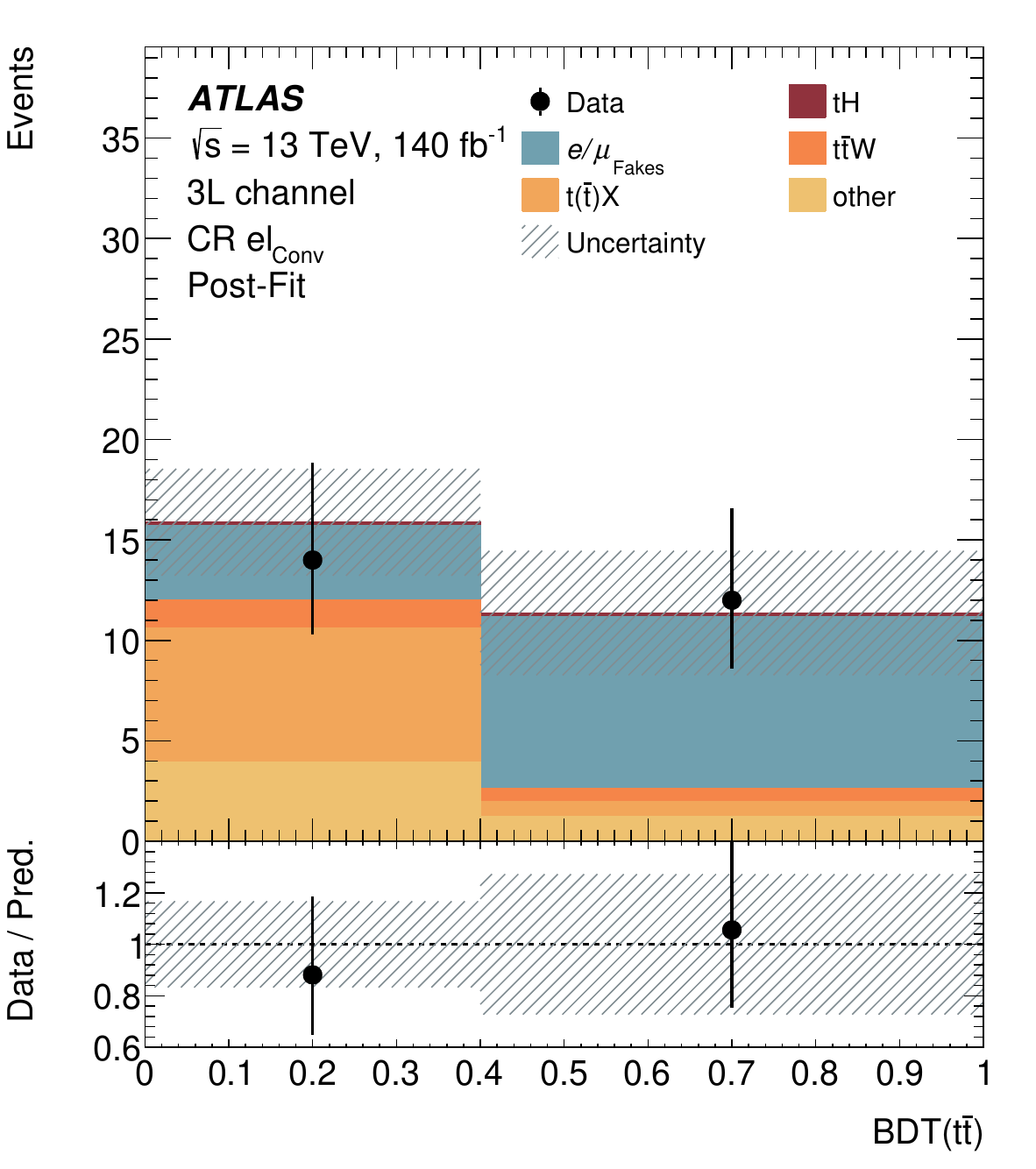}} \\
\caption{Comparison between data and the signal-plus-background prediction for the event yields in the nine CRs.
The term $e/\mu_\text{Fakes}$ refers to backgrounds from non-prompt or misidentified leptons, namely contributions from heavy-flavour decays and photon conversions.
The signal and background contributions after the likelihood fit to data (\enquote{Post-Fit}) under the signal-plus-background hypothesis are shown as filled histograms.
The ratio of the data to the total prediction (\enquote{Pred.}) is shown in the lower panel. %
The combined statistical and systematic uncertainty in the prediction is indicated by the grey hatched band. }
\label{fig:summary_CR}
\end{figure}

The observed and expected yields in the three SRs and the nine CRs after the combined likelihood fit under the signal-plus-background hypothesis are compared with data in \cref{fig:summary_SR,fig:summary_CR}.
Detailed tables showing pre-fit and post-fit yields for all regions are reported in~\cref{app:yields}.
A good agreement between data and prediction is found in all the regions.

Additional measurements are performed using the ITC hypothesis ($\ITC$) as signal model.
The best-fit value of the \tH signal strength under the ITC hypothesis is:
\begin{equation*}
\mutHresultITC\,.
\label{eq:mu_tH_IC}
\end{equation*}

\begin{figure}[t!]
\centering
\subfloat[]{\includegraphics[width=0.33\textwidth]{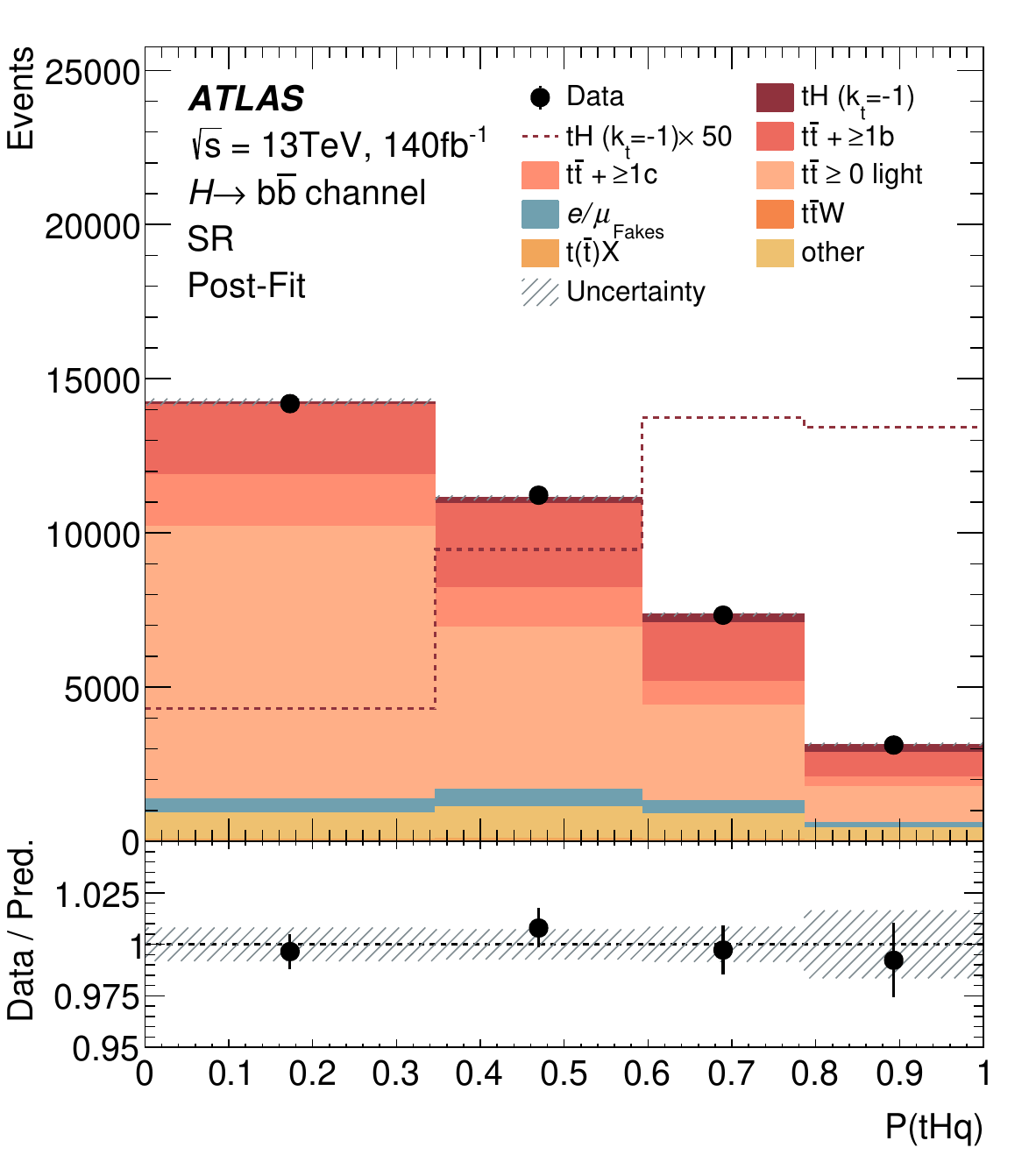}}
\subfloat[]{\includegraphics[width=0.33\textwidth]{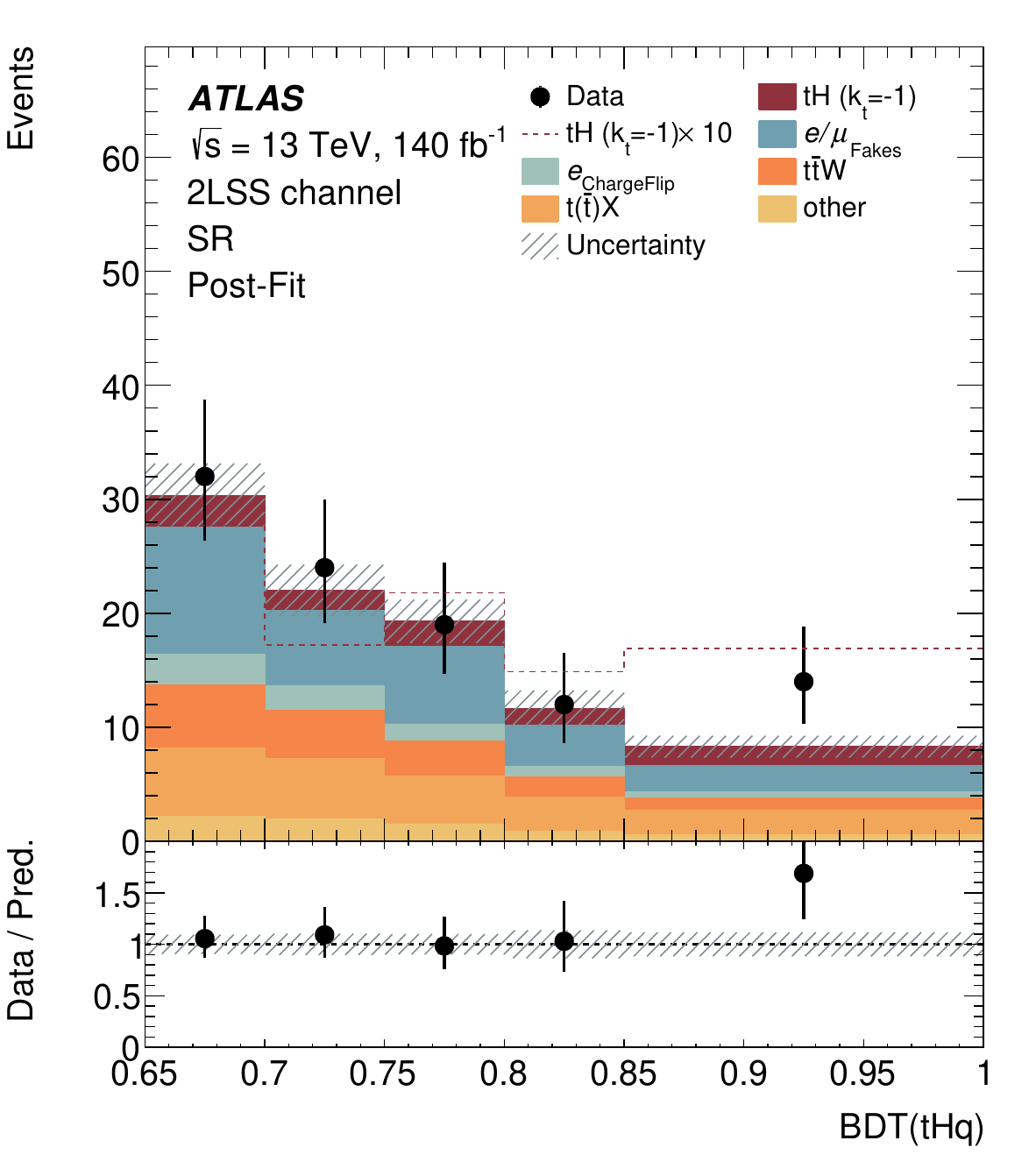}}
\subfloat[]{\includegraphics[width=0.33\textwidth]{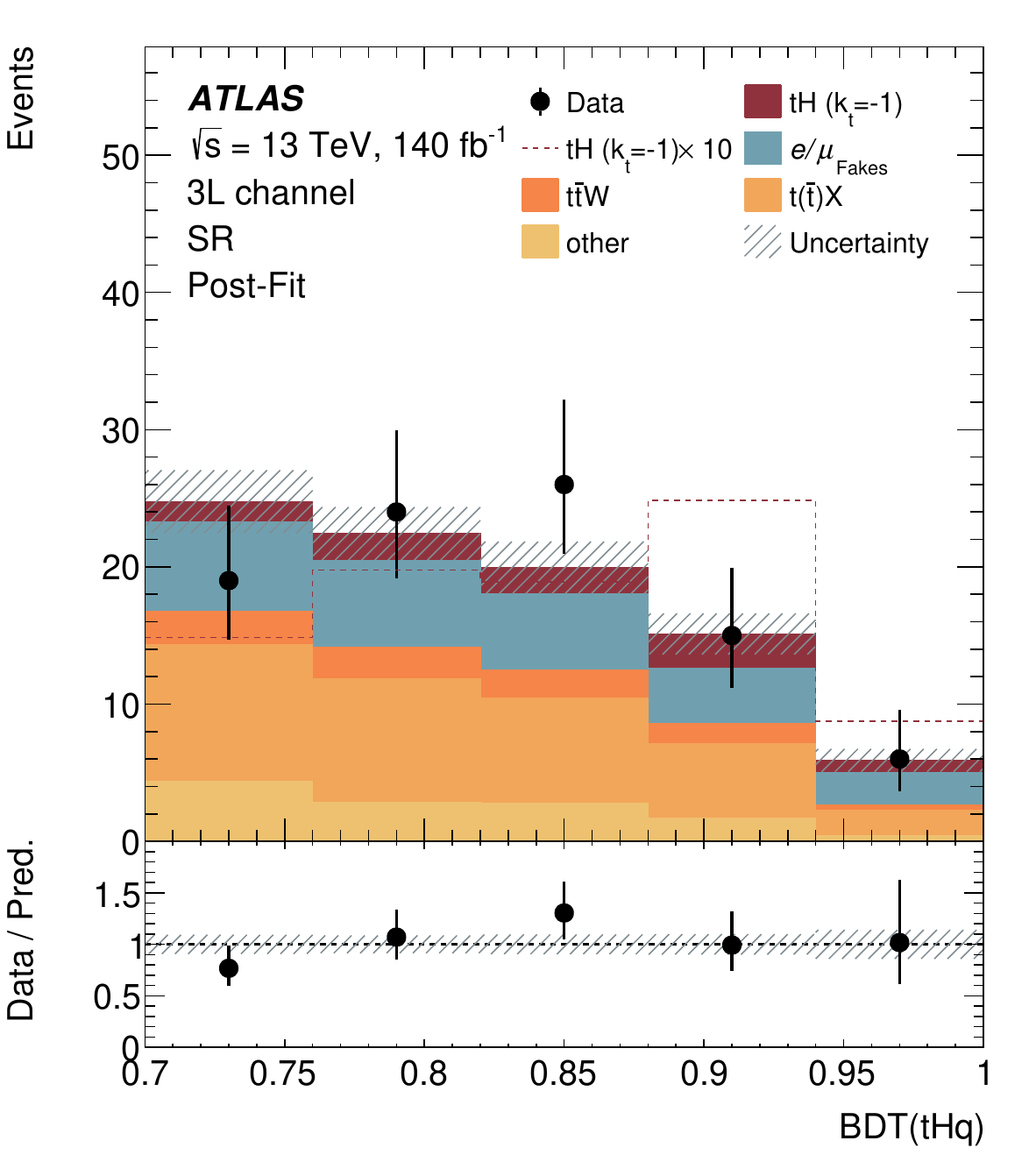}}
\caption{Comparison between data and the signal-plus-background prediction for the event yields in the (a) \onelep, (b) \dilepSS and (c) \trilep SRs in the ITC hypothesis.
The term $e/\mu_\text{Fakes}$ refers to backgrounds from non-prompt or misidentified leptons, namely contributions from heavy-flavour decays and photon conversions.
The ITC signal and the background contributions after the likelihood fit to data (\enquote{Post-Fit}) under the signal-plus-background hypothesis are shown as filled histograms.
A dashed red line is added to the plot to show the signal contribution normalised to the measured value multiplied by 50 for the \onelep channel and by 10 for the \dilepSS and \trilep channels.
The ratio of the data to the total post-fit prediction (\enquote{Pred.}) is shown in the lower panel.
The combined statistical and systematic uncertainty in the prediction is indicated by the grey hatched band.
}
\label{fig:summary_SR_ITC}
\end{figure}

The observed and expected yields in the three SRs after the combined likelihood fit under the signal-plus-background hypothesis are compared with data in \cref{fig:summary_SR_ITC} for the ITC hypothesis.

\begin{figure}[htbp]
\centering
\subfloat[\label{fig:IC_poi}]{\includegraphics[TRExPOITrim, height=0.29\textheight]{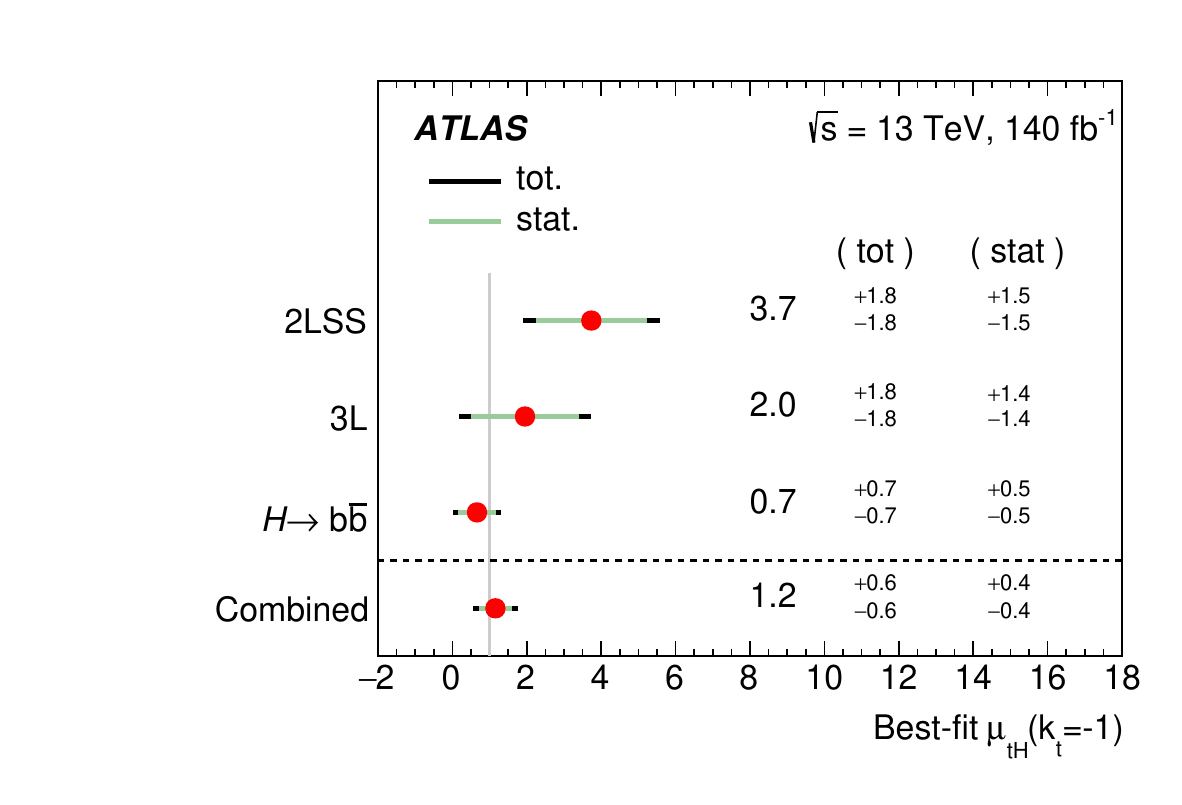}}
\subfloat[\label{fig:IC_limit}]{\includegraphics[TRExNPTrim, height=0.29\textheight]{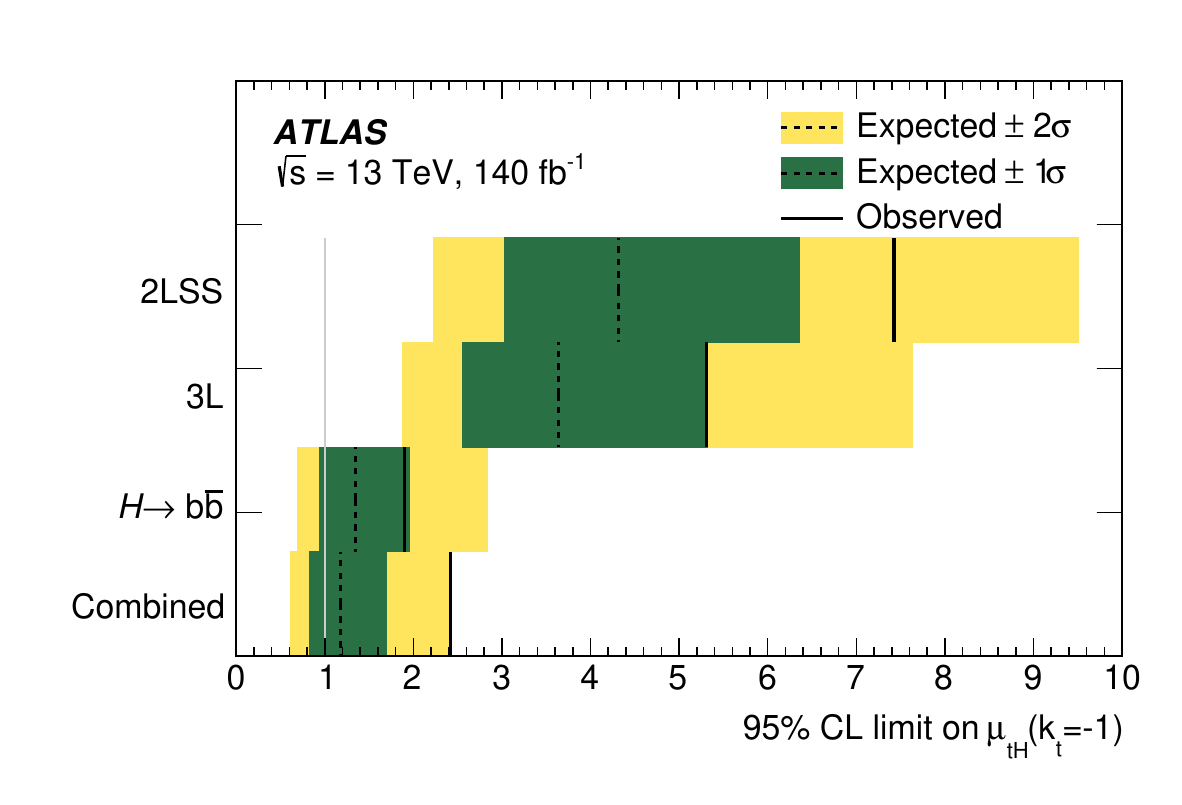}}
\caption{Fitted values of (a) $\mu_{\tH}({\ITC})$ and (b) the 95\% CL upper limits, shown for the individual channels and the combined measurement.
The values reported for the individual channels are obtained by fitting $\mu_{\tH}({\ITC})$ separately for each channel, whereas the combined measurement is derived by simultaneously fitting all channels together.
The observed limits are shown (solid black lines), together with the expected limits  under the ITC hypothesis (dotted black lines).
A grey line is added in correspondence of the $\mu_{\tH}({\ITC})$ equal to one.
In the case of the expected limits, the one- and two-standard-deviation uncertainty bands are also shown.}
\label{fig:IC_poi_limit}
\end{figure}

The measured inverted-coupling signal strengths in the individual channels and the combined result are summarised in~\cref{fig:IC_poi}.
The corresponding observed (expected) 95\% CL exclusion limit is \upperLimitObsITC~(\upperLimitExpITC), which is shown in~\cref{fig:IC_limit} for both the individual channels and the combined measurement.
The goodness-of-fit test probability for the $\ITC$ scenario fit is 95$\%$.
In contrast to the SM case, under the ITC hypothesis the channel with the highest sensitivity is the \onelep channel, overtaking the multi-lepton channels despite its modest signal-to-background ratio in the SM case (see~\cref{app:yields}). This inversion arises from the strong enhancement of the \tHq cross-section together with the discriminating power of key variables in the \onelep channel, such as the
the number of jets and number of jets associated with the reconstructed hadronic top quark from different \ttbar reconstruction techniques. By contrast, variables used in training the BDT(\tHq) classifiers in the \dilepSS and \trilep channels, such as invariant masses or angular distances of leptons, show a similar behaviour as the background under the $\kappa_t=-1$ hypothesis, leading to a loss of signal-to-background discrimination power.
The uncertainty on the measured signal strength for the ITC hypothesis is dominated by systematic sources.
In particular, the absolute uncertainty on $\mu_{\tH}(\ITC)$ from the theoretical uncertainty in the \ttb category is 0.25, the JES and JER together contribute 0.20, the uncertainty from the signal theoretical prediction is 0.18, and that from the modelling of \ttbar in the \dilepSS and \trilep channels is 0.17.

%

%
%
\FloatBarrier

%

%
\section{Conclusion}%
\label{sec:conclusion}
The search for the production of a Higgs boson in association with a
single top quark (\tH) is presented.
The analysis is based on \textit{pp} collision data at $\SI{13}{\TeV}$ with an integrated luminosity of \lumi,
recorded from $2015$ to $2018$ with the ATLAS detector at the LHC.
The signal regions are split into final states with one, two same-sign or three isolated leptons (electrons or muons) probing the \Hbb, \HWW, \HZZ and \Htautau decays of the Higgs boson.
Additional background-enriched regions are used in the fit to improve the modelling of several leading backgrounds, such as \ttb, \ttW and three different categories of the fake and non-prompt leptons background.
The \tH signal strength is found to be \mutHresult~in the SM hypothesis.
The measured value of the SM cross-section is $\sigma(\tH) = \measxsectH\,\mathrm{fb}$.
The main source of uncertainty for this measurement is the one associated with the available experimental data, therefore a better constrain can be obtained analysing a larger event sample.
Despite the analysis being optimised to set the best \tH limit under the SM hypothesis, it is possible to also test the ITC hypothesis.
The signal strength value in this case is found to be \mutHresultITC.
The excess in data favours the ITC hypothesis, but both hypotheses are compatible with the data.


%
\section*{Acknowledgements}
%

%
%

%
%

We thank CERN for the very successful operation of the LHC and its injectors, as well as the support staff at
CERN and at our institutions worldwide without whom ATLAS could not be operated efficiently.

The crucial computing support from all WLCG partners is acknowledged gratefully, in particular from CERN, the ATLAS Tier-1 facilities at TRIUMF/SFU (Canada), NDGF (Denmark, Norway, Sweden), CC-IN2P3 (France), KIT/GridKA (Germany), INFN-CNAF (Italy), NL-T1 (Netherlands), PIC (Spain), RAL (UK) and BNL (USA), the Tier-2 facilities worldwide and large non-WLCG resource providers. Major contributors of computing resources are listed in Ref.~\cite{ATL-SOFT-PUB-2025-001}.

We gratefully acknowledge the support of ANPCyT, Argentina; YerPhI, Armenia; ARC, Australia; BMWFW and FWF, Austria; ANAS, Azerbaijan; CNPq and FAPESP, Brazil; NSERC, NRC and CFI, Canada; CERN; ANID, Chile; CAS, MOST and NSFC, China; Minciencias, Colombia; MEYS CR, Czech Republic; DNRF and DNSRC, Denmark; IN2P3-CNRS and CEA-DRF/IRFU, France; SRNSFG, Georgia; BMFTR, HGF and MPG, Germany; GSRI, Greece; RGC and Hong Kong SAR, China; ICHEP and Academy of Sciences and Humanities, Israel; INFN, Italy; MEXT and JSPS, Japan; CNRST, Morocco; NWO, Netherlands; RCN, Norway; MNiSW, Poland; FCT, Portugal; MNE/IFA, Romania; MSTDI, Serbia; MSSR, Slovakia; ARIS and MVZI, Slovenia; DSI/NRF, South Africa; MICIU/AEI, Spain; SRC and Wallenberg Foundation, Sweden; SERI, SNSF and Cantons of Bern and Geneva, Switzerland; NSTC, Taipei; TENMAK, T\"urkiye; STFC/UKRI, United Kingdom; DOE and NSF, United States of America.

Individual groups and members have received support from BCKDF, CANARIE, CRC and DRAC, Canada; CERN-CZ, FORTE and PRIMUS, Czech Republic; COST, ERC, ERDF, Horizon 2020, ICSC-NextGenerationEU and Marie Sk{\l}odowska-Curie Actions, European Union; Investissements d'Avenir Labex, Investissements d'Avenir Idex and ANR, France; DFG and AvH Foundation, Germany; Herakleitos, Thales and Aristeia programmes co-financed by EU-ESF and the Greek NSRF, Greece; BSF-NSF and MINERVA, Israel; NCN and NAWA, Poland; La Caixa Banking Foundation, CERCA Programme Generalitat de Catalunya and PROMETEO and GenT Programmes Generalitat Valenciana, Spain; G\"{o}ran Gustafssons Stiftelse, Sweden; The Royal Society and Leverhulme Trust, United Kingdom.

In addition, individual members wish to acknowledge support from CERN: European Organization for Nuclear Research (CERN DOCT); Chile: Agencia Nacional de Investigaci\'on y Desarrollo (FONDECYT 1230812, FONDECYT 1240864, Fondecyt 3240661); China: Chinese Ministry of Science and Technology (MOST-2023YFA1605700, MOST-2023YFA1609300), National Natural Science Foundation of China (NSFC - 12175119, NSFC 12275265); Czech Republic: Czech Science Foundation (GACR - 24-11373S), Ministry of Education Youth and Sports (ERC-CZ-LL2327, FORTE CZ.02.01.01/00/22\_008/0004632), PRIMUS Research Programme (PRIMUS/21/SCI/017); EU: H2020 European Research Council (ERC - 101002463); European Union: European Research Council (BARD No. 101116429, ERC - 948254, ERC 101089007), European Regional Development Fund (SMASH COFUND 101081355, SLO ERDF), Horizon 2020 Framework Programme (MUCCA - CHIST-ERA-19-XAI-00), European Union, Future Artificial Intelligence Research (FAIR-NextGenerationEU PE00000013), Italian Center for High Performance Computing, Big Data and Quantum Computing (ICSC, NextGenerationEU); France: Agence Nationale de la Recherche (ANR-21-CE31-0022, ANR-22-EDIR-0002); Germany: Baden-Württemberg Stiftung (BW Stiftung-Postdoc Eliteprogramme), Deutsche Forschungsgemeinschaft (DFG - 469666862, DFG - CR 312/5-2); China: Research Grants Council (GRF); Italy: Istituto Nazionale di Fisica Nucleare (ICSC, NextGenerationEU), Ministero dell'Università e della Ricerca (NextGenEU 153D23001490006 M4C2.1.1, NextGenEU I53D23000820006 M4C2.1.1, NextGenEU I53D23001490006 M4C2.1.1, SOE2024\_0000023); Japan: Japan Society for the Promotion of Science (JSPS KAKENHI JP22H01227, JSPS KAKENHI JP22H04944, JSPS KAKENHI JP22KK0227, JSPS KAKENHI JP24K23939, JSPS KAKENHI JP24KK0251, JSPS KAKENHI JP25H00650, JSPS KAKENHI JP25H01291, JSPS KAKENHI JP25K01023); Norway: Research Council of Norway (RCN-314472); Poland: Ministry of Science and Higher Education (IDUB AGH, POB8, D4 no 9722), Polish National Science Centre (NCN 2021/42/E/ST2/00350, NCN OPUS 2023/51/B/ST2/02507, NCN OPUS nr 2022/47/B/ST2/03059, NCN UMO-2019/34/E/ST2/00393, UMO-2022/47/O/ST2/00148, UMO-2023/49/B/ST2/04085, UMO-2023/51/B/ST2/00920, UMO-2024/53/N/ST2/00869); Portugal: Foundation for Science and Technology (FCT); Spain: Ministry of Science and Innovation (MCIN \& NextGenEU PCI2022-135018-2, MICIN \& FEDER PID2021-125273NB, RYC2019-028510-I, RYC2020-030254-I, RYC2021-031273-I, RYC2022-038164-I); Sweden: Carl Trygger Foundation (Carl Trygger Foundation CTS 22:2312), Swedish Research Council (Swedish Research Council 2023-04654, VR 2021-03651, VR 2022-03845, VR 2022-04683, VR 2023-03403, VR 2024-05451), Knut and Alice Wallenberg Foundation (KAW 2018.0458, KAW 2022.0358, KAW 2023.0366); Switzerland: Swiss National Science Foundation (SNSF - PCEFP2\_194658); United Kingdom: Royal Society (NIF-R1-231091); United States of America: U.S. Department of Energy (ECA DE-AC02-76SF00515), Neubauer Family Foundation.

%
%


%
\clearpage
\appendix
\part*{Appendix}
\addcontentsline{toc}{part}{Appendix}

\section{List of input variables used to train the different BDTGs}
\label{appendix}

For the \dilepSS and \trilep channels, leptons are ordered in three different ways:
\begin{itemize}
\item \emph{Transverse momentum} ordering: the lepton with the highest \pT is called $\ell_A$, the one with the second highest \pT is called $\ell_B$
while the lepton with the softest \pT is called $\ell_C$;
\item \emph{Charge and $\Delta R$ between leptons} ordering: the lepton with the charge opposite to the sum of the three lepton charges is called $\ell_0$,
the lepton with the smallest $\Delta R$ to $\ell_0$ is called $\ell_1$,
the remaining lepton is called $\ell_2$;
\item \emph{Charge and $\Delta R$ with leading \bjet} ordering: the lepton with the charge opposite to the sum of the three lepton charges is called $\hat{\ell}_0$,
the lepton with the smallest $\Delta R$ with the leading \bjet is called $\hat{\ell}_1$ while the remaining lepton is called $\hat{\ell}_2$.
\end{itemize}
The jet satisfying the 70$\%$ \btag requirement and having the highest \pT is called leading \bjet.
The jet that fails the \btag requirement and maximises the invariant mass with the leading \bjet is called spectator jet, \js.
Every jet having $|\eta|>2.5$ is called forward jet, \jf.
A forward jet is defined as leading if it has the highest \pT among all the forward jets.
Every jet with $|\eta|<2.5$ is called central jet.
Using the aforementioned leptons and jet definitions, several variables are constructed.

\begin{landscape}
\tablesetup
\begin{longtable}{l | c | c c c c | c c c | p{7.5cm}}
\caption{List of variables considered in the training of the BDTGs in each channel. The $\times$ symbol marks in which BDTGs the variable is used.
Variables labelled with ``custom'' are better explained in the description column.}
\label{tab:variables}\\
\toprule
\multirow{2}{*}{Variable} & \multirow{2}{*}{\onelep} & \multicolumn{4}{c|}{\dilepSS}                      & \multicolumn{3}{c|}{\trilep}        & \multirow{2}{*}{Description} \\
&                         & {\scriptsize BDT(\ttbar)} & {\scriptsize BDT(VV)} & {\scriptsize BDT(\ttW)} & {\scriptsize BDT(\tHq)} & {\scriptsize BDT(\tHq)} & {\scriptsize BDT(\ttbar)} & {\scriptsize BDT(\ttW)} &  \\
\midrule\endfirsthead
\caption{List of variables considered in the training of the BDTs in each channel. The $\times$ symbol marks in which BDTs the variable is used.
Variables labelled with ``custom'' are better explained in the description column.} \\
\toprule
\multirow{2}{*}{Variable} & \multirow{2}{*}{\onelep} & \multicolumn{4}{c|}{\dilepSS}                      & \multicolumn{3}{c|}{\trilep}        & \multirow{2}{*}{Description} \\
&                         & {\scriptsize BDT(\ttbar)} & {\scriptsize BDT(VV)} & {\scriptsize BDT(\ttW)} & {\scriptsize BDT(\tHq)} & {\scriptsize BDT(\tHq)} & {\scriptsize BDT(\ttbar)} & {\scriptsize BDT(\ttW)} &  \\
\midrule\endhead
$p^{*}_{\ell_A}(\bjet)$ & -   & - & - & - & -                           & - & - & $\times$     & \multirow[t]{3}{7.5cm}{Momentum of lepton with respect to the leading \bjet rest frame.} \\
$p^{*}_{\ell_B}(\bjet)$ & -   & - & - & - & -                           & $\times$ & - & -     & \\
$p^{*}_{\ell_C}(\bjet)$ & -   & - & - & - & -                           & - & - & $\times$     & \\
\midrule
\pTX[][\text{min}]      & -   & $\times$  & - & $\times$  & $\times$    & - & - & -            & \pT of softest lepton.\\
\midrule
$\Delta R(\ell_A,\ell_B+\ell_C)$ & -  & - & - & - & -    & - & - & $\times$     & \multirow[t]{2}{7.5cm}{$\Delta R$ between lepton and the system of the other two leptons.} \\
$\Delta R(\ell_B,\ell_A+\ell_C)$ & -  & - & - & - & -    & - & - & $\times$     & \\
\midrule
$\Delta R(\ell_A,\bjet)$ & -  & - & - & - & -            & $\times$ & - & $\times$     & \multirow[t]{2}{7.5cm}{$\Delta R$ between lepton and leading \bjet.} \\
$\Delta R(\ell_B,\bjet)$ & -  & - & - & - & -            & - & - & $\times$            & \\
\midrule
$\Delta R(\text{non-}\bjet,\ell_1)$   & -   & - & - & - & $\times$            & - & - & -      & \multirow[t]{6}{7.5cm}{$\Delta R$ or $\Delta\eta$ between lepton and its closest \bjet or non-\bjet.}\\
$\Delta R(\text{non-}\bjet,\ell_2)$   & -   & - & - & - & $\times$            & - & - & -      & \\
$\Delta R(\ell_1,\bjet)$              & -   & $\times$ & $\times$ & - & -     & - & - & -      & \\
$\Delta R(\ell_2,\bjet)$              & -   & $\times$ & $\times$ & - & -     & - & - & -      & \\
$\Delta\eta(\ell_1,\bjet)$            & -   & $\times$ & $\times$ & - & -     & - & - & -      & \\
$\Delta\eta(\ell_2,\bjet)$            & -   & $\times$ & $\times$ & - & -     & - & - & -      & \\
\midrule
$m(\ell,\ell)_{\SS}$          & -   & - & - & - & -                  & $\times$ & $\times$ & $\times$     & \multirow[t]{4}{7.5cm}{Invariant mass, $\Delta R$, $\Delta \phi$ or $\Delta\eta$ of same-sign leptons.} \\
$\Delta\eta(\ell,\ell)_{\SS}$ & -   & - & - & - & $\times$           & $\times$ & - & -                   & \\
$\Delta\phi(\ell,\ell)_{\SS}$ & -   & - & - & $\times$  & -          & - & - & -                          & \\
$\Delta R(\ell,\ell)_{\SS}$   & -   & $\times$ & - & $\times$ & -    & - & - & -                          & \\
\midrule
$m(\ell_0,\ell_2)$        & -   & - & - & - & -                                 & $\times$ & $\times$ & -            & \multirow[t]{13}{7.5cm}{Invariant mass and $\Delta R$ of leptons, jets and/or leading \bjet. The lepton ordering is charge and $\Delta R$ based.} \\
$m(\ell_0,\ell_1)$        & -   & - & - & - & -                                 & $\times$ & $\times$ & $\times$     & \\
$m(\ell_1,\bjet)$         & -   & - & - & - & -                                 & $\times$ & - & $\times$            & \\
$m(\ell_2,\bjet)$         & -   & - & - & - & -                                 & $\times$ & $\times$ & -            & \\
$\Delta R(\ell_0,\ell_1)$ & -   & - & - & - & -                                 & - & - & $\times$                   & \\
$m(\ell_2,\text{j}_2)$    & -   & $\times$ & $\times$ & $\times$ & $\times$     & - & - & -                          & \\
$m(\ell_2,\text{j}_1)$    & -   & $\times$ & $\times$ & - & $\times$            & - & - & -                          & \\
$m(\ell_1,\text{j}_2)$    & -   & - & - & - & $\times$                          & - & - & -                          & \\
$m(\ell_1+\ell_2,\text{j}_2)$             & -   & - & - & $\times$ & -                        & - & - & -                          & \\
$m(\ell_1+\ell_2,\text{j}_1+\text{j}_2)$  & -   & - & - & $\times$ & $\times$                 & - & - & -                          & \\
$m(\ell_1+\ell_2,\text{j}_1)$             & -   & $\times$ & - & - & -                        & - & - & -                          & \\
$m(\text{j}_1+\text{j}_2)$                & -   & - & $\times$ & $\times$ & $\times$          & - & - & -                          & \\
$m(\ell_1+\ell_2)$                        & -   & $\times$ & $\times$ & $\times$ & $\times$   & - & - & -                          & \\
\midrule
$m_{H}(WW)$                             & -   & $\times$ & - & $\times$ & $\times$              & - & - & -       & Mass of a candidate Higgs Boson when decays into a pair of W bosons.\\
$m(\text{jet},\text{jet})_{W}$          & -   &  - & $\times$ & $\times$ & $\times$             & - & - & -       & Mass of the best two jet candidates for the W boson.\\
$m(\text{jet},\text{jet})_{\text{top}}$ & -   & $\times$ & $\times$ & $\times$ & $\times$       & - & - & -       & Mass of the best top-quark candidate from jets.\\
\midrule
$m(\hat{\ell}_0,\hat{\ell}_1)$ & -  & - & - & - & -    & - & $\times$ & -     & Invariant mass of $\hat{\ell}_0$ and $\hat{\ell}_1$.\\
\midrule
$m(\hat{\ell}_0,\hat{\ell}_2)$& -         & - & - & - & -            & $\times$ & - & $\times$     & \multirow[t]{2}{7.5cm}{Invariant mass and $\Delta R$ of $\hat{\ell}_0$ and $\hat{\ell}_2$.} \\
$\Delta R(\hat{\ell}_0,\hat{\ell}_2)$& -  & - & - & - & -            & - & - & $\times$            & \\
\midrule
$m(\hat{\ell}_2,\bjet)$        & -   & - & - & - & -            & $\times$ & - & $\times$     & \multirow[t]{2}{7.5cm}{Invariant mass and $\Delta R$ of leading \bjet and $\hat{\ell}_1$ and $\hat{\ell}_2$.} \\
$\Delta R(\hat{\ell}_2,\bjet)$ & -   & - & - & - & -            & - & $\times$ & -            & \\
\midrule
$m(\ell,\bjet)_{\text{top}}$   & -   & - & - & - & -            & - & $\times$ & $\times$     & Invariant mass of lepton and \bjet giving best top-quark visible mass.\\
\midrule
$m(\ell,\ell)_{\text{top}}$    & -   & - & - & - & -            & $\times$ & $\times$ & -     & Invariant mass of two leptons giving best top-quark visible mass.\\
\midrule
$p_{\text{FW}1}$        & -         & - & - & $\times$ & -           & - & - & -                          & First Fox--Wolfram moment~\cite{Fox:1978vu,Bernaciak:2012nh}. \\
$p_{\text{FW}4}$        & -         & $\times$ & $\times$ & - & -    & - & - & $\times$                   & Fourth Fox--Wolfram moment. \\
$\pTX[][\text{min}]$    & -         & - & - & - & -                  & $\times$ & $\times$ & $\times$     & \pT of the softest lepton.\\
$m(\js)$                & -         & - & - & - & $\times$           & - & - & -                          & Mass of the spectator jet.\\
$E(\js)$                & -         & - & - & - & $\times$           & - & - & -                          & Energy of the spectator jet.\\
$\pT(\js)$            & -         & - & - & - & $\times$           & - & - & -                          & \pT of the spectator jet.\\
$\MET/\HT$              & -         & - & $\times$ & - & -           & - & - & $\times$                   & Ratio between \MET and \HT.\\
$\text{N}(\text{jet})$         & $\times$  & - & - & - & -                  & - & - & -                   & Number of non-\btagged jets.\\
$\text{N}(\text{non-}\bjet)$   & -         & - & - & - & -                  & - & - & $\times$                   & Number of non-\btagged jets.\\
$\text{N}(\text{central-jet})$ & -         & - & - & - & -                  & $\times$ & $\times$ & $\times$     & Number of central jets.\\
$\text{N}(\bjet)$              & -         & - & - & - & -                  & - & $\times$ & $\times$            & Number of \btagged jets.\\
$\Delta m(Z,\ell\ell_{\SF})_{\text{min}}$ & -   & $\times$ & $\times$ & $\times$ & -            & - & $\times$ & $\times$            & Minimum difference between the reconstructed invariant mass of two leptons with the same flavour and \PZ-boson mass.\\
$\Delta m(Z,\ell\ell)_{\text{min}}$       & -   & - & - & - & -                                 & $\times$ & $\times$ & $\times$     & Minimum difference between the reconstructed invariant mass of two leptons and \PZ-boson mass.\\
\MET                                      & -   & $\times$ & $\times$ & $\times$ & $\times$     & $\times$ & $\times$ & $\times$     & \MET.\\
$\Sigma_i \text{charge}(\ell_i)$          & -   & $\times$ & $\times$ & $\times$ & $\times$     & - & - & -                          & Sum of lepton charges.\\
\midrule
$m(\bjet,\js)$          & -             & - & - & $\times$ & $\times$                   & $\times$ & - & -                   & \multirow[t]{3}{7.5cm}{Invariant mass, $\Delta\phi$, $\Delta\eta$ and $\Delta R$ of leading \bjet and spectator jet.} \\
$\Delta\phi(\bjet,\js)$ & -             & - & - & - & -                                 & $\times$ & - & -                   & \\
$\Delta\eta(\bjet,\js)$ & -             & $\times$ & - & - & -                          & - & - & -                          & \\
$\Delta R(\bjet,\js)$   & -             & - & - & $\times$ & $\times$                   & - & - & $\times$                   & \\
\midrule
$\Delta R(\ell,\jf)$    & -             & $\times$ & - & - & -      & - & - & -     & \multirow[t]{2}{7.5cm}{$\Delta R$  and $\Delta\eta$ between the leading forward jet and its closest lepton.} \\
$\Delta\eta(\ell,\jf)$  & -             & - & - & - & $\times$      & - & - & -     & \\
\midrule
$\Sigma_i q(\ell_i)$    & -             & - & - & - & -                                 & $\times$ & $\times$ & $\times$     & Sum of lepton charges.\\
Lepton flavours         & -             & $\times$ & - & - & -                & - & - & -         & Identification of the flavour of the leptons.\\
$id(\ell_0)$            & $\times$      & - & - & - & -     & - & - & -    & identification number of the lepton combining charge and flavour information. \\
$e\mu$-events           & -             & $\times$ & $\times$ & - & -         & - & - & -         & Identification of the events $e\mu$.\\
$ee$-events             & -             & - & $\times$ & $\times$ & $\times$  & - & - & -         & Identification of the events $ee$.\\
\HT                     & -             & - & $\times$ & - & $\times$                   & $\times$ & $\times$ & $\times$     & \HT.\\
$\HT(\ell)$         & -             & $\times$ & - & $\times$ & $\times$            & - & - & -                          & Sum of the transverse momentum of the leptons.\\
$\HT(\text{jets})$      & -             & $\times$ & - & $\times$ & -                   & - & - & -                          & Sum of the transverse momentum of the jets.\\
$\tilde{m}_{lep}$       & -             & $\times$ & $\times$ & $\times$ & $\times$     & $\times$ & - & -                   & $\sqrt{\Sigma_i (E_{\ell_i}^2+p_{T,\ell_i}^2})$.\\
$\Delta R(\ell,\ell)_{\text{min}}$          & -   & - & - & - & -           & - & - & $\times$                    & Minimum $\Delta R$ between two leptons.\\
$\Delta\phi(\phi_{T}^{\text{miss}},\ell_2)$ & -   & - & - & $\times$ & -    & - & - & -                           & $\Delta \phi$ between $\phi_{T}^{\text{miss}}$ and $\ell_2$.\\
\midrule
``custom'' $\chi^2$     & -   & $\times$ & $\times$ & $\times$ & -       & - & - & -       & $\chi^2(m(\ell_i,\ell_j)_{\text{top}})_{\text{min}}$, minimum $\chi^2$ value between the masses of top-quark candidates given by the combination of leptons.\\
$m(\ell,\ell)_{\text{top}}$                           & -   & $\times$ & $\times$ & $\times$ & -       & - & - & -       & Invariant mass of two leptons giving best top-quark visible mass.\\
\midrule
$\eta(\text{No-}\bjet)_{\text{max}}$ & -     & - & - & $\times$ & -   & - & - & -                          & Maximum $\eta$ value among non-\btagged jets.\\
$\Delta\eta(\ell_2,\text{non-}\bjet)$& -     & - & - & - & -          & - & $\times$ & -                   & $\Delta \eta$ between $\ell_2$ and closest non-\btagged jets.\\
\midrule
$\Delta\eta(\ell,\jf)$ & -    & - & - & - & -     & $\times$ & - & -                  & $\Delta \eta$ between leading forward jet and closest lepton.\\
\midrule
$\text{N}(\text{non-}\bjet)$   & -    & - & $\times$ & - & $\times$                   & - & - & -     & Number of non \btagged jets.\\
$\text{N}(\text{central-jet})$ & -    & $\times$ & $\times$ & $\times$ & $\times$     & - & - & -     & Number of central jets.\\
$\text{N}(\bjet)$              & -    & $\times$ & $\times$ & $\times$ & -            & - & - & -     & Number of \btagged jets.\\
\midrule
$b\text{-score}_1$ & -      & $\times$ & $\times$ & $\times$ & $\times$     & $\times$ & $\times$ & $\times$    & Binned DL1r score (calibrated) of leading \bjet.\\
$b\text{-score}_2$ & -      & $\times$ & $\times$ & $\times$ & -            & - & - & $\times$                  & Binned DL1r score (calibrated) of second \bjet.\\
$b\text{-score}_3$ & -      & $\times$ & $\times$ & $\times$ & -            & - & - & $\times$                  & Binned DL1r score (calibrated) of third \bjet.\\
\midrule
``custom'' $n_{j}$           & $\times$        & - & - & - & -     & - & - & -    & $n_{j,t_{\text{had}}}(\chi^2_{\text{min, ttl}}, \text{PCBT-bin } 4)$, number of jets that belong to the hadronic top-quark found from reconstructing \ttbar events but allowing only Hypothesis $I$  within the 4$^{\text{th}}$ PCBT bin. \\
``custom'' $n_{j}$           & $\times$        & - & - & - & -     & - & - & -    & $n_{j,t_{\text{had}}}(\chi^2_{\text{min, ttAll}}, \text{PCBT-bin } 4)$, number of jets that belong to the hadronic top-quark found from reconstructing \ttbar events  within the 4$^{\text{th}}$ PCBT bin. \\
``custom'' $n_{j}$           & $\times$        & - & - & - & -     & - & - & -    & $n_{j, t_{\text{had}}}(\chi^2_{\text{min, ttAll}}, \text{PCBT-bin }\in\{1,2,3\})$, number of jets that belong to the hadronic top-quark found from reconstructing \ttbar events  within the 1$^{\text{st}}$, 2$^{\text{nd}}$ or 3$^{\text{rd}}$ PCBT bin. \\
$p_{\text{T}}^{\text{fwd},0}$         & $\times$        & - & - & - & -     & - & - & -    &  Transverse momentum of the leading forward-jet. \\
Sphericity                   & $\times$        & - & - & - & -     & - & - & -    &  as defined in~Ref.~\cite{STDM-2011-33}.  \\
$\chi^2_{\text{min,ttAll}}$                                                                  & $\times$        & - & - & - & -     & - & - & -    & The minimum $\chi^2$ value found from reconstructing \ttbar events.   \\
``custom'' $n_{j}$           & $\times$        & - & - & - & -     & - & - & -    & $n_{j, \text{not-}t_{\text{had}}}(\chi^2_{\text{min, ttl}}, \text{PCBT-bin}\in\{1,2,3\})$, number of jets that do not belong to the hadronic top-quark found from reconstructing \ttbar events but allowing only Hypothesis $I$  within the 1$^{\text{st}}$, 2$^{\text{nd}}$ or 3$^{\text{rd}}$ PCBT bin.     \\
$p_{\text{T}}^{\text{light},1}$                                & $\times$        & - & - & - & -     & - & - & -    & Transverse momentum of the 2$^{\text{nd}}$ leading jet failing to meet the $b$-tagging requirement. \\
``custom'' $\chi^2$          & $\times$        & - & - & - & -     & - & - & -    & $\chi^2_{\text{top-higgs}}(m_{H}=111.5,m_{t}=168)$, the minimum $\chi^2$ value found from reconstructing \tHbb events, where the leptonic $W$ from the top-quark decay is reconstructed using the neutrino reconstruction technique. \\
$\chi^2_{\text{min,ttl}}$                             & $\times$        & - & - & - & -     & - & - & -    & The minimum $\chi^2$ value found from reconstructing \ttbar events allowing only Hypothesis $I$.  \\
$\text{DL1r}_{\text{light},0}$                        & $\times$        & - & - & - & -     & - & - & -    & Pseudo-continuous $b$-tagging distribution for the leading jet failing to meet the b-tagging requirement. \\
$p_{\text{T}}^{\text{light},2}$                                & $\times$        & - & - & - & -     & - & - & -    & Transverse momentum of the 3$^{\text{rd}}$ leading jet failing to meet the $b$-tagging requirement. \\
$\eta_{\text{light},0}$                               & $\times$        & - & - & - & -     & - & - & -    & Pseudo-rapidity of the leading jet failing to meet the $b$-tagging requirement. \\
``custom'' $\chi^2$                                   & $\times$        & - & - & - & -     & - & - & -    & $\chi^2_{\text{top-higgs}}(m_{H}=113,m_{t}=165)$, the minimum $\chi^2$ value found from reconstructing \tHbb events, where the leptonic $W$ from the top-quark decay is reconstructed using the neutrino reconstruction technique. \\
$p_{\text{T}}^{\text{light},0}$                                                 & $\times$        & - & - & - & -     & - & - & -    & Transverse momentum of the leading jet failing to meet the b-tagging requirement. \\
$\Delta\eta_{\text{top-higgs}}$                                        & $\times$        & - & - & - & -     & - & - & -    & The $\Delta\eta$ between the reconstructed top quark and the reconstructed Higgs boson according to the minimum $\chi^2$ method. \\
``custom'' $n_{j}$                                                     & $\times$        & - & - & - & -     & - & - & -    & $n_{j,t_{\text{had}}}(\chi^2_{\text{min, ttl}}, \text{PCBT-bin } 0)$, number of jets that belong to the hadronic top-quark found from reconstructing \ttbar events but allowing only Hypothesis $I$  within the 0$^{\text{th}}$ PCBT bin. \\
$m_{t_{\text{had}}}(\chi^2_{\text{min, ttAll}})$                       & $\times$        & - & - & - & -     & - & - & -    & The mass of the reconstructed hadronic top-quark found from reconstructing \ttbar events.  \\
$m_{t_{\text{had}}}(\chi^2_{\text{min, ttl}})$                         & $\times$        & - & - & - & -     & - & - & -    & The mass of the reconstructed hadronic top-quark found from reconstructing \ttbar events but allowing only Hypothesis $I$.  \\
$\eta_{\text{light},2}$                                                & $\times$        & - & - & - & -     & - & - & -    & Pseudo-rapidity of the 3$^{\text{rd}}$ leading jet failing the $b$-tagging requirement. \\
$\Delta R(q^{W}_{1},q^{W}_{2})$                                        & $\times$        & - & - & - & -     & - & - & -    & The minimum $\Delta R$ between the two jets assigned to the hadronic $W$ decay from reconstructing \ttbar events.  \\
$m(b_{\text{top}}, j_{\text{tag}})$                                    & $\times$        & - & - & - & -     & - & - & -    & The invariant mass  of the $b$-jet from top-quark decay found using a reconstruction of \tHbb~ events and the light-flavour jet most de-correlated from the top-quark system. The light jet is chosen such that it maximises the invariant mass $m(b_{\text{top}}, j_{\text{tag}})$. \\
$\eta_{\text{tag}}$                                                    & $\times$        & - & - & - & -     & - & - & -    & The $\eta$  of the light-flavour jet, $\eta_{\text{tag}}$ most de-correlated from the top-quark system. The light jet is chosen such that it maximises the invariant mass $m(b_{\text{top}}, j_{\text{tag}})$. \\
\end{longtable}
\end{landscape}


\section{Pre-fit and post-fit yields tables}
\label{app:yields}
Pre-fit and post-fit yields of the analysis regions are reported in~\cref{tab:onelep_yields} for the \onelep channel.
\Cref{tab:dilepSS_yields,tab:dilepSS_postyields} show the pre- and post-fit yields for the \dilepSS channel, respectively.
\Cref{tab:trilep_yields,tab:trilep_postyields} show the pre- and post-fit yields for the \trilep channel, respectively.

\begin{table}[htbp]
\centering
{
\tablesetup
\newcolumntype{C}{>{\centering\arraybackslash}S[table-format=6.4]@{\(\,\pm\,\)}S[table-format=3.4]}
\caption{Pre-fit and post-fit event yields in the SR and CR(\ttb) of the \onelep channel. Post-fit yields are after the inclusive fit in all channels. All uncertainties are included, taking into account correlations in the post-fit case.}
\label{tab:onelep_yields}
\begin{adjustbox}{max width=0.999\textwidth}
\begin{tabular}{l|
*{1}{r@{\(\,\pm\,\)}l }
*{1}{r@{\(\,\pm\,\)}l |}
*{1}{r@{\(\,\pm\,\)}l }
*{1}{r@{\(\,\pm\,\)}l }}

\toprule
\multicolumn{9}{c}{\onelep channel} \\
\midrule
& \multicolumn{4}{c|}{Pre-fit} & \multicolumn{4}{c}{Post-fit} \\
Process & \multicolumn{2}{c}{SR} & \multicolumn{2}{c|}{CR(\ttb)} & \multicolumn{2}{c}{SR} & \multicolumn{2}{c}{CR(\ttb)}\\
\midrule
\tHq   & 43 & 7 & 9.1 & 2.0 & 350 & 130 & 72 & 29 \\
\tWH   & 3.4 & 2.0 & 14 & 8 & 28 & 20 & 110 & 80 \\
\ttb   & 6800 & 1100 & 20100 & 2600 & 8100 & 1000 & 25600 & 1900 \\
\ttc   & 4000 & 2400 & 11000 & 6000 & 3900 & 1500 & 10200 & 3300 \\
\ttlight & 19300 & 3300 & 23100 & 3500 & 18500 & 1100 & 23300 & 2100 \\
$e/\mu_{\text{Fakes}}$   & 1700 & 500 & 1600 & 500 & 1700 & 500 & 1600 & 500 \\
\ttW & 20.1 & 3.0 & 134 & 19 & 20.2 & 2.9 & 133 & 19 \\
\ttH & 120 & 15 & 683 & 70 & 120 & 15 & 683 & 70 \\
\ttZ & 68 & 9 & 340 & 40 & 68 & 9 & 340 & 40 \\
\tZq & 69 & 7 & 20.4 & 1.9 & 69 & 6 & 20.4 & 1.9 \\
\tWZ & 0.20 & 0.10 & 0.9 & 0.5 & 0.20 & 0.11 & 0.9 & 0.5 \\
$Wt$ channel & 970 & 260 & 1500 & 500 & 970 & 250 & 1600 & 500 \\
$t$-channel  & 1070 & 180 & 350 & 120 & 1060 & 160 & 350 & 120 \\
$s$-channel  & 43 & 8 & 40 & 11 & 42 & 7 & 39 & 10 \\
$W$ + jets   & 740 & 310 & 900 & 400 & 750 & 300 & 890 & 350 \\
$Z$ + jets   & 140 & 50 & 100 & 40 & 140 & 50 & 100 & 40 \\
$VV$         & 48 & 25 & 70 & 40 & 41 & 22 & 63 & 33 \\
other Higgs  & 6 & 5 & 5.3 & 2.9 & 7 & 5 & 5.2 & 2.8 \\
Rare top     & 0.08 & 0.04 & 3.3 & 1.7 & 0.08 & 0.04 & 3.4 & 1.7 \\
\midrule
Total  & 35000 & 5000 & 60000 & 8000 & 35870 & 210 & 65000 & 340 \\
Data   & \multicolumn{2}{c}{35869} & \multicolumn{2}{c|}{65002} & \multicolumn{2}{c}{35869} & \multicolumn{2}{c}{65002}\\
\bottomrule
\end{tabular}
\end{adjustbox}
}
\end{table}


\begin{table}[htbp]
\centering
{
\tablesetup
\newcolumntype{C}{>{\centering\arraybackslash}S[table-format=6.4]@{\(\,\pm\,\)}S[table-format=3.4]}
\caption{Pre-fit event yields in the SR and CRs of the \dilepSS channel. The `others` process includes \tWZ, Tri-boson, s-channel, 3-tops and VH processes.}
\label{tab:dilepSS_yields}
\begin{adjustbox}{max width=0.999\textwidth}
\begin{tabular}{l|
*{5}{r@{\(\,\pm\,\)}l }}

\toprule
\multicolumn{11}{c}{Pre-fit \dilepSS channel}\\
\midrule
Process & \multicolumn{2}{c}{SR} & \multicolumn{2}{c}{CR(\muHF)} & \multicolumn{2}{c}{CR(\elHF)} & \multicolumn{2}{c}{CR(\elCo)} & \multicolumn{2}{c}{CR(\ttW)} \\
\midrule
\tHq   & 3.1 & 0.5 & 0.65 & 0.11 & 2.5 & 0.4 & 0.085 & 0.024 & 0.30 & 0.06 \\
\tWH   & 0.09 & 0.04 & 0.20 & 0.04 & 0.89 & 0.13 & 0.016 & 0.006 & 0.39 & 0.05 \\
\muHF   & 16 & 5 & 59 & 11 & 19 & 4 & 0.24 & 0.24 & 3.0 & 2.0 \\
\elHF   & 4.9 & 3.1 & \multicolumn{2}{c}{-} & 86 & 17 & 1.3 & 0.8 & 0.9 & 0.8 \\
\elCo   & 3.4 & 2.7 & \multicolumn{2}{c}{-} & 80 & 15 & 47 & 15 & 2.8 & 1.4 \\
other mis-id leptons   & 4 & 4 & 15 & 12 & 41 & 28 & 3.0 & 3.0 & 3.7 & 2.6 \\
$e_\text{ChargeFlip}$   & 8.3 & 1.9 & \multicolumn{2}{c}{-} & 220 & 50 & 37 & 11 & 4.4 & 1.0 \\
\ttW   & 14.9 & 2.1 & 35 & 4 & 160 & 8 & 5.0 & 0.4 & 131 & 5 \\
\ttZ   & 3.7 & 0.6 & 8.0 & 1.4 & 38 & 6 & 1.12 & 0.21 & 15.3 & 2.2 \\
\ttH   & 2.4 & 0.4 & 7.2 & 1.0 & 30 & 4 & 0.77 & 0.12 & 13.5 & 2.0 \\
\tZq   & 14.2 & 1.9 & 3.9 & 0.5 & 22.0 & 2.0 & 0.68 & 0.09 & 1.05 & 0.18 \\
$VV$ & 7.3 & 3.3 & 9 & 5 & 50 & 15 & 1.8 & 1.5 & 7.2 & 3.1 \\
others & 0.45 & 0.13 & 1.1 & 0.4 & 7.9 & 2.1 & 0.28 & 0.10 & 2.0 & 0.5 \\
4-tops & 0.042 & 0.014 & 0.050 & 0.015 & 0.54 & 0.15 & 0.016 & 0.007 & 0.83 & 0.24 \\
\midrule
Total  & 83 & 10 & 138 & 16 & 760 & 80 & 98 & 21 & 187 & 8 \\
Data   & \multicolumn{2}{c}{101} & \multicolumn{2}{c}{137} & \multicolumn{2}{c}{756} & \multicolumn{2}{c}{110} & \multicolumn{2}{c}{200} \\
\bottomrule
\end{tabular}
\end{adjustbox}
}
\end{table}


\begin{table}[htbp]
\centering
{
\tablesetup
\newcolumntype{C}{>{\centering\arraybackslash}S[table-format=6.4]@{\(\,\pm\,\)}S[table-format=3.4]}
\caption{Post-fit event yields in the SR and CRs of the \dilepSS channel. Post-fit yields are after the inclusive fit in all channels. All uncertainties are included, taking into account correlations in the post-fit case. The `others` process includes \tWZ, Tri-boson, s-channel, 3-tops and VH processes.}
\label{tab:dilepSS_postyields}
\begin{adjustbox}{max width=0.999\textwidth}
\begin{tabular}{l|
*{5}{r@{\(\,\pm\,\)}l }}

\toprule
\multicolumn{11}{c}{Post-fit \dilepSS channel}\\
\midrule
Process & \multicolumn{2}{c}{SR} & \multicolumn{2}{c}{CR(\muHF)} & \multicolumn{2}{c}{CR(\elHF)} & \multicolumn{2}{c}{CR(\elCo)} & \multicolumn{2}{c}{CR(\ttW)} \\
\midrule
\tHq   & 25 & 9 & 5.3 & 2.1 & 21 & 8 & 0.67 & 0.30 & 2.4 & 0.9 \\
\tWH   & 0.7 & 0.4 & 1.6 & 0.7 & 7.3 & 3.2 & 0.13 & 0.07 & 3.2 & 1.4 \\
\muHF   & 13 & 6 & 47 & 18 & 15 & 7 & 0.22 & 0.17 & 2.9 & 1.9 \\
\elHF   & 2.6 & 3.4 & \multicolumn{2}{c}{-} & 50 & 60 & 0.7 & 0.9 & 0.5 & 0.7 \\
\elCo   & 3.8 & 3.1 & \multicolumn{2}{c}{-} & 98 & 35 & 58 & 15 & 3.4 & 1.6 \\
other mis-id leptons  & 4 & 4 & 15 & 11 & 38 & 25 & 3.3 & 2.8 & 3.2 & 2.1 \\
$e_\text{ChargeFlip}$   & 8.0 & 1.8 & \multicolumn{2}{c}{-} & 210 & 50 & 36 & 10 & 4.2 & 1.0 \\
\ttW   & 15.0 & 2.4 & 41 & 5 & 178 & 17 & 5.7 & 0.6 & 144 & 13 \\
\ttZ   & 3.7 & 0.6 & 7.9 & 1.3 & 37 & 6 & 1.07 & 0.20 & 15.1 & 2.1 \\
\ttH   & 2.4 & 0.4 & 7.3 & 0.9 & 30 & 4 & 0.78 & 0.11 & 13.5 & 1.9 \\
\tZq   & 14.1 & 1.9 & 3.9 & 0.5 & 22.1 & 2.0 & 0.68 & 0.09 & 1.03 & 0.17 \\
$VV$   & 6.0 & 2.4 & 7.1 & 3.5 & 45 & 13 & 2.0 & 1.6 & 6.0 & 2.4 \\
others   & 0.44 & 0.13 & 1.1 & 0.4 & 7.9 & 2.1 & 0.28 & 0.11 & 1.9 & 0.5 \\
4-tops  & 0.043 & 0.014 & 0.050 & 0.015 & 0.55 & 0.15 & 0.017 & 0.007 & 0.85 & 0.24 \\
\midrule
Total  & 99 & 8 & 138 & 11 & 762 & 27 & 109 & 10 & 202 & 12 \\
Data   & \multicolumn{2}{c}{101} & \multicolumn{2}{c}{137} & \multicolumn{2}{c}{756} & \multicolumn{2}{c}{110} & \multicolumn{2}{c}{200} \\
\bottomrule
\end{tabular}
\end{adjustbox}
}
\end{table}


\begin{table}[htbp]
\centering
{
\tablesetup
\newcolumntype{C}{>{\centering\arraybackslash}S[table-format=6.4]@{\(\,\pm\,\)}S[table-format=3.4]}
\caption{Pre-fit event yields in the SR and CRs of the \trilep channel. The `others` process includes gluon--gluon fusion and vector-boson fusion Higgs boson production, Tri-boson, s-channel and t-channel single-top, 3-tops and VH processes.}
\label{tab:trilep_yields}
\begin{adjustbox}{max width=0.999\textwidth}
\begin{tabular}{l|
*{5}{r@{\(\,\pm\,\)}l }}

\toprule
\multicolumn{11}{c}{Pre-fit \trilep channel}\\
\midrule
Process & \multicolumn{2}{c}{SR} & \multicolumn{2}{c}{CR(\muHF)} & \multicolumn{2}{c}{CR(\elHF)} & \multicolumn{2}{c}{CR(\elCo)} & \multicolumn{2}{c}{CR(\ttW)} \\
\midrule
\tHq   & 1.45 & 0.24 & 0.24 & 0.06 & 0.159 & 0.030 & 0.025 & 0.007 & 0.26 & 0.06 \\
\tWH   & 0.29 & 0.04 & 0.24 & 0.04 & 0.150 & 0.023 & 0.019 & 0.004 & 0.64 & 0.07 \\
\muHF   & 9 & 4 & 84 & 11 & 3.2 & 1.8 & 0.08 & 0.11 & 10 & 7 \\
\elHF   & 5.3 & 2.3 & 2.7 & 2.2 & 47 & 9 & 0.16 & 0.04 & 6.9 & 3.0 \\
\elCo   & 4.2 & 1.8 & 3.5 & 3.2 & 14 & 6 & 14 & 9 & 11 & 8 \\
other mis-id leptons   & 4.9 & 3.0 & 14 & 10 & 8 & 5 & 1.0 & 1.0 & 10 & 8 \\
\ttW   & 7.9 & 0.9 & 13.6 & 0.6 & 8.1 & 0.8 & 1.97 & 0.33 & 118 & 4 \\
\tWZ   & 0.8 & 0.5 & 1.4 & 0.7 & 0.8 & 0.5 & 0.6 & 0.4 & 3.5 & 1.9 \\
\ttZ   & 14.3 & 2.1 & 15.6 & 2.4 & 8.2 & 1.3 & 4.9 & 1.1 & 37 & 5 \\
\ttH   & 7.7 & 1.0 & 7.6 & 1.0 & 4.7 & 0.7 & 0.51 & 0.16 & 21.0 & 2.8 \\
\tZq   & 10.8 & 1.2 & 9.2 & 1.1 & 4.8 & 0.6 & 1.8 & 0.4 & 6.0 & 0.7 \\
tW   & 1.1 & 0.7 & 6 & 4 & 3.6 & 2.5 & 0.9 & 0.7 & 1.6 & 1.2 \\
$VV$   & 12 & 5 & 22 & 7 & 11.0 & 3.5 & 4.8 & 1.7 & 13 & 5 \\
others & 0.24 & 0.08 & 0.44 & 0.12 & 0.16 & 0.05 & 0.040 & 0.012 & 1.0 & 0.6 \\
4-tops & 0.14 & 0.04 & 0.15 & 0.05 & 0.085 & 0.026 & 0.037 & 0.013 & 1.7 & 0.5 \\
\midrule
Total & 80 & 11 & 181 & 17 & 113 & 15 & 31 & 10 & 242 & 14 \\
Data   & \multicolumn{2}{c}{90} & \multicolumn{2}{c}{175} & \multicolumn{2}{c}{134} & \multicolumn{2}{c}{26} & \multicolumn{2}{c}{268} \\
\bottomrule
\end{tabular}
\end{adjustbox}
}
\end{table}


\begin{table}[htbp]
\centering
{
\tablesetup
\newcolumntype{C}{>{\centering\arraybackslash}S[table-format=6.4]@{\(\,\pm\,\)}S[table-format=3.4]}
\caption{Post-fit event yields in the SR and CRs of the \dilepSS channel. Post-fit yields are after the inclusive fit in all channels. All uncertainties are included, taking into account correlations in the post-fit case. The `others` process includes gluon--gluon fusion and vector-boson fusion Higgs boson production, Tri-boson, s-channel and t-channel single-top, 3-tops and VH processes.}
\label{tab:trilep_postyields}
\begin{adjustbox}{max width=0.999\textwidth}
\begin{tabular}{l|
*{5}{r@{\(\,\pm\,\)}l }}

\toprule
\multicolumn{11}{c}{Post-fit \trilep channel}\\
\midrule
Process & \multicolumn{2}{c}{SR} & \multicolumn{2}{c}{CR(\muHF)} & \multicolumn{2}{c}{CR(\elHF)} & \multicolumn{2}{c}{CR(\elCo)} & \multicolumn{2}{c}{CR(\ttW)} \\
\midrule
\tHq   & 12 & 4 & 2.0 & 0.9 & 1.3 & 0.5 & 0.20 & 0.09 & 2.2 & 0.9 \\
\tWH   & 2.3 & 0.9 & 2.0 & 0.9 & 1.2 & 0.5 & 0.15 & 0.07 & 5.2 & 2.2 \\
\muHF   & 8 & 4 & 77 & 18 & 3.0 & 1.5 & 0.07 & 0.09 & 9 & 6 \\
\elHF   & 8 & 4 & 3.8 & 2.9 & 68 & 14 & 0.22 & 0.08 & 10 & 6 \\
\elCo   & 3.1 & 2.1 & 3 & 4 & 11 & 6 & 11 & 5 & 9 & 11 \\
other mis-id leptons   & 3.8 & 2.2 & 13 & 8 & 7 & 4 & 1.0 & 0.9 & 9 & 7 \\
\ttW   & 8.3 & 1.3 & 15.0 & 1.6 & 9.3 & 1.2 & 2.0 & 0.4 & 131 & 12 \\
\tWZ   & 0.9 & 0.5 & 1.5 & 0.7 & 0.9 & 0.5 & 0.62 & 0.33 & 3.6 & 1.9 \\
\ttZ   & 14.2 & 2.0 & 15.5 & 2.3 & 8.4 & 1.3 & 4.6 & 0.9 & 38 & 5 \\
\ttH   & 7.7 & 1.0 & 7.6 & 1.0 & 4.7 & 0.7 & 0.48 & 0.14 & 21.4 & 2.8 \\
\tZq   & 10.7 & 1.2 & 9.3 & 1.1 & 4.9 & 0.6 & 1.74 & 0.32 & 6.0 & 0.7 \\
tW   & 1.1 & 0.7 & 5 & 4 & 4.0 & 2.7 & 0.9 & 0.7 & 1.4 & 1.1 \\
$VV$   & 10 & 4 & 21 & 6 & 10.6 & 3.4 & 4.2 & 1.3 & 11 & 4 \\
others   & 0.23 & 0.08 & 0.43 & 0.12 & 0.15 & 0.05 & 0.037 & 0.011 & 0.9 & 0.4 \\
4-tops   & 0.14 & 0.04 & 0.15 & 0.05 & 0.087 & 0.026 & 0.035 & 0.012 & 1.7 & 0.5 \\
\midrule
Total & 90 & 7 & 177 & 13 & 135 & 11 & 27 & 5 & 260 & 11 \\
Data   & \multicolumn{2}{c}{90} & \multicolumn{2}{c}{175} & \multicolumn{2}{c}{134} & \multicolumn{2}{c}{26} & \multicolumn{2}{c}{268} \\
\bottomrule
\end{tabular}
\end{adjustbox}
}
\end{table}


\newpage


%

%
%

%
%
%
\printbibliography
\clearpage
 
\begin{flushleft}
\hypersetup{urlcolor=black}
{\Large The ATLAS Collaboration}

\bigskip

\AtlasOrcid[0000-0002-6665-4934]{G.~Aad}$^\textrm{\scriptsize 104}$,
\AtlasOrcid[0000-0001-7616-1554]{E.~Aakvaag}$^\textrm{\scriptsize 17}$,
\AtlasOrcid[0000-0002-5888-2734]{B.~Abbott}$^\textrm{\scriptsize 123}$,
\AtlasOrcid[0000-0002-0287-5869]{S.~Abdelhameed}$^\textrm{\scriptsize 119a}$,
\AtlasOrcid[0000-0002-1002-1652]{K.~Abeling}$^\textrm{\scriptsize 55}$,
\AtlasOrcid[0000-0001-5763-2760]{N.J.~Abicht}$^\textrm{\scriptsize 49}$,
\AtlasOrcid[0000-0002-8496-9294]{S.H.~Abidi}$^\textrm{\scriptsize 30}$,
\AtlasOrcid[0009-0003-6578-220X]{M.~Aboelela}$^\textrm{\scriptsize 45}$,
\AtlasOrcid[0000-0002-9987-2292]{A.~Aboulhorma}$^\textrm{\scriptsize 36e}$,
\AtlasOrcid[0000-0001-5329-6640]{H.~Abramowicz}$^\textrm{\scriptsize 157}$
% ...
\bigskip
\\

% ...
$^{2}$Department of Physics, University of Alberta, Edmonton AB; Canada.\\
% ...
$^{17}$Department for Physics and Technology, University of Bergen, Bergen; Norway.\\
% ...
$^{30}$Physics Department, Brookhaven National Laboratory, Upton NY; United States of America.\\
% ...
$^{36}$$^{(a)}$Facult\'e des Sciences Ain Chock, Universit\'e Hassan II de Casablanca;$^{(b)}$Facult\'{e} des Sciences, Universit\'{e} Ibn-Tofail, K\'{e}nitra;$^{(c)}$Facult\'e des Sciences Semlalia, Universit\'e Cadi Ayyad, LPHEA-Marrakech;$^{(d)}$LPMR, Facult\'e des Sciences, Universit\'e Mohamed Premier, Oujda;$^{(e)}$Facult\'e des sciences, Universit\'e Mohammed V, Rabat;$^{(f)}$Institute of Applied Physics, Mohammed VI Polytechnic University, Ben Guerir; Morocco.\\
% ...
$^{45}$Physics Department, Southern Methodist University, Dallas TX; United States of America.\\
% ...
$^{49}$Fakult\"{a}t Physik , Technische Universit{\"a}t Dortmund, Dortmund; Germany.\\
% ...
$^{55}$II. Physikalisches Institut, Georg-August-Universit\"{a}t G\"ottingen, G\"ottingen; Germany.\\
% ...
$^{104}$CPPM, Aix-Marseille Universit\'e, CNRS/IN2P3, Marseille; France.\\
% ...
$^{119}$$^{(a)}$New York University Abu Dhabi, Abu Dhabi;$^{(b)}$United Arab Emirates University, Al Ain; United Arab Emirates.\\
% ...
$^{123}$Homer L. Dodge Department of Physics and Astronomy, University of Oklahoma, Norman OK; United States of America.\\
% ...
$^{157}$Raymond and Beverly Sackler School of Physics and Astronomy, Tel Aviv University, Tel Aviv; Israel.\\
% ...
$^{a}$ Also at Affiliated with an institute formerly covered by a cooperation agreement with CERN.\\
$^{b}$ Also at An-Najah National University, Nablus; Palestine.\\
$^{c}$ Also at Borough of Manhattan Community College, City University of New York, New York NY; United States of America.\\
$^{d}$ Also at Center for Interdisciplinary Research and Innovation (CIRI-AUTH), Thessaloniki; Greece.\\
$^{e}$ Also at Centre of Physics of the Universities of Minho and Porto (CF-UM-UP); Portugal.\\
$^{f}$ Also at CERN, Geneva; Switzerland.\\
% ...

\end{flushleft}


%

%
%
%
%

%
%
%
%

\end{document}